\documentclass[11pt]{article}
\pdfoutput = 1

\usepackage[utf8]{inputenc}
\usepackage{color,graphicx}
\usepackage{verbatim}
\usepackage{enumerate}
\usepackage{amssymb}
\usepackage{stmaryrd}
\usepackage{physics}
\usepackage{amsfonts}
\usepackage{cite}
\usepackage{array}
\usepackage{setspace}
\usepackage{float}
\usepackage{url}
\usepackage{mathtools}
\usepackage{bbold}
\usepackage{tikz}
\usepackage{graphicx}
\usepackage{wrapfig}
\usepackage{subfigure}
\usepackage[labelfont=bf,font={rm},font={small}]{caption}
\usepackage{tikz}
\usepackage{cite}
\usepackage{cancel}
\usepackage{amsmath}
\usepackage{nicematrix}
\usepackage{xcolor}
\allowdisplaybreaks
\usepackage[most]{tcolorbox}

\usepackage[T1]{fontenc} 

\usepackage{braket}
\usepackage{todonotes}
\usepackage{braket}
\usepackage{appendix}
\usepackage{float}
\usepackage{array}
\usepackage{comment}
\usepackage{xcolor}
\usepackage{amsmath}

\usepackage{amsmath, amssymb, graphics, setspace}
\usepackage{tcolorbox} 
\usepackage[margin = 2.2cm]{geometry}
\newcommand{\mathsym}[1]{{}}
\newcommand{\unicode}[1]{{}}

\usepackage[margin = 2.2cm]{geometry}
\usepackage[ragged]{footmisc}
\usepackage{hyperref}
\hypersetup{
    colorlinks,
    citecolor=blue,
    filecolor=black,
    linkcolor=blue,
    urlcolor=blue,
    linktocpage=true
}

\def\e{{\rm e}}

\newcommand{\be}{\begin{equation}}
\newcommand{\ee}{\end{equation}}
\newcommand{\bea}{\begin{align}}
\newcommand{\eea}{\end{align}}
\newcommand{\bi}{\begin{itemize}}
\newcommand{\ei}{\end{itemize}}

\def\XXint#1#2#3{{\setbox0=\hbox{$#1{#2#3}{\int}$}
     \vcenter{\hbox{$#2#3$}}\kern-.5\wd0}}

\numberwithin{equation}{section}

\title{Template}

\begin{document}

\thispagestyle{empty}
\begin{center}
\vspace*{.4cm}
 
    {\LARGE \bf 
Weil-Petersson volumes for extended JT  supergravity\\\bigskip from ordinary differential equations
  }
    
    \vspace{0.4in}
    {\bf Wasif Ahmed, Clifford V. Johnson, 
      Krishan Saraswat
    }

\bigskip\bigskip

    {
Physics Department, Broida Hall, University of California, Santa Barbara, CA 93106, USA}
    \vspace{0.1in}

    {\tt wasif@ucsb.edu}, 
    {\tt cliffordjohnson@ucsb.edu},  {\tt ksaraswat@ucsb.edu}
\end{center}

\vspace{0.4in}
\begin{abstract}
\noindent 
Recent work~\cite{Johnson:2024bue} produced an efficient method for computing Weil-Petersson volumes using two ordinary differential equations (ODEs)  that appear naturally in double scaled random matrix models. One is the defining string equation 
of the model and the other is the Gel'fand-Dikii equation satisfied by  the diagonal resolvent of an auxiliary  Hamiltonian used to compute 
correlators of macroscopic loops. In concert, when applied to Jackiw-Teitelboim gravity, the recursive expansion of these  two ODEs efficiently define, order by order in genus, the  Weil-Petersson volumes $V_{g,1}(b)$ for bordered hyperbolic Riemann surfaces with one geodesic boundary (length $b$) and  genus $g$. The method works equally well for both ordinary and ${\cal N}{=}1$ supersymmetric JT gravity cases. This paper explores the method to higher genus, verifying some conjectures of ref.~\cite{Johnson:2024bue}, and deriving several useful recursive formulae for general use. The method is then applied to the new examples furnished by recent matrix model definitions of JT supergravity with extended supersymmetry, and several example expressions for the volumes are derived, confirming and extending ${\cal N}{=}2$ examples of Turiaci and Witten, and furnishing new formulae for the cases with small and large ${\cal N}{=}4$ supersymmetry. The prospects for extending the ODE method to the full set of $V_{g,n}(\{ b_i\})$, ($i=1,\ldots ,n$), are discussed.

\end{abstract}
\pagebreak
\setcounter{page}{1}
\tableofcontents

\newpage

\section{Introduction}

\subsection{Background}

A wide class of random matrix models, after taking  the ``double scaling limit''\cite{Brezin:1990rb,Douglas:1990ve,Gross:1990vs},  capture  the physics  of Euclidean 2D quantum gravity, in the sense that (at least perturbatively) they compute the path integral over 2D surfaces of all geometries and topologies, weighted by an appropriate gravitational action. The matrix models typically start out being  of Hermitian matrices $H$ of size $N{\times} N$, with some  potential $N{\rm Tr}(V(H))$ which is polynomial in~$H$. Treating the defining partition function as a toy field theory, the 't Hooftian double-line (Feynman) diagrammatic expansion can be interpreted as a tesslation/discretization of all such surfaces, and in the large $N$ limit, the expansion in $1/N$ is a topological expansion~\cite{'tHooft:1973jz,Brezin:1978sv,Bessis:1980ss}. Diagrams that tesselate surfaces of Euler number $\chi{=}2{-}2g{-}b$ ({\it i.e.,} $g$ handles, $b$ boundaries) come with a factor $N^{-\chi}$.

The double scaling limit is the process of taking the large $N$ limit while also tuning parameters in the potential to approach critical values where universal physics representing the sum over {\it smooth} random 2D surfaces is extracted. 
Which  particular class of 2D surfaces is being summed over, and with what action,  depends upon the type of matrix model, and the potential $V(H)$.

There is also an holographic perspective on the role of the random matrix models: An analogy with higher dimensional gravity suggests that a complete 2D theory of gravity should be dual to a 1D non-gravitational theory. As such one might expect it to have a specific Hamiltonian~$H$. However, the Euclidean path integral  defines the gravity theory only through the dynamics of  geometry. As such it is  a coarse-grained definition, and indeed many different~$H$ are seen to be  good approximate microscopic candidates for the physics underlying a given macroscopic geometry. Instead, sampling from the statistical ensemble of Hamiltonians $H$ naturally gives smooth geometries. The potential $V(H)$  is now to be thought of as determining the probability of a draw of $H$ from the ensemble, and  the path integral over surfaces is dual to the statistical sampling, in the sense of Wigner~\cite{10.2307/1970079}, of  the spectral properties of the $H$s.

Given a particular $H$, one might compute its eigenvalues $\{ E_i\}$, and construct the usual canonical partition sum $Z(\beta)=\sum_i\e^{-\beta E_i}={\rm Tr}(\e^{-\beta H})$, for some inverse temperature $\beta$. Without a definite $H$ one computes instead the average of this across the ensemble. There is a very convenient and powerful language for computing this average within the matrix model, and the 't Hooftian diagrammatics shows that it is the expectation value of a loop of length $\beta$ in the sum over geometries. It is computed as follows~\cite{Banks:1990df}:
\begin{equation}
    \label{eq:basic-loop}
 \langle Z(\beta)\rangle\equiv   \langle {\rm Tr}(\e^{-\beta{ H}})\rangle =\int_{-\infty}^\mu \langle x|\e^{-\beta{\cal H}} |x\rangle \,\,dx\ ,
\end{equation}
where the constant $\mu$ will be defined below, and:
\begin{equation}
    \label{eq:schrodinger}
    {\cal H} = -\hbar^2\frac{\partial^2}{\partial x^2}+u(x)\ ,
\end{equation}
is the   Hamiltonian  of an auxiliary quantum mechanical system, defined on the real line $x{\in}\mathbb{R}$.\footnote{\label{fn:HvsH}The Hamiltonian ${\cal H}$ should not be confused with the ensemble of Hamiltonians, for which $H$ is a representative. Any particular $H$ has a discrete spectrum, while that of ${\cal H}$ is continuous. The quantum mechanics of ${\cal H}$ gives {\it statistical} information about the $H$s, so it is natural for it to know about all possible energies. This auxiliary system arises from re-writing the random matrix model in terms of an infinite family of orthogonal polynomials, $P_i(\lambda)$, $n\in\mathbb{Z}$, $N$ of which can be used as a basis for computing statistical quantities in the random matrix model. The polynomials are defined through a recursion relation of standard form, with  recursion coefficients ultimately determined by $V(H)$. (In the simplest Gaussian case, the polynomials are the Hermite polynomials, for example.) In the double scaling limit, the index $i$ becomes a continuous variable $x$, and the recursion coefficients become the function $u(x)$. Correspondingly,  the process of determining the orthogonal polynomials is the process of computing the spectrum of eigenfunctions of ${\cal H}$. Sums over the $N$ polynomials to build physical quantities in the random matrix model become integrals over $x$ from $-\infty$ to some value $\mu$, the precise value of which depends upon the model.}  The parameter $\hbar$ is the surviving part of the $1/N$ parameter.

 The function~$u(x)$, playing the role of a potential, is the solution to some ordinary differential equation called a ``string equation''. 
 In the bosonic case, the string  equation can be written as~\cite{Brezin:1990rb,Douglas:1990ve,Gross:1990vs}:
 \begin{equation}
 \label{eq:little-string-equation}
     {\cal R}\equiv\sum_{k=1}^\infty t_k R_k[u]{+}x  = 0\ ,
 \end{equation}
 where $R_k[u]$ is the $k$th Gel'fand-Dikii polynomial in $u(x)$ and its $x$-derivatives, normalized here so that the purely polynomial part is unity. The first few are~\cite{Gelfand:1975rn}:
 \begin{equation}
 R_1[u]{=}u\ ,\quad  R_2[u]{=}u^2{-}\frac{\hbar^2}{3}u^{\prime\prime}\ , \quad \text{and}\quad  R_3[u]{=}u^3{-}\frac{\hbar^2}{2}(u^\prime)^2{-} {\tiny \hbar^2}uu^{\prime\prime}{+}\frac{\hbar^4}{10}u^{\prime\prime\prime\prime}\ .
 \end{equation} A prime denotes an $x$-derivative, and each comes with an~$\hbar$. More generally, $R_k{=}u^k+\cdots\# \hbar^{2k-2}u^{(2k-2)}$, where $u^{(m)}$ means the $m$th derivative. The intermediate terms involve mixed orders of derivatives. The~$R_k[u]$ may be readily determined for any $k$ using a recursion relation. (Appendix \ref{GelfandPolyAppendix} explores their properties, including their organization in terms of the genus expansion in $\hbar$.). 
%

The integer $k$ in (\ref{eq:little-string-equation}) labels the $k$th multicritical~\cite{Kazakov:1989bc} model, which can be used as a basis for building (by choosing a set of $t_k$s) a wide range of 2D gravity theories. For example, the ``minimal string'' theories {\it i.e.,} the $(2,2l-1)$ conformal minimal models coupled to gravity, correspond to the set~$\{ t_k\}$ with $k=1,2,\ldots,l$, where~\cite{Mertens:2020hbs}:
\begin{equation}
    t_k=\frac{\pi^{2k-2}}{2k!(k-1)!}\frac{4^{k-1}(l+k-2)!}{(l-k)!(2l-1)^{2k-2}}\ .
    \label{eq:minimal-model-teekay}
\end{equation}
For~(\ref{eq:little-string-equation}), perturbation theory is obtained by expanding around the leading $x<0$ solution. The leading ($\hbar\to0$) part of $u(x)$ will be denoted as $u_0(x)$, and ${\cal R}_0$ will be used for the leading part of $\cal R$, and the leading equation is:
\begin{equation}
\label{eq:basic-leading}
    {\cal R}_0\equiv\sum_{k=1}^\infty
t_k u_0^k+x= 0\ .
\end{equation} 
The perturbative expansion $u(x)=u_0(x)+\sum_{g=1}^\infty u_{2g}(x)\hbar^{2g}+\cdots$ will be discussed shortly. 
The integral in (\ref{eq:basic-loop}) is up to some $\mu\leq0$, where $\mu$ is a parameter of the model. In the minimal string model $\mu$ is the coupling to a particular pointlike operator, (the cosmological constant for the $(2,3)$ model). 

There is an intuitive understanding of the integral that is worth keeping in mind: Recalling the remarks in footnote~\ref{fn:HvsH}, summing over the $N$ sectors labelled by orthogonal polynomial index $i$ became an integral over $x$ from $-\infty$ to $\mu$, and (because of the  Vandermonde determinant that appears as a Jacobian in going from $H$ to its eigenvalues) this is equivalent to a fermionic description where the Fermi sea is filled up to Fermi level $x{=}\mu$. Configuration $u_0(x)$ as defined in~(\ref{eq:basic-leading}) sets up (through the integral in~(\ref{eq:basic-loop})) the basic filled Fermi sea, and the subsequent perturbation theory then has a good local description in terms of corrections to $u_0(x)$ at the Fermi surface. Those can, through use of the full string equation~(\ref{eq:little-string-equation}), be written in terms of $u_0(x)$ and its derivatives, and so the physics to come will all depend upon $u_0(x)$ and its derivatives evaluated at $x{=}\mu$.

For the Jackiw-Teitelboim-like 2D gravity models~\cite{Jackiw:1984je,Teitelboim:1983ux} of interest in this paper, the set $\{ t_k,\mu\}$ are fixed to particular values by matching to the leading spectral density of the model.
In the prototype case,  (bosonic) JT gravity, which will  be discussed  in some detail below, they are~\cite{Dijkgraaf:2018vnm,Okuyama:2019xbv,Johnson:2019eik}:
\begin{eqnarray}
t_k=\frac{\pi^{2k-2}}{2k!(k-1)!}\ ,\,\,\,\mu=0\ .\quad\text{(JT)}
    \label{eq:leading-string-JT}
\end{eqnarray}
These  specific values for the set $(\{ t_k\}$ and the parameter $\mu$ arise from matching the leading spectral density of JT gravity with the leading spectral density of the random matrix model. Alternatively~\cite{Saad:2019lba}
it may be thought of as the large $l$ limit of the minimal model set given in~(\ref{eq:minimal-model-teekay}). The two ways of approaching JT gravity (large~$l$ limit of minimal string model and multicritical building block approach) are equivalent~\cite{Johnson:2020heh}.  

The ``extremal entropy'' parameter $S_0$ of JT gravity is related to $\hbar$  according to $\hbar{=}\e^{-S_0}$. 
Since $\hbar$ is the (renormalized) $1/N$ of the random matrix model,  this gives $S_0{\sim}\ln N$, consistent with the picture that the matrices $H$ are candidate  Hamiltonians for the effective dynamics of the black holes (in the near-horizon limit where JT gravity theories arise). 
 
 A more intricate string equation with wider applications will be of interest here. It is~\cite{Morris:1990bw,Dalley:1992qg,Dalley:1992br}:
\begin{equation}
\label{eq:big-string-equation}
u{\cal R}^2-\frac{\hbar^2}2{\cal R}{\cal R}^{\prime\prime}+\frac{\hbar^2}4({\cal R}^\prime)^2=\hbar^2\Gamma^2\ , 
\end{equation}
where ${\cal R}$ is defined as in~(\ref{eq:little-string-equation}), and the $t_k$s again label a family of multicritical building blocks out of which gravity theories can be built. Parameter $\Gamma$ naturally takes values in the integers or half integers, and will be discussed further below.  Taking the case $\Gamma=0$ for now, notice that the $x<0$ regime obeys, to all orders in perturbation theory, the same equation~(\ref{eq:little-string-equation}).  Equation~(\ref{eq:big-string-equation}) has a nice perturbative solution in the $x>0$ regime as well. It starts as $u_0(x)=0$. Expanding about that perturbative solution is most naturally associated with random matrix models with positive Hamiltonians, and as such are useful for defining supersymmetric systems.

 In the example of  ${\cal N}{=}1$ JT supergravity, the leading string equation solution satisfies~\cite{Johnson:2020heh,Johnson:2020exp}: 
\begin{eqnarray}
    \sum_{k=1}^\infty t_k u_0^k+x &=& 0\ ,\quad (x<0)\nonumber\\
    u_0(x)&=&0\ , 
\quad(x>0)\quad \text{with}\quad t_k=\frac{\pi^{2k}}{(k!)^2}\ ,\,\,\,\mu=1\ .\quad\quad\quad\quad\text{(SJT)}
    \label{eq:leading-string-SJT}
\end{eqnarray}
The integral in~(\ref{eq:basic-loop}) runs into the positive $x$ regime up to $\mu=1$. Perturbation theory beyond this leading solution is developed around the $x>0$ behaviour. The first few terms are listed for later use:
\begin{eqnarray}
u(x)&=&0+\hbar^2\left(\Gamma^2-\frac{1}{4}\right)\frac{1}{x^2}+\hbar^4\left(\Gamma^2-\frac{1}{4}\right)\left(\Gamma^2-\frac{9}{4}\right)\left(\frac{-2t_1}{x^5}\right)\nonumber\\&&\hskip2cm+\hbar^6\left(\Gamma ^2-\frac{9}{4}\right) \left(\Gamma ^2-\frac{1}{4}\right)
   \left(\frac{7 \left(\Gamma ^2-\frac{21}{4}\right) t_1^2}{x^8}-\frac{2
   \left(\Gamma ^2-\frac{25}{4}\right) t_2}{x^7}\right)+\cdots
   \label{eq:super-JT-perturbative}
\end{eqnarray}
The $\Gamma=0$ case will be discussed at length below. (In fact, for the $\Gamma=\pm\frac12$ cases, perturbation theory vanishes to all orders.)



\begin{figure}[h]
    \centering
    \includegraphics[width=0.5\linewidth]{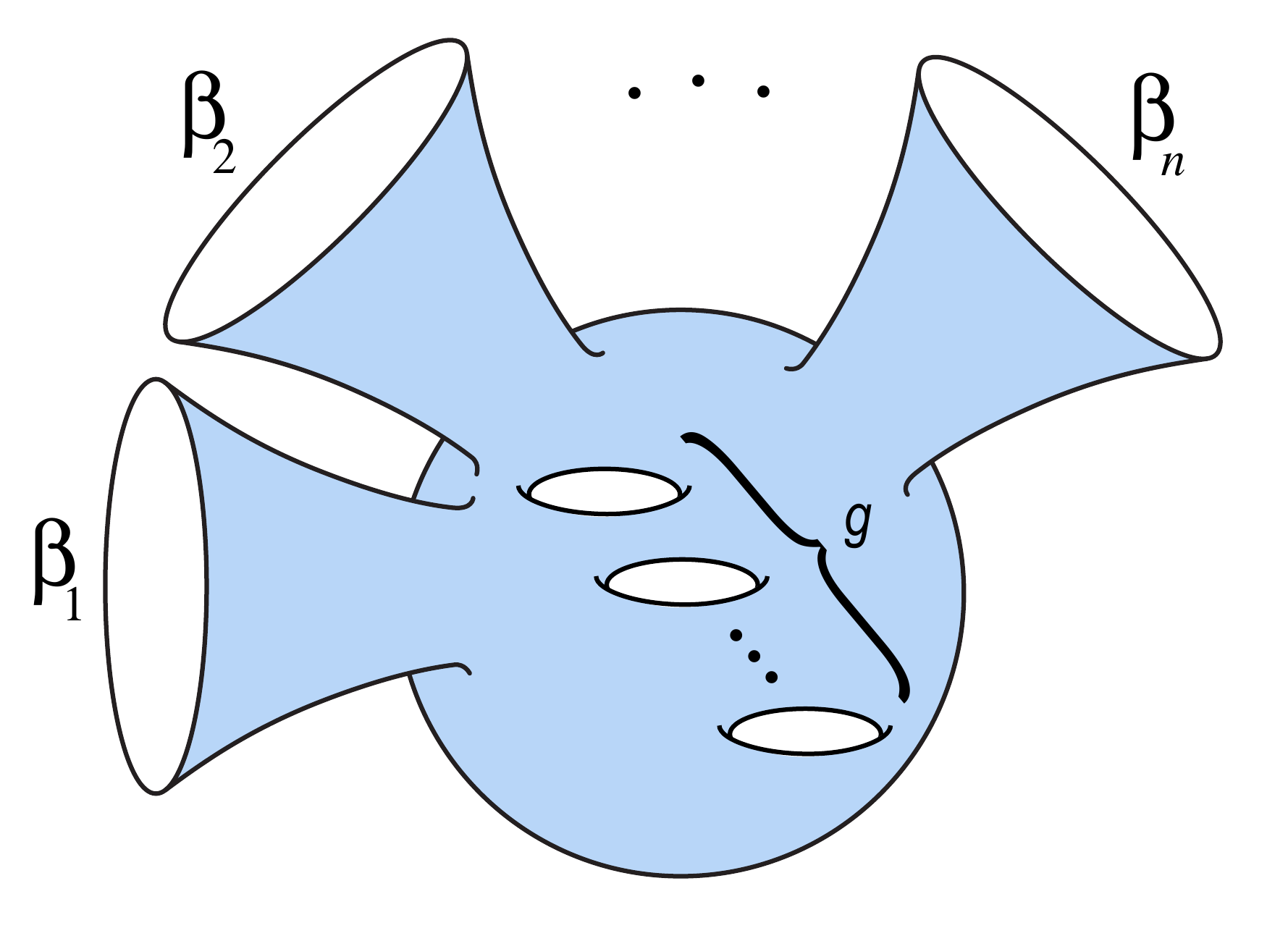}
    \caption{The connected correlator of multiple loops, $\langle Z(\beta_1)\cdots Z(\beta_n)\rangle$, can be written in terms of~$V_{g,n}(\{b_i\})$, the volume of the moduli space of a prototype surface of genus $g$ with $n$ boundaries.}
    \label{fig:multi-trumpets}
\end{figure}

Our interest here is in the fact that through the path integral, the natural 2D gravity observables are not properties of individual surfaces, but rather  features of the {\it moduli space} of surfaces in a given  topological class. The connection of these matters to random matrix models has of course has been an area of active interest for more than 30 years, starting with refs.~\cite{Witten:1989ig,Distler:1989ax,Witten:1990hr} (see ref.~\cite{Dijkgraaf:2018vnm} for a swift overview), but in recent times, a great deal of activity has centered around JT gravity (and various supersymmetric variants to be discussed). There, the surfaces of interest are hyperbolic Riemann surfaces of genus $g$ and some number, $n$, of boundaries of lengths $\{ \beta_i\}$ ($i=1,\ldots ,n$), which arise from the correlator $\langle Z(\beta_1)\cdots Z(\beta_n)\rangle$, generalizing~(\ref{eq:basic-loop}). See figure~\ref{fig:multi-trumpets} for a schematic representation. 
These can in turn~\cite{Saad:2019lba} be written in terms of the Weil-Petersson volume $V_{g,n}(\{b_i\})$ of bordered hyperbolic Riemann surfaces with geodesic boundaries of lengths $\{ b_i\}$. Much has been learned about these volumes in recent times due to the work of Mirzakhani~\cite{Mirzakhani:2006fta}, who showed that they were all polynomial in the $b_i$ and who derived a recursive formula for their explicit computation.
The matrix model turns out to be a natural place for computing these volumes, as observed by Eynard and Orantin~\cite{Eynard:2007fi}. They showed that the recursion relations of Mirzakhani are equivalent to matrix model loop equations. In particular, a  mathematical procedure, ``topological recursion'' 
 can be defined for working out the volumes recursively. The basic input is a spectral curve that is equivalent to an analytic continuation of the leading (large $N$) spectral density of the random matrix model.
 For ${\cal N}{=}1$ supersymmetric  JT gravity, there is a natural generalization involving the moduli space of super-Riemann surfaces of the appropriate type~\cite{Stanford:2019vob}, and there are matrix models within  which the Weil-Petersson volumes can be computed as well (discussed below).

In refs.~\cite{Johnson:2024bue,Johnson:2024fkm} it was noticed that, at least for the subset of surfaces with one boundary (but arbitrary genus), there is an extremely natural definition of the volumes (for ordinary JT gravity and ${\cal N}{=}1$ JT supergravity) entirely through  ordinary differential equations (ODEs). Such a definition is interesting since on the one hand it gives a very efficient way of determining the volumes (by recursively solving the ODEs), and on the other hand the differential equations contain more information than just the individual volumes themselves. The equations can give useful asymptotic information about the volumes as a function of $g$ and $n$, as well as non-perturbative information that perhaps goes beyond the geometrical/topological language, in a sense that may transcend the path integral definition of the 2D gravity theory.

The purpose of this paper is to apply this ``ODE method'' to the classes of gravity theory that arise when considering JT supergravity with $\cal N$ amounts of extended supersymmetry~\cite{Forste:2017apw,Stanford:2017thb,Mertens:2017mtv,Heydeman:2020hhw,Heydeman:2025vcc}. The Weil-Petersson volumes that arise in such cases have yet to be studied in much detail. In the case of ${\cal N}{=}2$,  ref.~\cite{Turiaci:2023jfa} defined them through loop equations (and computed some examples) and noticed some interesting properties (some to be discussed below). Our methods will reproduce those examples, and readily produce more examples, and many of the properties of the volumes have a ready interpretation from the ODE approach. Random matrix model realizations go  beyond ${\cal N}{=}2$, as discussed in refs.~\cite{Turiaci:2023jfa,Johnson:2024tgg}, and shown in detail for a variety of models recently in ref.~\cite{Johnson:2025oty}. We use our methods to explore the recent cases with both small~\cite{Heydeman:2020hhw,Turiaci:2023jfa} and large~\cite{Heydeman:2025vcc} ${\cal N}=4$, observing a number  of   interesting features, as well as producing  new examples of the volumes.

\subsection{Review of the method}
\label{sec:method}
The core  relationship at the heart of the method of ref.~\cite{Johnson:2024bue} is that the basic loop expectation value~(\ref{eq:basic-loop})  can be written as the Laplace transform (with respect to $(-)E$) of the inverse of ${\cal H}{-}E$:
\begin{eqnarray}
\label{eq:loop-development}
    \langle {\rm Tr}(\e^{-\beta{ H}})\rangle &\equiv&\int_{-\infty}^\mu \langle x|\e^{-\beta{\cal H}} |x\rangle\, dx=\int \!dE\,\e^{\beta E} \,\int_{-\infty}^\mu\langle x|\frac{1}{{\cal H}-E}|x\rangle \, dx\nonumber\\
     &=&\frac{1}{\hbar}\int \!dE\,\e^{\beta E} \,\int_{-\infty}^\mu{\widehat R}(x,E) \, dx \quad 
     \ ,
    \end{eqnarray}
    and the last step indicates that this defines the diagonal resolvent ${\widehat R}(x,E)$, which is well known (through the classic work of Gel'fand and Dikii~\cite{Gelfand:1975rn}) to obey a differential equation:
\begin{equation}
\label{GDResDiffEq}
    4(u-E)\widehat{R}^2-2\hbar^2\widehat{R}\widehat{R}''+\hbar^2\widehat{R}'^2=1\ ,
\end{equation}
for which there is the following series solution:
\begin{equation}
\label{SeriesRepofRhat}
    \widehat{R}(x,E)=\sum_{j=0}^\infty\frac{-(2j-1)!}{(-4)^jj!(j-1)!}\frac{R_j[u]}{(-E)^{j+\frac12}}\ . 
\end{equation}
 Here, $R_j[u]$ are the $j$th  Gel'fand-Dikii polynomials in $u$ and its derivatives mentioned below~(\ref{eq:little-string-equation}). Every order in $j$ here has several different powers of $\hbar$. Since we are interested in $\hbar$ (topological) perturbation theory, it is of interest to seek instead  the following  expansion of ${\widehat R}(x,E)$:
 \begin{equation}
 \label{eq:R-expansion}
     \widehat{R}(x,E)=\sum_{g=0}^\infty \hbar^{2g}\widehat{R}_g(x,E)+\cdots\ ,
 \end{equation}
(the ellipsis denotes non-perturbative parts), which comes from two sources: the $\hbar$ organization within the $R_k[u]$, as well as the fact that $u(x)$ itself has an expansion:
 \begin{equation}
 \label{eq:u-expansion}
     u(x){=}\sum_{g}u_{2g}(x)\hbar^{2g}+\cdots\ . 
 \end{equation}
 The leading terms in the expansion for ${\widehat R}_g(x,E)$ may be efficiently developed by simply recursively solving equation~(\ref{GDResDiffEq}):\footnote{The coefficient of the penultimate term corrects a typographical error in ref.~\cite{Johnson:2024bue}.}
\begin{eqnarray}
&&
{\widehat R}(x,E)
= -\frac12\frac{1}{[u_0(x)-E]^{1/2}}
+
\frac{\hbar^2}{64}\left\{
\frac{16u_2(x)}{[u_0(x)-E]^{3/2}}
+\frac{4u^{\prime\prime}_0(x)}{[u_0(x)-E]^{5/2}}-\frac{5(u^{\prime}_0(x))^2}{[u_0(x)-E]^{7/2}}\right\}
\label{eq:gelfand-dikii-A}
\\
&&\hskip+1cm
+\frac{\hbar^4}{4096}\left\{
\frac{1024u_4(x)}{[u_0(x)-E]^{3/2}}
-\frac{256[3u_2(x)^2-u^{\prime\prime}_2(x)]}{[u_0(x)-E]^{5/2}}
+\frac{64[u^{(4)}_0(x) -10u_2(x) u^{''}_0(x)-10u^\prime_0(x)u^\prime_2(x)]}{[u_0(x)-E]^{7/2}}\right. \nonumber \\
&&\hskip+0.5cm
\left.
-\frac{16[28u^{(3)}_0(x)u^\prime_0(x)+21u^{\prime\prime}_0(x)^2-70u_2(x)u_0(x)^2]}{[u_0(x)-E]^{9/2}}
+\frac{{1848}u^{\prime}_0(x)^2u^{\prime\prime}_0(x)}{[u_0(x)-E]^{11/2}}-\frac{1155u^{\prime}_0(x)^4}{[u_0(x)-E]^{13/2}}\right\} +\cdots\ .   \nonumber
\end{eqnarray} 
Given expansion~(\ref{eq:R-expansion}), one can write:
\begin{eqnarray}
\label{eq:loop-development}
    \langle {\rm Tr}(\e^{-\beta{ H}})\rangle 
       &=&\frac{1}{\hbar}\int \!dE\,\e^{\beta E} \,\int_{-\infty}^\mu\sum_{g=0}^\infty{\widehat R}_g(x,E) dx\,\hbar^{2g} +\cdots \nonumber \\
    &=&\langle Z(\beta)\rangle_0+\int \!dz\,\e^{-\beta z^2} \, 
    \sum_{g=1}^\infty W_{g,1}(z)\hbar^{2g-1} +\cdots \ ,
    \end{eqnarray}
where:
\begin{equation}
    \langle Z(\beta)\rangle_0 =\frac{1}{2\hbar\sqrt{\pi\beta}}\int_{-\infty}^\mu\! dx\, \e^{-\beta u_0(x)}\ ,
\end{equation}
is the leading piece, which won't be the concern of this paper (its Laplace transform gives the leading spectral density), 
and we have defined the object: 
\begin{equation}
\label{eq:W-definition}
    W_{g,1}(z) = -2z\int_{-\infty}^\mu\!\! {\widehat R}_g(x,E)\,dx\ ,
\end{equation}  moving to working 
on the double cover of the (cut) $E$ plane using  $z^2=-E$.

\begin{figure}[h]
    \centering
\includegraphics[width=0.5\linewidth]{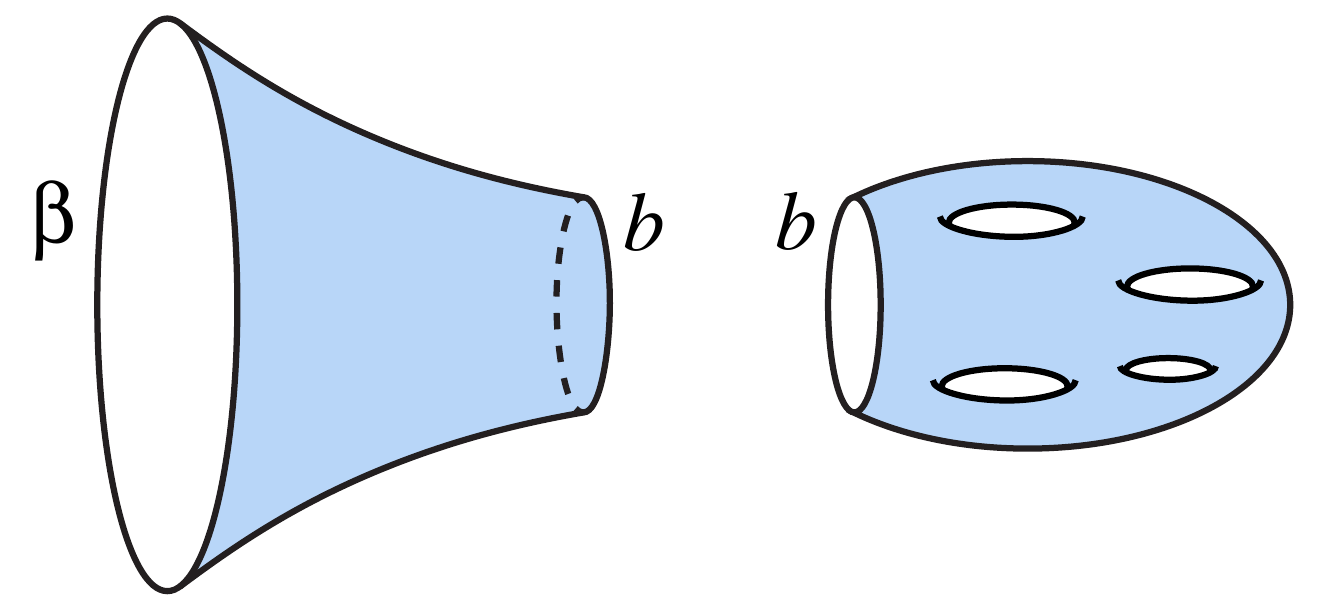}
    \caption{Schematic depiction of gluing a trumpet to a bordered higher genus Riemann surface.}
    \label{fig:fig:trumpet-recital}
\end{figure}

It is  remarkable  (observed in ref.~\cite{Johnson:2024bue}, and explored further in ref.\cite{Lowenstein:2024gvz} and  this paper) that beyond zeroth order in $\hbar$, upon using the string equation to relate the higher order $u_i(x)$ to~$u_0(x)$ and its derivatives, the quantity ${\widehat R}_{g}(x,E)$ {\it is in fact a total derivative}. So this means that the quantities $W_{g,1}(z)$ can be readily computed and are polynomials in inverse powers of $[u_0+z^2]^{1/2}$, and  have coefficients that come from $u_0$ and its derivatives evaluated at $x=\mu$.
To get a feel for this, a few examples should be explored. For JT gravity, where perturbation theory comes from expanding around the $x<0$ solution, it turns out that:
\begin{equation}
\label{eq:u2}
    u_2(x)
    =-\frac{1}{12}\frac{d^2}{dx^2}\ln(u_0')\ ,
\end{equation}
and hence, focusing on order $\hbar^2$, one sees that~\cite{Johnson:2024bue}:
\begin{equation}
\label{eq:totally-awesome}
    {\widehat R}_1(x,E) = \frac{d}{dx}\left(-\frac{u_0^{\prime\prime}(x)}{48 u_0^\prime(x)[u_0(x)-E]^{3/2}}+\frac{u_0^\prime(x)}{32[u_0(x)-E]^{5/2}}\right)\ ,
\end{equation}
so this means that the quantity $W_{1,1}(z)$ is entirely  determined by the behaviour of $u_0(x)$ its derivatives at the boundary $x=\mu$. Contributions from $x=-\infty$ boundary all vanish since $u_0^{(n)}\to0$ there for all~$n$, as follows from the form of the leading string equation~(\ref{eq:leading-string-JT}). That equation shows that 
$u_0(0)=0$, and differentiating and using the chain rule gives further that: \begin{equation}
    \label{eq:some-derivatives}
    u_0^\prime(0)=-\frac{1}{t_1}{=}{-}2 \ , \quad\text{and}\quad u^{\prime\prime}(0)=-\frac{2t_2}{t_1^3}=-4\pi^2\ .
\end{equation} So using (\ref{eq:W-definition}), and (\ref{eq:totally-awesome})) gives:
\begin{equation}
\label{eq:W11-JT}
    W_{1,1}=\frac{\pi^2}{12z^2}+\frac{1}{8z^4}\ ,\quad\text{(JT)}\ .
\end{equation}
Meanwhile in ${\cal N}{=}$ JT supergravity, $\mu{=}1$ and the perturbative expansion comes from the $x>0$ region. Equation~(\ref{eq:super-JT-perturbative}) shows that  the  leading ($\hbar^0$) piece is $u_0(x)=0$, followed by corrections that are simply inverse powers of $x$ at each order. These are of course readily integrated. At $O(\hbar^2)$,  $u_2(x)=-\frac{1}{4x^2}$, giving (directly from~(\ref{eq:gelfand-dikii-A}) and~(\ref{eq:W-definition})):
\begin{equation}
    W_{1,1}=-\frac{1}{8z^2}\ .\quad\text{(SJT)}
\end{equation}
The $W_{g,1}(z)$ just derived are in fact the Laplace transforms of the Weil-Petersson volumes $V_{g,1}(b)$:
\begin{eqnarray}
    {W}_{g,1}&\equiv&-{2}z\int_{-\infty}^\mu{\widehat R}_g(x,z)\, dx
    =\int_0^\infty \! bdb\,\e^{-b z}\, {\widehat V}_{g,1}(b)\ ,
\label{eq:W-vs-V}
\end{eqnarray}
and so this procedure yields:
\begin{equation}
\label{eq:V11-basics}
    V_{1,1}=\frac{(4\pi^2+b^2)}{48}\ ,\quad\text{(JT)}
\quad
\text{and}
\quad
    V_{1,1}=-\frac{1}{8}\ ,\quad\text{(SJT)}
\end{equation}
familiar expressions for the Weil-Petersson volumes with one geodesic boundary of length $b$ and one handle~\cite{Wolpert1983,Mirzakhani:2006fta,Stanford:2019vob}.  The full partition function $\langle Z(\beta)\rangle$ depends on the length, $\beta,$ of the asymptotic boundary however. The connection is through sewing on a ``trumpet'', whose form is readily derivable by  putting~(\ref{eq:W-vs-V}) into~(\ref{eq:loop-development}) and doing the $z$-integral to  give:
\begin{eqnarray}
\label{eq:loop-and-volumes}
   && \langle Z(\beta)\rangle 
   =
    \langle {\rm Tr}(\e^{-\beta{\cal H}})\rangle_0 
    +\frac{1}{2\sqrt{\pi\beta}}\int_0^\infty\!\! bdb\,\e^{-\frac{ b^2}{4\beta}} \sum_{g=1}^\infty {\widehat V}_{g,n}(b) \hbar^{2g-1}+\cdots\nonumber\\
   &&\hskip2.0cm =\langle Z(\beta)\rangle_0+\sum_{g=1}^\infty \int_0^\infty\!\! bdbZ_{\rm tr}(\beta,b) {\widehat V}_{g,n}(b) \hbar^{2g-1}+\cdots \ ,   
\end{eqnarray}
where:
\begin{equation}
\label{eq:trumpet-voluntary}
    Z_{\rm tr}(\beta,b) = \frac{1}{2\sqrt{\pi\beta}}\e^{-\frac{ b^2}{4\beta}}\ ,
\end{equation}
is the trumpet partition function. In general, this factor, and the gluing procedure {\it via} integral~(\ref{eq:loop-and-volumes}), appears for each boundary  as needed to convert $V_{g,n}$ to a correlation function of $\langle Z(\beta_i)\cdots Z(\beta_n) \rangle_g$~\cite{Saad:2019lba}. Here there is always a single factor since our exploration incorporates only one boundary, but arbitrary genus $g$.
See figure~\ref{fig:fig:trumpet-recital} for an $n=1$ example.

In fact, for genus $g=0$, where $n$-point connected correlators of $Z(\beta)$ are given by the simple formula~\cite{Banks:1990df,Ginsparg:1993is}:
\begin{equation}
    \langle Z(\beta_1)\cdots Z(\beta_n)\rangle = \frac{\sqrt{\beta_1\cdots\beta_n}}{2\pi^\frac{n}{2}\beta_T}\left[\partial_x^{n-2}\e^{-\beta_T u_0(x)}\right]_{x=\mu}\ ,
    \label{eq:correlator}
\end{equation}
where $\beta_T\equiv\sum_i\beta_i$,
it is straightforward~\cite{Mertens:2020hbs} to use the trumpet to convert this into expressions for~$V_{0,n}(\{ b_i\})$, for $n{>}2$. The $n{=}2$ case yields~\cite{Saad:2019lba} the classic universal connected correlator result~\cite{Brezin:1993qg,BEENAKKER1,BEENAKKER3,Forrester1994}:
\begin{equation}
    \label{eq:universal}
    \langle Z(\beta_1) Z(\beta_2)\rangle =
    \int_0^\infty bdb Z_{\rm tr}(\beta_1,b)Z_{\rm tr}(\beta_2,b)
    =\frac{1}{4\pi\sqrt{\beta_1\beta_2}}\int_0^\infty bdb\,\e^{-\frac{b^2}{4}\left(\frac{1}{\beta_1}+\frac{1}{\beta_2}\right)}=\frac{\sqrt{\beta_1\beta_2}}{2\pi(\beta_1+\beta_2)}\ ,
\end{equation}
for which there is no volume to glue to, but for $n{=}3$, we have:
\begin{eqnarray}
    \label{eq:3-point}
    \langle Z(\beta_1) Z(\beta_2)Z(\beta_3)\rangle 
    &=&-\frac{\sqrt{\beta_1\beta_2\beta_3}}{2\pi^{\frac32}}u^\prime_0(0)\e^{-\beta_T u_0(0)}
    =\frac{\sqrt{\beta_1\beta_2\beta_3}}{\pi^{\frac32}}\ ,\nonumber\\
    &=&\int_0^\infty b_1db_1b_2db_2b_3db_3  
    Z_{\rm tr}(\beta_1,b_1)Z_{\rm tr}(\beta_2,b_2)Z_{\rm tr}(\beta_2,b_3)V_{0,3}(b_1,b_2,b_3)\ .\nonumber\\
    &\Longrightarrow& V_{0,3}(b_1,b_2,b_3) = 1\ , 
\end{eqnarray}
and more non-trivially, for $n{=}4$~\cite{nakanishi2001areas}:
\begin{eqnarray}
    \label{eq:4-point}
    \langle Z(\beta_1) Z(\beta_2)Z(\beta_3)Z(\beta_4)\rangle 
    &=&\frac{\sqrt{\beta_1\beta_2\beta_3\beta_4}}{2\pi^{2}}\e^{-\beta_T u_0(0)}\left[\beta_T(u^\prime_0)^2-u_0^{\prime\prime}\right]_{x=0}\ \nonumber\\
    &=&\frac{\sqrt{\beta_1\beta_2\beta_3\beta_4}}{\pi^{2}}\left[2\beta_T+2\pi^2\right]_{x=0}\ .\nonumber\\
    &\Longrightarrow& V_{0,4}(b_1,b_2,b_3,b_4) = \frac{1}{2}\left(b_1^2+b_2^2+b_3^2+b_4^2+4\pi^2\right)\ ,
\end{eqnarray}
where in both cases, relations~(\ref{eq:some-derivatives}) were used. In summary, equation~(\ref{eq:correlator}) together with the trumpet operation using~(\ref{eq:trumpet-voluntary}) constitutes a simple general generator for the $V_{0,n}(\{b_i\})$, reducing to Zograf's result  for~$V_{0,n}({\bf 0})$~\cite{Zograf1993WeilPetersson}.

Returning  to the cases with one boundary, the continuation to higher $g$ is interesting to explore further, and will be the main focus. At next order, we have (see computations in Section~\ref{sec:perturbing-string-equation}):
\begin{equation}
\label{u4N=2Sol}
    u_4(x)=\frac{d^2}{dx^2}\left[\frac{u_0''(x)^3}{90 u_0'(x)^4}-\frac{7 u_0^{(3)}(x)
   u_0''(x)}{480 u_0'(x)^3}+\frac{u_0^{(4)}(x)}{288
   u_0'(x)^2}\right]\ ,
\end{equation}
and  a bit of experimentation reveals:
\begin{equation}
    {\widehat R}_2(x,E) = \frac{d}{dx}\left({\widehat Q}_2(x)\right)\ ,
\end{equation}
where:
\begin{eqnarray}
   &&\hskip-1.1cm \widehat{Q}_2(x)=\frac{105u_0'^3}{2048[u_0-E]^{11/2}}-\frac{203u_0'u_0''}{3072[u_0-E]^{9/2}}+\frac{29u_0'u_0'''-3u_0''^2}{1536u_0'[u_0-E]^{7/2}} \\
        &&\hskip-1.0cm-\frac{17u_0''^3-34u_0'u_0''u_0'''+15u_0'^2u_0^{(4)}}{3840u_0'^3[u_0-E]^{5/2}}-\frac{64u_0''^4-111u_0'u_0''^2u_0'''+21u_0'^2u_0'''^2+31u_0'^2u_0''u_0^{(4)}-5u_0'^3u_0^{(5)}}{5760u_0'^5[u_0-E]^{3/2}}\ .
        \nonumber
\label{eq:queue-two}
\end{eqnarray}
 To handle this order for JT gravity one needs to supplement relations~(\ref{eq:some-derivatives}) with a few more derivatives, worked out from~(\ref{eq:leading-string-JT}):
 \begin{eqnarray}
      && u_0^{(3)}(0)= \frac{6 t_3 t_1-12 t_2^{2}}{t_1^{5}}=-20 \pi^{4}\ , \quad u_0^{(4)}(0)= \frac{-24 t_1^{2} t_4+120 t_3 t_1 t_2-120 t_2^{3}}{t_1^{7}}=-\frac{488 \pi^{6}}{3}\ ,\nonumber\\ && u_0^{(5)}(0)=\frac{120 \left(-6 t_1^{2} t_2 t_4+21 t_1 t_2^{2} t_3-3 t_1^{2} t_3^{2}+t_1^{3} t_5-14 t_2^{4}\right)}{t_1^{9}}= -\frac{5516 \pi^{8}}{3}\ .
      \label{eq:some-more-derivatives}
 \end{eqnarray} Hence, one sees that:
\begin{equation}
    W_{2,1} = \frac{105}{128 z^{10}}+\frac{203 \pi^{2}}{192 z^{8}}+\frac{139 \pi^{4}}{192 z^{6}}+\frac{169 \pi^{6}}{480 z^{4}}+\frac{29 \pi^{8}}{192 z^{2}}\ . \qquad\text{(JT)}
\end{equation}
Finally, after inverse Laplace transform, one gets the known Weil-Petersson volume~\cite{penner1992weil,Mirzakhani:2006fta}:
\begin{equation}
\label{eq:V21-basics-bos}
    V_{2,1} = \frac{\left(4 \pi^{2}+b^{2}\right) \left(12 \pi^{2}+b^{2}\right) \left(6960 \pi^{4}+384 \pi^{2} b^{2}+5 b^{4}\right)}{2211840} \ . \qquad\text{(JT)}
\end{equation}
For ${\cal N}{=}1$ JT supergravity, at this order, in addition to $u_0(x){=}0, u_2(x){=}{-}\frac{1}{4x^2}$ used at the  previous order,  we have from~(\ref{eq:super-JT-perturbative}) that $u_4(x){=}-\frac{9t_1}{8x^5}$, and with $t_1=\pi^2,\mu=1$ (from (\ref{eq:leading-string-SJT})) this all yields:
\begin{equation}
\label{eq:V21-basics-Neq1}
    W_{2,1}= -\frac{9\pi^2}{64 z^{2}}-\frac{9}{128 z^{4}}\quad\Longrightarrow\quad V_{2,1}=-\frac{3}{256}(b^2+12\pi^2)\ ,\qquad \text{(SJT)}\ ,
\end{equation}
the Weil-Petersson volume for ${\cal N}{=}1$ JT supergravity~\cite{Stanford:2019vob}.

An essential feature here, as emphasized in ref.~\cite{Johnson:2024bue}, is the fact at each successive order, ${\widehat R}_g(x,E)$ becomes a total derivative. This not a general feature of the object ${\widehat R}(x,E)$, but comes from inputting the fact that {\it $u(x)$ perturbatively solves the string equation}. There's both a physical and mathematical component to why it must be true. The physical one, on the matrix model side, is that the macroscopic loop is described in the orthogonal polynomial formalism in terms of a filled Fermi sea with energy levels described by $x$ running from $-\infty$ to $\mu$. The physics is  controlled by the behaviour of $u(x)$ and its derivatives at $x{=}\mu$. This is precisely what is afforded by the total derivative in the computational scheme outlined above, since only the boundary term at $x{=}\mu$ contributes. (The $x{=}{-}\infty$ boundary produces vanishing or non-universal contributions.) The mathematical reason is one of simplicity. The Weil-Petersson volumes  $V_{g,n}(\{ b_i\})$ were shown by Mirzakhani~\cite{Mirzakhani:2006fta} to all be polynomials in the geodesic lengths~$b_i$. This translates into the $W_{g,n}(\{z_i\})$ being polynomials in the $1/z_i$. It is hard to see how this simplicity could arise from the $x$-integral of increasingly complex expressions for ${\widehat R}_g(x,E)$ other than if they are total derivatives.

Note that the manner in which the total derivatives appear seems quite different in the ordinary JT case (using the $x<0$ expansion) as compared to the SJT case (using the $x>0$ expansion), which overall is much simpler. One of the things that will emerge later in the paper is a more universal procedure that is the same for both the $x<0$ and the $x>0$ expansions.

A remarkable feature of the method is that it {\it does not care} about the details of the system in hand, {\it e.g.}  whether the system is supersymmetric or not, or JT-gravity-like or not (there are applications to full worldsheet string theory too, such as the Virasoro minimal string and supersymmetric variants\cite{Collier:2023cyw,Johnson:2024bue,Johnson:2024fkm,Johnson:2025vyz})-- all those details are hidden inside the leading behaviour of the function $u(x)$, which is determined by the string equation. Beyond that,  the Gel'fand-Dikii equation just takes $u(x)$  as input and  the machinery  works fluidly.

  It is worth remarking here that there are a number of reasons why it is particularly advantageous  to have a method like this. On the practical side, it is considerably more efficient, as a method for generating volumes, than other approaches.  It is possible that this is simply because it is (so far) restricted to one boundary, but even with that restriction it is useful to have. Having an ODE be the ``container'' for the volumes also makes it very amenable to extracting general features about asymptotic behavior (such as in the large $g$ limit), and even non-perturbative (in $g$) behaviour.  A natural question that springs to mind is whether  this method can be extended to all $V_{g,n}$, {\it i.e.,} whether multiple boundaries and arbitrary genus can be captured by a differential equation. The ODE system recently proposed in Ref~\cite{Johnson:2025dyb} as a natural generalization of the Gel'fand-Dikii equation is promising in this regard.

These are all interesting subjects for future investigation. The purpose of this paper is to explore the method more and apply it to the newer application of defining volumes for JT supergravity with extended supersymmetry.

\subsection{This paper's results}
This paper further explores the ODE method reviewed above, deriving recursion relations for the key quantities, the  ${\widehat R}_g(x,E)$, that follow from combining recursive properties ({\bf Section~\ref{sec:perturbing-string-equation}}) of the string equation with those of the Gel'fand Dikii equation ({\bf Section~\ref{sec:perturbing-GD-equation}}). 
The fact that they can be written as a total derivative at any genus, ${\widehat R}_g(x,E)=d_x\widehat{Q}_g(x,E)$, is confirmed at several orders. (That it is generally true is beyond doubt, esecially given the reasoning below equation~(\ref{eq:V21-basics-Neq1}), although we do not have a closed-form proof.) In addition to the calculations presented in the main body of the paper,  several Appendices are provided with supporting computations and useful schemes for constructing the~${\widehat Q}_g(x,E)$ at any desired order.
{\bf Section~\ref{sec:defining-volumes}} generally discusses the construction of the volumes that result from having the~${\widehat Q}_g(x,E)$, tailoring conventions for the applications to come later.

Recent work~\cite{Turiaci:2023jfa,Johnson:2023ofr,Johnson:2024tgg,Johnson:2025oty} has shown that the string equation approach is extremely powerful for defining models of JT supergravity with extended supersymmetry~\cite{Forste:2017apw,Stanford:2017thb,Mertens:2017mtv,Heydeman:2020hhw,Heydeman:2025vcc}. Associated to these should be Weil-Petersson volumes for the underlying super-Riemann surfaces. For the case of ${\cal N}{=}2$, some of these were worked out by Turiaci and Witten~\cite{Turiaci:2023jfa}, along with topological recursion equations following from the  underlying loop equations. In {\bf Section~\ref{sec:ODE-approach-Neq2}} we compute several volumes for that case with the ODE method, confirming the examples given in ref.~\cite{Turiaci:2023jfa}, but also  ({\bf Section~\ref{sec:TW-approach}}) using their recursion approach to generate more examples to compare (successfully) with volumes computed in this paper.  This also allows us to observe the relative efficiency of the ODE approach at obtaining the~$V_{g,1}(b)$. There are a number of useful Appendices
containing computational details that the reader may find helpful.
We then go on to study, in {\bf Section~\ref{sec:ODE-approach-Neq4}}, the cases of small and large ${\cal N}{=}4$ JT supergravity  that have recently~\cite{Johnson:2025oty} been shown to also have a multicritical matrix model description using string equations. We exhibit several new examples of volumes.  We also observe that there is again (as already noted for the ${\cal N}{=}2$ case in ref.~\cite{Turiaci:2023jfa}) a decomposition such that  bosonic Weil-Petersson volumes appear as a subsector. There is a natural explanation for this within the current framework, which we discuss.
{\bf Section~\ref{sec:closing-remarks}} concludes the paper with a few closing remarks.

\section{Unified perturbative treatment of the  string equation}
\label{sec:perturbing-string-equation}
The general form of the string equation that we wish to solve is given by starting with~(\ref{eq:big-string-equation}) but treating~$\Gamma$ differently~\cite{Johnson:2023ofr}. It counts degenerate zero energy states, but the goal here is to be able to study systems that have a number of BPS states that scales comparably to the overall number of microstates, given by $\hbar^{-1}=\e^{S_0}$. This is the case for the extended JT gravity systems resulting from the near-horizon geometries of near-BPS black holes~\cite{Heydeman:2020hhw,Heydeman:2025vcc}. Therefore we hold fixed finite ${\widetilde\Gamma}{=}\hbar\Gamma$ in studying perturbation theory around small $\hbar$, and rewrite the string equation as:
\begin{equation}
\label{eq:scaled-big-string-equation}
u\mathcal{R}^2-\frac{\hbar^2}{2}\mathcal{R}\mathcal{R}''+\frac{\hbar^2}{4}(\mathcal{R}')^2=\widetilde{\Gamma}^2\ ,\quad\text{with}\quad{\cal R}\equiv\sum_{k=1}^\infty t_k R_k[u]+x\ .
\end{equation}
Since the right hand side is now classical, where previously it was subleading, it will result in a  structure for perturbation theory that is rather different  from before, although it should reproduce our previous results when $\widetilde{\Gamma}\to 0$.

\begin{figure}[t]
    \centering
    \includegraphics[width=0.85\textwidth]{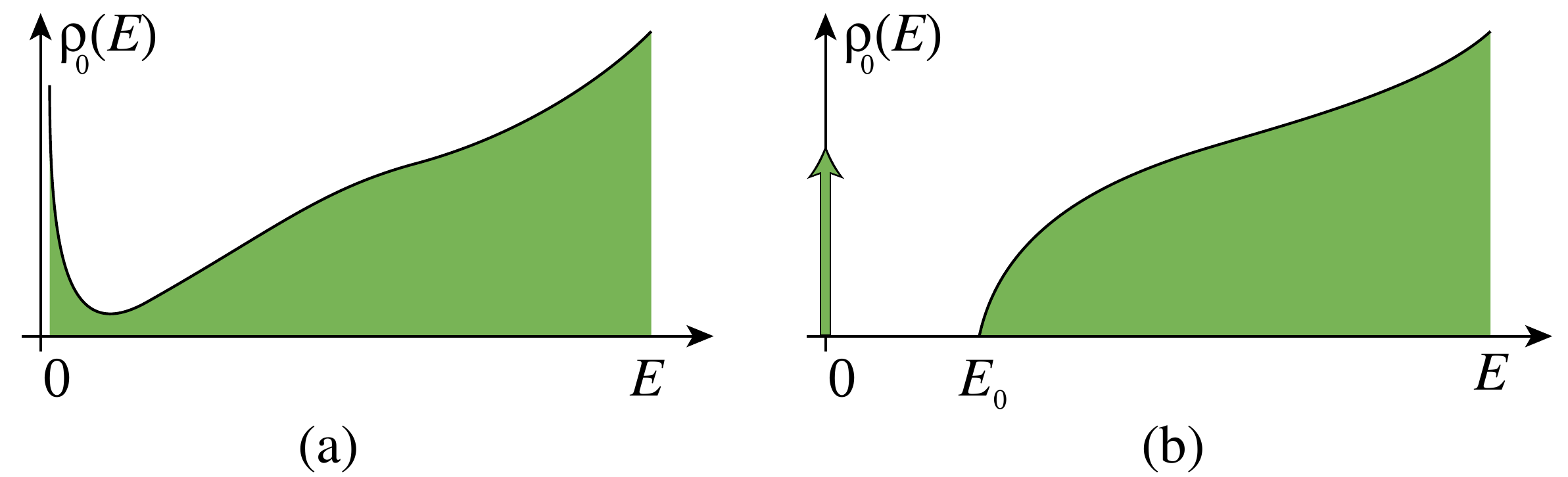}
    \caption{(a) The typical ${\cal N}{=}1$ behaviour of a density. (b) The possible ${\cal N}{>}1$ situation with some BPS states at $E{=}0$ and a non-BPS sector beginning at some threshold energy $E_0$.}
    \label{fig:super-density}
\end{figure}

For a particular supergravity, the $t_k$ will have a specific form, as well as $\widetilde\Gamma$.
An example of this is ${\cal N}{=}2$ JT supergravity, where it was found in ref.~\cite{Johnson:2023ofr} that:
\begin{equation}
 {t}_k=\frac{\pi^{k-1}J_k(2\pi\sqrt{E_0})}{2(2k+1)k!E_0^{k/2}}\ ,\quad\text{with}\quad    \widetilde{\Gamma}=\frac{\sin(2\pi\sqrt{E_0})}{4\pi^2}\ .
\end{equation}
where $E_0$ is a threshold energy at which the continuous non-BPS spectrum starts. See  figure~\ref{fig:super-density}(b), and compare to the typical ${\cal N}{=}1$ behaviour in  figure~\ref{fig:super-density}(a), with no non-zero threshold energy and generically a $1/\sqrt{E}$ behaviour at $E{=}0$.

The next step is to substitute the expansion into the string equation:
\begin{equation}
\begin{split}
&u=u_0+\hbar^2u_2+\hbar^4u_4+\cdot\cdot\cdot\\
&\mathcal{R}=\mathcal{R}_0+\hbar^2\mathcal{R}_2+\hbar^4\mathcal{R}_4+\cdot\cdot\cdot\\
\end{split}
\end{equation}
As before, the goal is to write all higher order parts in terms of the zeroth order solution, $u_0(x)$,  and its derivatives, which solves the leading equation (obtained by setting $\hbar{=}0$ in~(\ref{eq:scaled-big-string-equation})):
\begin{equation}
\label{eq:leading-string-alt}
u_0\mathcal{R}_0^2=\widetilde{\Gamma}^2\ ,
\end{equation}
where:
\begin{equation}
\label{RsExpansion-a}
\mathcal{R}_0\equiv G_0+x\ , \quad\text{with}\quad
    G_0\equiv\sum_{k=1}^\infty t_ku_0^k\ .
\end{equation}
Working to order $\mathcal{O}(\hbar^2)$ to begin with, we have:
\begin{equation}
\label{RsExpansion}
\mathcal{R}_2=u_2\dot{G}_0-\frac{u_0''}{6}\ddot{G}_0-\frac{u_0'^2}{12}\dddot{G}_0\ ,
\end{equation} where an overdot is a $u_0$ derivative,
and at second order in $\hbar$ we need to solve:
\begin{equation}
\label{SecondOrderStringEq}
\mathcal{R}_0'^2+2\mathcal{R}_0(4\mathcal{R}_2u_0+2\mathcal{R}_0u_2-\mathcal{R}_0'')=0\ .
\end{equation}
We can rearrange the zeroth order solution and taking $x$-derivatives with respect to $x$ to obtain the following expressions:
\begin{equation}
\label{R0Derivatives}
    \mathcal{R}_0'=-w\frac{\widetilde{\Gamma}}{2u_0^{3/2}}u_0'\ ,\qquad
    \mathcal{R}_0''=w\frac{\widetilde{\Gamma}}{4u_0^{5/2}}\left[3u_0'^2-2u_0u_0''\right]\ ,
\end{equation}
where $w=\pm1$ ($w=1$ is the branch we want for $\mathcal{N}=2$.\footnote{{In general both the plus and minus branch can be important.\cite{Johnson:2025oty} An example of the importance of carefully choosing the sign of $w$ appears in our later discussion of small $\mathcal{N}=4$ where the sign of $w$ needs to be adjusted according to $w=\text{sgn}(J)(-1)^{2J+1}$.}}) Next note the zeroth order sting equation reads:
\begin{equation}
\label{eq:tree-string-equation}
    G_0[u_0(x)]=-x+\frac{w\widetilde{\Gamma}}{u_0^{1/2}}\ .
\end{equation}
Taking a derivative  with respect to $x$ gives:
\begin{equation}
    \dot{G}_0u_0'=-1-\frac{w\widetilde{\Gamma}u_0'}{2u_0^{3/2}}\ ,
\end{equation}
leading to  the following expressions:
\begin{equation}
\label{dotGs}
    \dot{G}_0=-\frac{w\widetilde{\Gamma}}{2u_0^{3/2}}-\frac{1}{u_0'}\ , \qquad
    \ddot{G}_0=\frac{3w\widetilde{\Gamma}}{4u_0^{5/2}}+\frac{u_0''}{u_0'^3}\ ,\quad\text{and}\quad
    \dddot{G}_0=-\frac{15w\widetilde{\Gamma}}{8u_0^{7/2}}+\frac{-3u_0''^2+u_0'u_0'''}{u_0'^5}\ .
\end{equation}
This allows us to express $\mathcal{R}_2$ in equation~(\ref{RsExpansion}) entirely in terms of $u_0$ and its derivatives. Solving for $u_2$ explicitly yields:
\begin{equation}
\label{Solutionu2unotzero}
    u_2(x)=\frac{u_0''^2-u_0'u_0'''}{12u_0'^2}=-\frac{1}{12}\frac{d^2}{dx^2}\ln(u_0'(x))\ .
\end{equation}
which was used earlier in Section~\ref{sec:method} when explaining the method of ref.~\cite{Johnson:2024bue}.
There it was used as a relation for $x<0$ perturbation theory, while here no such specification was made. The key point is that now that the string equation has a classical piece on the right hand side, perturbation theory for $x<0$ or $x>0$ has the same structure. More will be said about this below.

Moving to general orders, the string equation at genus $g\geq 1$ can be written as:
\begin{equation}
    \begin{split}
&2u_0\mathcal{R}_0\mathcal{R}_{2g}+\mathcal{R}_0^2u_{2g}+\frac{1}{2}\left[\delta_{g,1}\left(\mathcal{R}_0\mathcal{R}_0''-\frac{1}{2}\mathcal{R}_0'^2\right)+\mathcal{R}_0'\mathcal{R}_{2(g-1)}'-\mathcal{R}_0\mathcal{R}_{2(g-1)}''-\mathcal{R}_0''\mathcal{R}_{2(g-1)}\right]\\
&+2\sum_{n=1}^{g-1}u_{2(g-n)}\mathcal{R}_{2n}\mathcal{R}_0+\sum_{n=1}^{g-1}u_0\mathcal{R}_{2n}\mathcal{R}_{2(g-n)}+\frac{1}{4}\sum_{n=1}^{g-2}\left[\mathcal{R}_{2(g-n-1)}'\mathcal{R}_{2n}'-2\mathcal{R}_{2(g-n-1)}\mathcal{R}''_{2n}\right]\\
&+\sum_{L=1}^{g-1}\sum_{n=1}^{L-1}u_{2(g-L)}\mathcal{R}_{2(L-n)}\mathcal{R}_{2n}=0\ .\\
    \end{split}
\end{equation}
For example for genus 2 (i.e. $\mathcal{O}(\hbar^4)$) we set $g=2$ to obtain the following equation:
\begin{equation}
2u_0\mathcal{R}_0\mathcal{R}_4+\mathcal{R}_0^2u_4+\frac{1}{2}\left[\mathcal{R}_0'\mathcal{R}_2'-\mathcal{R}_0\mathcal{R}_2''-\mathcal{R}_0''\mathcal{R}_2\right]+2u_2\mathcal{R}_2\mathcal{R}_0+u_0\mathcal{R}_2^2=0\ ,
\end{equation}
where $\mathcal{R}_4$ is given by:
\begin{equation}
\begin{split}
    &\mathcal{R}_4=u_4\dot{G}_0+\frac{1}{288}G_0^{(6)} u_0'^4+\frac{11}{360}G_0^{(5)}u_0'^2u_0''-\frac{1}{120}G_0^{(4)}\left(10u_2u_0'^2-3u_0''^2-4u_0'u_0'''\right)\\
    &-\frac{1}{60}G_0^{(3)}(10u_0'u_2'+10u_2u_0''-u_0'''')-\frac{1}{6}G_0^{(2)}(-3u_2^2+u_2'')\ ,
\end{split}
\end{equation}
and using these together with some more $u_0$-derivatives of $G_0$ we can arrive at the following expression for $u_4(x)$:
\begin{equation}
\label{u4N=2Sol}
    u_4(x)=\frac{d^2}{dx^2}\left[\frac{u_0''(x)^3}{90 u_0'(x)^4}-\frac{7 u_0^{(3)}(x)
   u_0''(x)}{480 u_0'(x)^3}+\frac{u_0^{(4)}(x)}{288
   u_0'(x)^2}\right]\ .
\end{equation}
also used in Section~{\ref{sec:method}}.


\section{Perturbative treatment of the Gel'fand-Dikii equation}
\label{sec:perturbing-GD-equation}
As discussed in Section~\ref{sec:method}, the resolvent $\widehat{R}(x,E)$ satisfies the  (Gel'fand-Dikii) equation~\cite{Gelfand:1975rn}, repeated for convenience:
\begin{equation}
\label{GDResDiffEq}
    4(u-E)\widehat{R}^2-2\hbar^2\widehat{R}\widehat{R}''+\hbar^2\widehat{R}'^2-1=0\ .
\end{equation}
The goal is develop a series expansion for $\widehat{R}(x,E)$ ordered perturabtively in $\hbar$, since the topological organization is our interest:
 \begin{equation}
 \label{eq:resolvent-expansion}
     \widehat{R}(x,E)=\sum_{g=0}^\infty \hbar^{2g}\widehat{R}_g(x,E)+\cdots\ ,
 \end{equation}
 The leading term, $\widetilde{R}_0(x,E)$ solves~(\ref{GDResDiffEq}) with $\hbar=0$:
 \begin{equation}
     \label{eq:leadingGD}
      \widehat{R}_0(x,E)=-\frac{1}{2X^{1/2}}\quad \text{with}\quad X\equiv u_0(x)-E\ ,
 \end{equation}
 where a sign choice has been made, matching earlier conventions.
Going beyond the  terms developed in~ref.~\cite{Johnson:2024bue},  we can write a recursion relation for $\widehat{R}_g$   by inserting~(\ref{eq:resolvent-expansion})  and $u(x){=}\sum\hbar^{2g}u_{2g}(x)+\cdots$ into equation~(\ref{GDResDiffEq}) and organizing in powers of $\hbar$. It is derived in  Appendix \ref{RecursionRelDerRhatsAppendix}, and the result is ({\it c.f.} equation~(\ref{eq:gelfand-dikii-A})):
\begin{equation}
    \begin{split}
&8X\widehat{R}_0\widehat{R}_g+2\left[\widehat{R}_0'\widehat{R}_{g-1}'-\widehat{R}_0''\widehat{R}_{g-1}-\widehat{R}_{g-1}''\widehat{R}_0+2u_{2g}\widehat{R}_0^2+\delta_{g,1}\left(\widehat{R}_0''\widehat{R}_0-\frac{1}{2}(\widehat{R}_0')^{2}\right)\right]+8\sum_{L=1}^{g-1}u_{2(g-L)}\widehat{R}_0\widehat{R}_L\\
        &+4X\sum_{p=1}^{g-1}\widehat{R}_p\widehat{R}_{g-p}+\sum_{p=1}^{g-2}\left[\widehat{R}_p'\widehat{R}'_{g-p-1}-2\widehat{R}_p''\widehat{R}_{g-p-1}\right]+4\sum_{L=1}^{g-1}\sum_{\substack{p=1 \\ p\neq L}}^Lu_{2(g-L)}\widehat{R}_p\widehat{R}_{L-p}=0\ ,
    \end{split}
\end{equation}
where the shorthand $X$ was defined in equation~(\ref{eq:leadingGD}).
Here the recursion relation is organized so that all the $\widehat{R}_0$ and $\widehat{R}_g$ dependence is in the first line. In particular, we can explicitly substitute $\widehat{R}_0$ into the first set of terms and divide the entire recursion relation by $8X\widehat{R}_0$ to obtain the recursion relation in the following form:  
\begin{equation}
\label{RecursionforRAnyg}
\begin{split}
    &\widehat{R}_g-\frac{\widehat{R}_{g-1}''-4\sum_{L=1}^{g-1}u_{2(g-L)}\widehat{R}_L}{4X}-\frac{u_0'\widehat{R}'_{g-1}-u_0''\widehat{R}_{g-1}}{8X^2}-\frac{3u_0'^2\widehat{R}_{g-1}}{16X^3}-\frac{u_{2g}}{4X^{3/2}}+\delta_{g,1}\left[\frac{u_0''}{16X^{5/2}}-\frac{5u_0'^2}{64X^{7/2}}\right]\\
    &-X^{1/2}\sum_{p=1}^{g-1}\widehat{R}_p\widehat{R}_{g-p}-\frac{1}{4X^{1/2}}\sum_{p=1}^{g-2}\left[\widehat{R}_p'\widehat{R}'_{g-p-1}-2\widehat{R}_p''\widehat{R}_{g-p-1}\right]-\frac{1}{X^{1/2}}\sum_{L=1}^{g-1}\sum_{\substack{p=1}}^{L-1}u_{2(g-L)}\widehat{R}_p\widehat{R}_{L-p}=0\ .
\end{split}
\end{equation}
From this we can algebraically isolate  $\widehat{R}_g(x,E)$ for any desired $g$. For example for $\widehat{R}_1(x,E)$ we obtain the following expression:
\begin{equation}
\begin{split}
    &\widehat{R}_1(x,E)=\frac{\widehat{R}_{0}''}{4X}+\frac{u_0'\widehat{R}'_{0}-u_0''\widehat{R}_{0}}{8X^2}+\frac{3u_0'^2\widehat{R}_{0}}{16X^3}+\frac{u_{2}}{4X^{3/2}}-\left[\frac{u_0''}{16X^{5/2}}-\frac{5u_0'^2}{64X^{7/2}}\right]\\
    &=\frac{u_2}{4X^{3/2}}+\frac{u''_0}{16X^{5/2}}-\frac{5u_0'^2}{64X^{7/2}}\ ,
\end{split}
\end{equation}
and for $\widehat{R}_2(x,E)$ we have:
\begin{equation}
\begin{split}
     &\widehat{R}_2(x,E)=\frac{u_4}{4X^{3/2}}+\frac{u_2''-3u_2^2}{16X^{5/2}}+\frac{u_0^{(4)}-10(u_0'u_2'+u_2u_0'')}{64X^{7/2}}+\frac{7(10u_2u_0'^2-3u_0''^2-4u_0'u_0''')}{256X^{9/2}}\\
        &\hskip10.5cm+\frac{231u_0'^2u_0''}{512 X^{11/2}}-\frac{1155u_0'^4}{4096 X^{13/2}}\ .
\end{split}
\end{equation}
These  exactly  match the  expressions that can be obtained  by doing an $\hbar$ expansion of the asymptotic expansion of Gel'fand and Dikii:
\begin{equation}
    \widehat{R}(x,E)=\sum_{l=0}^\infty\frac{-(2l-1)!}{(-4)^ll!(l-1)!}\frac{R_k[u_0+\hbar^2u_2+\hbar^4u_4+\cdot\cdot\cdot]}{(-E)^{l+\frac{1}{2}}}\ .
\end{equation} 
Looking ahead, to obtain volumes at genus $g$ we must do integrals of the form: $\int^{{\mu}}\widehat{R}_g(x,E)dx$.
Recall that it was noticed that  when $u(x)$ satisfies the string equation then $\widehat{R}_{g}(x,E)$ becomes a total derivative:
\begin{equation}
    \widehat{R}_{g}(x,E)=\frac{d}{dx}\widehat{Q}_g(x,E)\ . 
\end{equation}This can be demonstrated  by construction. In particular, using the recursion relations for $\widehat{R}_g$ we can see that  it takes the form:
\begin{equation}
\widehat{R}_g(x,E)=\sum_{k=1}^{3g}\frac{\widehat{R}_{g,k}(x)}{(X)^{k+\frac{1}{2}}}\ ,\qquad X=u_0-E\ .
\end{equation}
We can then make an Anzatz for $\widehat{Q}$ of the form:
\begin{equation}
    \widehat{Q}_g=\sum_{k=1}^{3g-1}\frac{\widehat{Q}_{g,k}(x)}{(u_0-E)^{k+\frac{1}{2}}}\ ,
\end{equation}
which leads to:
\begin{equation}
\widehat{Q}'_g=\frac{\widehat{Q}'_{g,1}}{X^{3/2}}-\frac{\left(3g-\frac{1}{2}\right)u_0'\widehat{Q}_{g,3g-1}}{X^{3g+\frac{1}{2}}}+\sum_{k=2}^{3g-1}\frac{\widehat{Q}'_{g,k}-(k-\frac{1}{2})u_0'\widehat{Q}_{g,k-1}}{X^{k+\frac{1}{2}}}\ .
\end{equation}
Setting $\widehat{R}_g =\widehat{Q}'_g$ gives us the following recursion relation:
\begin{equation}
\label{RecursionForsmallq}
  \widehat{Q}'_{g,k}-\left(k-\frac{1}{2}\right)u_0'\widehat{Q}_{g,k-1}-\widehat{R}_{g,k}=0\ .
\end{equation}
We solve this with boundary conditions that, $\widehat{Q}_{g,0}=\widehat{Q}_{g,3g}=\widehat{R}_{g,0}=0$. Note that we already know what $\widehat{R}_{g,k}$ is\footnote{A quick way to isolate for a particular coefficient $\widehat{R}_{g,k}$ in the expansion of $\widehat{R}_g$ is to use the relation, $\widehat{R}_{g,k}=\lim_{X\to 0}\frac{1}{(3g-k)!}\frac{d^{3g-k}}{dX^{3g-k}}\left(X^{3g+\frac{1}{2}}\widehat{R}_g\right)$.}  by recursively solving for $\widehat{R}_g$ in equation (\ref{RecursionforRAnyg}). So we can recursively solve for $\widehat{Q}$ in terms of $\widehat{R}$. However, we will solve it starting at $k=3g$ and work our way down to $k=1$. The reason for this is because if we start at $k=1$ then we must integrate to solve for $\widehat{Q}$ whereas starting from $k=3g$ one only needs to differentiate\footnote{Of course one could also go the other way as well (i.e. start at $k=1$ and go toward $k=3g$) starting with the solution to the string equation and integrating repeatedly should give the same result. This direction, however, in practice is considerably harder to implement for any arbitrary genus $g$ calculation.}. In this case the final term at $k=1$ tells us that $u_{2g}, g\geq 1$ is a solution to the string equation in particular we have $u_{2g}=4\widehat{Q}'_{g,1}$. In Appendix \ref{AppendixConstructQ} we demonstrate the procedure to explicitly construct $\widehat{Q}_g$ for $g=1,2$  and we find:
\begin{equation}
\label{ExpressionForQs}
\begin{split}
    &\widehat{Q}_1(x,E)=\frac{u_0'}{32X^{5/2}}-\frac{u_0''}{48u_0'X^{3/2}}\\
    &\widehat{Q}_2(x,E)=\frac{105u_0'^3}{2048X^{11/2}}-\frac{203u_0'u_0''}{3072X^{9/2}}+\frac{29u_0'u_0'''-3u_0''^2}{1536u_0'X^{7/2}}-\frac{17u_0''^3-34u_0'u_0''u_0'''+15u_0'^2u_0^{(4)}}{3840u_0'^3X^{5/2}}\\
        &\hskip5.5cm -\frac{64u_0''^4-111u_0'u_0''^2u_0'''+21u_0'^2u_0'''^2+31u_0'^2u_0''u_0^{(4)}-5u_0'^3u_0^{(5)}}{5760u_0'^5X^{3/2}}\ .
\end{split}
\end{equation}
and, last but not least,  we display  $\widehat{Q}_3$:
\begin{equation}
\label{eq:Q3}
    \begin{split}
    &\widehat{Q}_3(x,E)=\frac{25025u_0'^5}{65536X^{17/2}}-\frac{77077u_0'^3u_0''}{98304X^{15/2}}+\frac{11u_0'(633u_0''^2+503u_0'u_0''')}{24576X^{13/2}}\\
    & \hskip1.5cm-\frac{57u_0''^3+656u_0'u_0''u_0'''+607u_0'^2u_0^{(4)}}{12288u_0'X^{11/2}}\\
    & \hskip1.5cm+\frac{-2385u_0''^4+4740u_0'u_0''^2u_0'''-1315u_0'^2u_0'''^2-1638u_0'^2u_0''u_0^{(4)}+1006u_0'^3u_0^{(5)}}{110592u_0'^3X^{9/2}}\\
    &\hskip1.5cm +\frac{\widehat{Q}_{3,3}}{X^{7/2}}+\frac{\widehat{Q}_{3,2}}{X^{5/2}}+\frac{\widehat{Q}_{3,1}}{X^{3/2}}\ , 
    \end{split}
\end{equation}
where:
\begin{equation}
\begin{split}
    &\widehat{Q}_{3,3}=-\frac{14055u_0''^5-34320u_0'u_0''^3u_0'''+11094u_0'^2u_0''^2u_0^{(4)}}{387072u_0'^5}\\
    &\hskip1.5cm-\frac{u_0'^2u_0''\left(16820u_0'''^2-2751u_0'u_0^{(5)}\right)+u_0'^3\left(-5413u_0'''u_0^{(4)}+539u_0'u_0^{(6)}\right)}{387072u_0'^5}\\
\end{split}
\end{equation}
\begin{equation}
\begin{split}
&\widehat{Q}_{3,2}=\frac{-54080u_0''^6+148710u_0'u_0''^4u_0'''-45462u_0'^2u_0''^3u_0^{(4)}}{967680u_0'^7}\\
    &\hskip1.5cm+\frac{9u_0'^2u_0''^2\left(-10965u_0'''^2+1124u_0'u_0^{(5)}\right)+14u_0'^3u_0''\left(2721u_0'''u_0^{(4)}-119u_0'u_0^{(6)}\right)}{967680u_0'^7}\\
    &\hskip1.5cm +\frac{9045u_0'''^3-2483u_0'(u_0^{(4)})^2-3764u_0'u_0'''u_0^{(5)}+175u_0'^2u_0^{(7)}}{967680u_0'^4}\\
\end{split}
\end{equation}
\begin{equation}
\begin{split}
&\widehat{Q}_{3,1}=-\frac{179200u_0''^7-530880u_0'u_0''^5u_0'''+152256u_0'^2u_0''^4u_0^{(4)}}{1451520u_0'^9}\\
    &\hskip1.5cm -\frac{3u_0'^2u_0''^3\left(142750u_0'''^2-10607u_0'u_0^{(5)}\right)+9u_0'^3u_0''^2\left(-19147u_0'''u_0^{(4)}+549u_0'u_0^{(6)}\right)}{1451520u_0'^9}\\
    &\hskip1.5cm -\frac{3u_0''\left[-26835u_0'''^3+7209u_0'u_0'''u_0^{(5)}+u_0'\left(4643(u_0^{(4)})^2-182u_0'u_0^{(7)}\right)\right]}{1451520u_0'^6}\\
    &\hskip1.5cm -\frac{\left[19422u_0'''^2u_0^{(4)}+35u_0'^2u_0^{(8)}-15u_0'\left(146u_0^{(4)}u_0^{(5)}+93u_0^{(3)}u_0^{(6)}\right)\right]}{1451520u_0'^5}\ .
\end{split}
\end{equation}
What these formulae strongly suggest  is that {\it if $u(x)$ satisfies the string equation}, at every order in the genus expansion the integral of $\widehat{R}_g(x,E)$  only depends on the $u_0$ and is derivatives at the endpoint of the domain of integration. The derivatives of $u_0(x)$ vanish at $-\infty$ and so it is only at $x{=}\mu$ that we get contributions. This elegantly defines a physical object at each genus that (after Laplace transform, see below) will be a (generalized) Weil-Petersson volume.

While it would be nice to have a proof that this happens to all orders (we have not found a complete argument), the recursive procedure that we have constructed is rather convincing, and we have constructed several examples at even higher genus. As emphasized in ref.~\cite{Johnson:2024bue}, it really {\it must} be true, because of the simple polynomial structure of the Weil-Petersson volumes in the classic case of bordered hyperbolic Riemann surfaces~\cite{Mirzakhani:2006fta}.

\section{Defining volumes}
\label{sec:defining-volumes}
Now, $\widehat{Q}_g(x,E)$ is a polynomial in $X^{-\frac12}$, where $X=u_0(x)-E$. Evaluating at $x=\mu$ and will therefore result in $(E_0-E)^{-\frac12}$ dependence. Given this new position of the cut in the $E$ plane, it is natural  to change to a uniformizing variable $z$ using $E_0-E=z^2$, ({\it c.f.} $-E=z^2$ previously). This still gives a factor of $-2z$ in the Jacobian in going from $E$ integration to $z$ integration and so the prescription for defining a volume should now be:
\begin{eqnarray}
    {W}_{g,1}(z)&\equiv&-{2}z\int_{-\infty}^\mu{\widehat R}_g(x,z^2+E_0)\, dx=-2z\widehat{Q}_g(\mu,z^2+E_0)
    =K_{g,1}\int_0^\infty \! bdb\,\e^{-b z}\,  V_{g,1}(b)\ ,
\label{eq:new:W-vs-V}
\end{eqnarray}
where $K_{g,1}$ is a normalization constant, previously set to unity ({\it c.f.} equation~(\ref{eq:W-vs-V})). To align with conventions to come~\cite{Turiaci:2023jfa} for ${\cal N}{=}2$ JT supergravity, we will choose $K_{g,1}{=}(2\pi)^{2g-1}$. We also find later that a natural normalization for (small) ${\cal N}{=}4$ is $K_{g,1}{=}4^{1-2g}$. Since the $W_{g,1}$ come multiplied by $\hbar^{2g-1}$ in the genus expansion, this amounts to a harmless rescaling of $\hbar$, or a shift in $S_0$. Hence more generally, the  normalization factors for volumes $V_{g,n}$ are $K_{g,n}=(2\pi)^{-\chi}$ (${\cal N}{=}2$) or $K_{g,n}=4^\chi$ (small ${\cal N}{=}4$), where the Euler number $\chi{=}2{-}2g{-}n$.
Put differently, we can extract the one-boundary volumes from:
\begin{eqnarray}
 &&V_{g,1}(b)=K_{g,1}^{-1}b^{-1}\mathcal{L}_z^{-1}(-2z\widehat{Q}_g({\mu},-z^2+E_0))[b]\ ,\nonumber\\
    &&\hskip1cm \text{with}\quad K_{g,1}= \left(\frac{1}{2\pi}\right)^{1-2g}\ , \quad \text{for}\quad {\cal N}=2 \,\,\text{SJT}\nonumber\\
    &&\hskip1cm\text{and}\,\,\quad K_{g,1}= 4^{1-2g}\ , \qquad \,\quad\,\,\text{for}\quad {\cal N}=4 \,\,\text{SJT}\ .\label{eq:Volume-Expression}
\end{eqnarray}
It is time to do some examples and demonstrate the method for extended superymmetric JT gravity.

\section{Volumes for ${\cal N}=2$ JT supergravity}

\subsection{ODE approach to computing volumes}
\label{sec:ODE-approach-Neq2}
The first thing to do is compute expressions for the derivatives of $u_0(x)$ when evaluated at the Fermi surface $x{=}\mu$.
Recall that the tree level string equation is given by equation~(\ref{eq:leading-string-alt}) (or~(\ref{eq:tree-string-equation})). As done in the prototype cases in Section~\ref{sec:method}, the $x$-derivatives of $u_0(x)$ are simply related, through the string equation, to $u_0$-derivatives of the equation itself, amounting to $u_0$-derivatives of $G_0(u_0)$. So a study of~$G_0(u_0)$ is warranted.
By definition, the leading string equation, when evaluated at $x{=}\mu$, should be solved by $u_0{=}E_0$:
\begin{equation}
    G_0(E_0)=-{\mu}+\frac{\widetilde{\Gamma}}{u_0^{1/2}}\ ,\quad\text{with}\quad \widetilde{\Gamma}=\frac{\sin(2\pi\sqrt{E_0})}{4\pi^2}\ .
\end{equation}
Now, explicitly there is the dependence:
\begin{equation}
    G_0(E_0)=\sum_{k=1}^\infty{t}_kE_0^k\ ,\quad\text{with}\quad
    {t}_k=\frac{\pi^{k-1}J_k(2\pi\sqrt{E_0})}{2(2k+1)k!E_0^{k/2}}\ .
\end{equation}
In the Appendix \ref{GExactatE0} it is  proven  that the expression for $G_0(E_0)$ can be resummed, giving:
\begin{equation}
    G_0(E_0)=\frac{1}{2\pi}\left[-J_0(2\pi\sqrt{E_0})+\frac{\sin(2\pi\sqrt{E_0})}{2\pi\sqrt{E_0}}\right]\ .
\end{equation}
This implies that:
\begin{equation}
    {\mu}=\frac{J_0(2\pi\sqrt{E_0})}{2\pi} = t_0\ ,
\end{equation}
which is a pleasing consistency check.
In Appendix \ref{GExactatE0}  the following identity for the $p$th $u_0$-derivative of $G_0$ is derived:
\begin{equation}
    G^{(p)}_0(E_0)=\frac{\pi^{2p-1}(2p)!}{2p!(1+2p)!}{}_1F_2(1;p+1,p+3/2;-\pi^2E_0)\ ,
\end{equation}
where the expression above is valid for $p=1,2,3,\ldots $ A useful alternative form is:
\begin{equation}
\label{StructureOfGDer}
    G_0^{(p)}(E_0)=\frac{\sqrt{\pi}}{\Gamma(\frac12-p)}\frac{\widetilde{\Gamma}}{E_0^{p+1/2}}+\sum_{k=1}^{p}\frac{c_k^{(p)}}{E_0^k}\ ,
\end{equation}
where  the branch and pole structure of the result are separated out. From  these results it can be demonstrated that the derivatives of $u_0$, when evaluated at the Fermi surface  $x{=}\mu$, is actually a polynomial in $E_0$ of the form:
\begin{equation}
\label{eq:u0s-derivatives}
    u_0^{(p)}({\mu})=\sum_{k=1}^{p} a_k^{(p)}E_0^k\ .
\end{equation}
For example, for the first 5 derivatives we  have:
\begin{equation}
\label{deru0atFermi}
\begin{split}
    &u_0'({\mu})=-4\pi E_0\\
    &u_0''({\mu})=24\pi^2E_0-16\pi^4E_0^2\\
    &u_0'''({\mu})=-192\pi^3E_0+416\pi^5E_0^2-160\pi^7E_0^3\\
    &u_0^{(4)}({\mu})=1920\pi^4E_0-8960\pi^6E_0^2+9408\pi^8E_0^3-\frac{7808\pi^{10}}{3}E_0^4\\
    &u_0^{(5)}({\mu})=-23040\pi^5E_0+192000\pi^7E_0^2-390016\pi^9E_0^3+270592\pi^{11}E_0^4-\frac{176512\pi^{13}}{3}E_0^5\ .
\end{split}
\end{equation}
Actually this is direct result of the structure we mention in equation (\ref{StructureOfGDer}), essentially the branch structure cancels in forming the $u_0^{(p)}(\mu)$ and we are left with a finite number of terms. 

Finally, notice that in the case of {\it e.g.,} ${\widehat R}_1(x,E)$ of Section~\ref{sec:method} one had expression~(\ref{eq:totally-awesome}) for the bosonic case, resulting from the special relation between $u_2$ and $u_0$ and its derivatives. For the ${\cal N}{=}1$ supersymmetric case, however, one had that all appearances of $u_0(x)$ and its derivatives must vanish, and so ${\widehat R}_1(x,E)=\frac14 \frac{u_2}{(-E)^{3/2}}=-\frac{1}{16}\frac{1}{x^2(-E)^{3/2}}$, which is quite different. The two kinds of results can be seen to be related to each other by taking the limit $E_0\to0$ and seeing that certain ratios of $u_0$ derivatives survive, even if the derivatives themselves do not. In particular, the second term in~(\ref{eq:totally-awesome}) vanishes, while the first is $-\frac{1}{48}\frac{u_0^{''}}{u_0^\prime}\frac{1}{(-E)^{3/2}}=-\frac{1}{48}\frac{(24\pi^2E_0 -16\pi^4 E_0^2)}{-4\pi E_0}\frac{1}{(-E)^{3/2}}$ which becomes $2\pi\times -\frac{1}{16}\frac{1}{(-E)^{3/2}}$, precisely the ($t_k$-independent) ${\cal N}{=}1$ result, as it should, up to the overall $2\pi$ that can be traced to a different value of $\mu$ between the two theories in the limit (unity vs $\frac{1}{2\pi}$). This is a nice example of how the universal treatment of the relation between the $u_{2g}(x)$ and $u_0(x)$ and its derivatives collapses back to the asymmetry between the $x<0$ and $x>0$ expansions  in the  case of vanishing ${\widetilde\Gamma}$.

 This also connects to the observation that there will always be structure such that there is a bosonic subsector to the volumes, appearing at highest order in $E_0$. (As observed in ref.~\cite{Turiaci:2023jfa} for ${\cal N}{=}2$.) This must follow from the fact that the highest order in~$E_0$ parts of the combinations of derivatives that appear in ${\widehat R}_g(x,E)$ {\it must be the same} as in the bosonic theory. Indeed, as an example the highest order in $E_0$ part of $u_0^{\prime\prime}/u_0^\prime$ has coefficient $(2\pi)\times 2\pi^2$, and that ratio is also $2\pi^2$ in the bosonic case (see equation~(\ref{eq:some-derivatives}). 
 
 This all can be made more general.
The highest component of $E_0$ in final expressions come with coefficient  $a^{(p)}_p$, (see equation~(\ref{eq:u0s-derivatives})), and so it is worth pausing to see if a pattern lurks in them. Derivatives of the tree level string equation will involve taking derivatives of two types of terms one involves $G_0(u_0(x))$ itself and the other will involve $\widetilde{\Gamma}/u_0^{1/2}$ derivatives of both terms will contain some explicit $\widetilde{\Gamma}$ dependence. However, the fact that all derivatives we computed are actually polynomials in~$E_0$ with no explicit $\widetilde{\Gamma}$ dependence motivates us to focus our attention on terms that are independent of $\widetilde{\Gamma}$. Such terms would be universally appearing regardless of if $\widetilde{\Gamma}$ is zero or non-zero. As we will demonstrate, such terms completely control the highest component of $E_0$ which will be proportional to the bosonic volume.     

Begin by considering the $p$th $x$-derivative of $G_0(u_0(x))$, for $p\geq 2$. It is straightforward to see that it will always take the form:
\begin{equation}
\label{derivativeOfu0AltRep}
u_0^{(p)}=\frac{1}{\dot{G_0}}\left[\sum_{k=2}^pG_0^{(k)}Y_{p,k}[u_0]\right]\ ,   
\end{equation}
where $Y_{p,k}[u_0]$ can only ever depend on products of lower order derivatives and at the Fermi surface $Y_{p,k}[u_0]$ is a polynomial of degree $p$ in $E_0$. In addition, $G_0^{(k)}$ itself will be a polynomial in $E_0^{-1}$ (ignoring the terms involving $\widetilde{\Gamma}$) . Based on these observations it follows that the highest power coefficient of the $p\geq 2$-th derivative is given by:
\begin{equation}
    a_p^{(p)}=\frac{1}{c_1^{(1)}}\left[\sum_{k=2}^pc_1^{(k)}Y_{p,k}[u_0]\right]\bigg\vert_{u_0^{(l)}\to a_l^{(l)}}\ ,
\end{equation}
where $c^{(k)}_1$ is defined in equation~(\ref{StructureOfGDer}). The expression above now defines a kind of recursion relation which relates $a_p^{(p)}$ to the $a_l^{(l)}$s for $l<p$. For example, for $\mathcal{N}{=}2$ SJT we have $c^{(k\geq 2)}_{1}=\frac{\pi^{2k-3}}{4(k-1)!}$ and $c_1^{(1)}{=}(4\pi)^{-1}$. We substitute that in and write:
\begin{equation}
\label{appRelations}
    a_p^{(p)}=\sum_{k=2}^p\frac{\pi^{2k-2}}{(k-1)!}(Y_{p,k}[u_0])\bigg\vert_{u_0^{(l)}\to a_l^{(l)}}\ ,
\end{equation}
For $\mathcal{N}=2$ we have $a_1^{(1)}=-4\pi$ and using this as an initial condition in equation~(\ref{appRelations}) we can recursively reproduce $a_p^{(p)}$ for any $p\geq 2$ as expected. However, we can actually take the exact same result and just substitute it into $a_1^{(1)}=-2$ for JT gravity and also recursively reproduce all the higher derivatives of JT gravity at the Fermi surface as well. It strongly suggests that the combination of terms in the sum of equation~(\ref{appRelations}) are universal with the only tunable parameter being the first derivative or initial condition of the recursion relation. In fact, one can check that the following relation holds between $a_p^{(p)}$ of $\mathcal{N}=2$ and the $p$-th derivative of JT solution at the Fermi surface:
\begin{equation}
    \frac{a_p^{(p)}\vert_{\mathcal{N}=2}}{u_0^{(p)}(\mu=0)|_{JT}}=(2\pi)^p\ .
\end{equation}
The origin of the $2\pi$ can be traced back to the ratio of the first derivatives  $u_0^\prime$ ({\it i.e.} the initial condition of the recursion relation in equation~(\ref{appRelations})) of the tree level solutions at the Fermi surface of two theories which is exactly $2\pi=\frac{-4\pi}{-2}$. 

Analyzing the expressions for $\widehat{Q}_g$ we see that the number of derivatives in the numerator minus the number of derivatives in the denominator is the same in every term and equal to $2g-1$. It follows that the coefficient multiplying the highest order in $E_0$ of $\widehat{Q}_g|_{\mathcal{N}=2}$ is exactly $(2\pi)^{2g-1}\widehat{Q}_g|_{\text{JT}}$. This explains the normalization factor $K_{g,1}$ in equation~(\ref{eq:Volume-Expression}) for $\mathcal{N}{=}2$. A similar structure of derivatives also appears in $\mathcal{N}{=}4$ with $J$ playing the role of $E_0$. In that case we  see that the appropriate relation will be:
\begin{equation}
    \frac{a_p^{(p)}\vert_{\mathcal{N}=4}}{u_0^{(p)}(\mu=0)|_{JT}}=\left(\frac{1}{4}\right)^p\ ,
\end{equation}
where the ratio of first derivatives $u_0^\prime$ gives the $\frac{1}{4}=\frac{-1/2}{-2}$. Therefore, to ensure we get exactly the bosonic volume at highest order in $J$ we must multiply the $\widehat{Q}_g$ by $4^{2g-1}$ which is precisely what $K_{g,1}$ does in equation~(\ref{eq:Volume-Expression}) in the $\mathcal{N}=4$ volume definition.

Already at tree level the effects of the normalization can be seen by computing $V_{0,n}$, which, as stated in Section~\ref{sec:method}, just comes from using standard formula~(\ref{eq:correlator}) and $n$ factors of the trumpet integral. The analogue of the computation in~(\ref{eq:3-point}) uses a rescaling of the trumpet~ (\ref{eq:trumpet-voluntary}) due to the threshold energy: $Z_{\rm tr}(\beta,b)\to Z_{\rm tr}(\beta,b)\e^{-\beta u_0(\mu)}$. Then since  $u_0(\mu)=E_0$, and $u_0^\prime(\mu)=-4\pi E_0$ from~(\ref{deru0atFermi}),  remembering to divide by  $(2\pi)^{{-}\chi}{=}2\pi$, as required by the comment above~(\ref{eq:Volume-Expression}) yields the result:
\begin{equation}
\label{eq:3-point-new}
    V_{0,3}(b_1,b_2,b_3)= E_0\ ,
\end{equation}
which agrees with the result of Turiaci and Witten in ref.~\cite{Turiaci:2023jfa}. Further agreement comes from working out $V_{0,4}(\{b_i\})$ along the lines given in~(\ref{eq:4-point}), where use of the derivatives in~(\ref{deru0atFermi}) gives, after dividing by~$(2\pi)^{{-}\chi}{=}(2\pi)^2$.
\begin{equation}
\label{eq:4-point-new}
      V_{0,4}(b_1,b_2,b_3,b_4)=-3E_0+\frac{E_0^2}{2}\left(4\pi^2+b^2_1+b^2_2+b^2_3+b_4^2\right)\ ,
\end{equation}
again in agreement with ref~\cite{Turiaci:2023jfa}. Given the procedures outlined here to generate derivatives of $u_0^{(p)}(\mu)$, equation~(\ref{eq:u0s-derivatives}), equation~(\ref{eq:correlator}) together with the trumpet factors (\ref{eq:trumpet-voluntary}) (times $\e^{-\beta E_0}$) amounts to  a simple way to generate all the $V_{0,n}(\{b_i\})$ for ${\cal N}{=}2$.

Of course, our goal is not really tree level results, but the higher genus results made readily available in the ODE approach, and so that is what we turn to next.
Substituting the results of equations (\ref{ExpressionForQs}) and~(\ref{deru0atFermi}) into equation~(\ref{eq:Volume-Expression}) we can now explicitly display the volumes that result from our procedures.  For genus 1 we have:
\begin{equation}
\label{eq:V11-Neq2}
    V_{1,1}(b)=-\frac{1}{8}+\frac{(b^2+4\pi^2)}{48}E_0\ ,
\end{equation}
which, happily, matches that derived by  Turiaci and Witten~\cite{Turiaci:2023jfa}.
Notice the part proportional to~$E_0$ is the bosonic volume for this order, while the $E_0$ independent piece is the ${\cal N}{=}1$ volume ({\it c.f.} equations~(\ref{eq:V11-basics})).
Meanwhile for genus 2 and 3, some swift computations give:
\begin{equation}
\label{eq:new-V21}
\begin{split}
&V_{2,1}^{(q)}(b)=-\frac{3}{256} \left(b^2 + 4\pi^2\right) \\
&\hskip1.5cm + \frac{1}{15360} \left(b^2 + 4\pi^2\right) \left(107b^2 + 1772\pi^2\right)E_0 \\
&\hskip1.5cm - \frac{1}{92160} \left(b^2 + 4\pi^2\right) \left(29b^4 + 1320b^2\pi^2 + 12592\pi^4\right)E_0^2 \\
&\hskip1.5cm + \frac{1}{2211840} \left(b^2 + 4\pi^2\right) \left(b^2 + 12\pi^2\right)\left(5b^4 + 384 b^2\pi^2 + 6960\pi^4\right)E_0^3\ ,
\end{split}
\end{equation}
and again, notice that the highest order in $E_0$ is the bosonic volume~(\ref{eq:V21-basics-bos}) for this order. Finally:
\begin{eqnarray}
\label{eq:new-V31}
    &&V_{3,1}(b)=-\frac{3(b^2+4\pi^2)(5b^2+92\pi^2)}{8192}+\frac{(b^2+4\pi^2)(4371b^4+258120\pi^2b^2+3131456\pi^4)}{3096576}E_0\\
    &&\hskip1.5cm-\frac{(b^2+4\pi^2)E_0^2}{185794560} \times A(b) +\frac{(b^2+4\pi^2)E_0^3}{5573836800} \times B(b) -\frac{(b^2+4\pi^2)E_0^4}{44590694400}\times C(b)\ +V^{\text{Bos.}}_{3,1}(b)E_0^5\ ,\nonumber
\end{eqnarray}
where:
\begin{eqnarray}
\label{eq:V31-bos}
   && V^{\text{Bos}}_{3,1}(b) = \frac{\left(b^2+4 \pi ^2\right)}{267544166400} \times\left(5 b^{12}+2136 \pi ^2 b^{10}+354432 \pi ^4 b^8+28796032 \pi ^6 b^6\right.\\
    &&\hskip5.5cm\left.
    +1187231232 \pi ^8 b^4
    +22848015360 \pi ^{10}
   b^2+152253906944 \pi ^{12}\right)\ . \nonumber
\end{eqnarray}
and we have defined:
\begin{eqnarray}
 &&   A(b)\equiv (20525b^6+2453260\pi^2b^4+84325760\pi^4b^2+801765504\pi^6) \ ,\\
   && B(b)\equiv (11733 b^8+2352568 \pi ^2 b^6+157864448 \pi ^4 b^4+4095016768 \pi ^6 b^2+33156774528 \pi^8)\ ,\nonumber\\
 && C(b)\equiv (539 b^{10}+163236 \pi ^2 b^8+18091216 \pi ^4 b^6+900291776 \pi ^6 b^4+19557966336 \pi
   ^8 b^2\nonumber\\
   &&\hskip11.5cm+141438717952 \pi ^{10})\ . \nonumber
\end{eqnarray}
These $V_{g,n}$ are polynomials in $E_0$ with coefficients that are polynomials in $b$.

\subsection{Turiaci-Witten  approach to computing volumes}
\label{sec:TW-approach}
While we see that $V_{1,1}$ agrees with the result of ref.~\cite{Turiaci:2023jfa}, it is important to check that the agreement persists to higher genus. Could the $V_{2,1}$ and $V_{3,1}$ examples of equations~(\ref{eq:new-V21}) and~(\ref{eq:new-V31}), which are new, be hallucinations brought about by a flaw in the ODE method? To check,  in this section we compute some examples using the methods of Turiaci and Witten. This also gives the opportunity to show how comparatively straightforward the new ODE method is.

\subsubsection*{$\bullet$ The recursion relation.}
Here we give an overview the recursion relation that was given in ref.~\cite{Turiaci:2023jfa}. It is given below:
\begin{equation}
\begin{split}
    &bV_{g,n+1}^{(q)}(b,B)=\frac{1}{2}\int_0^\infty b'db'b''db''\left[E_0D_2(b,b',b'')-D_0(b,b',b'')\right]\\
    &\hskip3.0cm\times\left[V^{(q)}_{g-1,n+2}(b',b'',B)+\sum_{stable}V^{(q)}_{h_1,|B_1|+1}(b',B_1)V^{(q)}_{h_2,|B_2|+1}(b'',B_2)\right]\\
    &\hskip3cm+\sum_{k=1}^{|B|}\int_{0}^\infty b'db'\left[E_0(b-T_2(b,b',b_k))+T_0(b,b',b_k)\right]V_{g,n}^{(q)}(b',B/b_k)\ ,
    \end{split}
    \end{equation}
    (here the superscript $(q)$ refers to the charge of the ${\cal N}{=}2$ multiplet, related to $E_0$. The details do not matter, as it is just notation here), and where: 
\begin{equation}
\begin{split}
    &D_0(b,b',b'')=\frac{1}{8}\left[\frac{1}{\cosh^2\left(\frac{b-b'-b''}{4}\right)}-\frac{1}{\cosh^2\left(\frac{b+b'+b''}{4}\right)}\right]\\
    &D_2(b,b',b'')=\ln\left[e^b\left(\frac{\cosh\left(\frac{b-b'-b''}{4}\right)}{\cosh\left(\frac{b+b'+b''}{4}\right)}\right)^2\right]\\
    &T_0(b,b',b'')=\frac{1}{16}\left[\frac{1}{\cosh^2\left(\frac{b+b'-b''}{4}\right)}+\frac{1}{\cosh^2\left(\frac{b+b'+b''}{4}\right)}-\frac{1}{\cosh^2\left(\frac{b-b'-b''}{4}\right)}-\frac{1}{\cosh^2\left(\frac{b-b'+b''}{4}\right)}\right]\\
    &T_2(b,b',b'')=\ln\left[\frac{\cosh\left(\frac{b+b'}{2}\right)+\cosh\left(\frac{b''}{2}\right)}{\cosh\left(\frac{b-b'}{2}\right)+\cosh\left(\frac{b''}{2}\right)}\right]\ .
\end{split}
\end{equation}
where $B=\{b_1,b_2,..,b_n\}$ and in the stable sum we sum over configurations that satisfy $B_1\cup B_2=B$ and $B_1\cap B_2=\varnothing$ with $g=h_1+h_2$ ignoring any terms involving genus zero volumes with less than three geodesic boundaries. Here the notation $B/b_k$ means the set $B$ with $b_k$ excluded. To explicitly evaluate the recursion relations we need to be able to perform the integrals. These integrals in general will simply be of the form:
\begin{equation}
\begin{split}
    &\mathcal{A}_{0}(b;p,q)=\int_0^{\infty}db'db''(b')^p(b'')^qD_0(b,b',b'')\ ,\quad\mathcal{A}_2(b;p,q)=\int_0^{\infty}db'db''(b')^p(b'')^qD_2(b,b',b'')\ ,\\
    &\mathcal{T}_0(b,b_1;p)=\int_0^\infty db'(b')^pT_{0}(b,b',b_1)\ , \quad
    \mathcal{T}_2(b,b_1;p)=\int_0^\infty db'(b')^p[b-T_{2}(b,b',b_1)]\ .
\end{split}
\end{equation}
In Appendix \ref{KernelIntegralsN=2} we explicitly give expressions for the integrals we need to do the computations that will follow. Another point to make is that we need to also provide initial conditions to the recursion relations. These are given by:
\begin{equation}
\label{initalConditionN=2Vol}
V^{(q)}_{0,3}(b_1,b_2,b_3)=E_0\quad \text{and}\quad 
    V^{(q)}_{1,1}(b)=-\frac{1}{8}+\frac{b^2+4\pi^2}{48}E_0\ .
\end{equation}
These were computed/confirmed in our approach in the previous subsection (\ref{sec:ODE-approach-Neq2}).
With these in hand we can begin our computations of various volumes we will need using the recursion approach of this section. To give the reader a sense of how the computations flow and what intermediate volumes are required for computations of a particular $V_{g,n}$ it is useful to consider the listing of volumes given in Table~\ref{tab:volume-table}.
%
%
\begin{table}[h]
    \centering
    \[
\begin{NiceMatrix}[columns-width=1.5cm,code-for-first-row=\scriptstyle,code-for-first-col=\scriptstyle,hvlines]
     & g = 0 & g = 1 & g = 2 & g = 3 \\
n = 1 & \cancel{V_{0,1}} & \fcolorbox{blue}{white}{$V_{1,1}$} & V_{2,1} & V_{3,1} \\
n = 2 & \cancel{V_{0,2}} & V_{1,2} & V_{2,2} & V_{3,2} \\
n = 3 & \fcolorbox{blue}{white}{$V_{0,3}$} & V_{1,3} & V_{2,3} & V_{3,3} \\
n = 4 & V_{0,4} & V_{1,4} & V_{2,4} & V_{3,4}
\CodeAfter
  \tikz \draw[->, thick] (4-2) -- (3-3);
  \tikz \draw[->, thick] (3-3) -- (2-4);
  \tikz \draw[->, thick] (5-2) -- (4-3);
  \tikz \draw[->, thick] (4-3) -- (3-4);
  \tikz \draw[->, thick] (3-4) -- (2-5);
\end{NiceMatrix}
\]
    \caption{A table showing the chain of volumes to be computed (running along the arrows) in order to compute a particular volume. Initial conditions are boxed in blue.}
    \label{tab:volume-table}
\end{table}
There, we crossed out the volumes of genus $0$ with fewer than three boundaries as those are excluded in the topological recursion formula. What makes this table useful is it gives a useful pictorial representation of what volumes we need to calculate before we compute any particular volume of choice. The first non-trivial volume we can calculate given the initial conditions $V_{0,3}$ and $V_{1,1}$ is $V_{1,2}$. Then we can compute $V_{2,1}$. Once we have filled the diagonal with $V_{0,3}, V_{1,2}, V_{2,1}$ we go back to the genus 0 column and start at $V_{0,4}$ which requires data about the preceding volumes. We populate the table diagonally upwards going to $V_{1,3}$, then $V_{2,2}$, then $V_{3,1}$ and repeat this prescription for higher genus as needed. The point to illustrate here is that higher genus computation require increasingly many intermediate volumes (the total number of preceding volumes roughly scaling like $\sim g^2/2$ for $g\gg 1$). Even to get $V_{3,1}$ one has to compute the intermediate volumes $V_{1,2},V_{2,1},V_{0,4},V_{1,3},V_{2,2}$. Furthermore, the computation of the these volumes involve doing several non-trivial integrals (whose values we list in Appendix \ref{KernelIntegralsN=2}).

\subsubsection*{$\bullet$ Computing $V^{(q)}_{1,2}(b,b')$.}
As we described above we must begin our computations at $V^{(q)}_{1,2}$. In this case in the recursion relation we set $n=1$ and $g=1$ with $B=b_1$ and $|B|=1$. We write:
\begin{equation}
\begin{split}
    &bV_{1,2}^{(q)}(b,b_1)=\frac{1}{2}\int_0^\infty b'db'b''db''\left[E_0D_2(b,b',b'')-D_0(b,b',b'')\right]\\
    &\hskip 4cm\times\left[V^{(q)}_{0,3}(b',b'',b_1)+\sum_{stable}V^{(q)}_{h_1,|B_1|+1}(b',B_1)V^{(q)}_{h_2,|B_2|+1}(b'',B_2)\right]\\
    &\hskip 2.5cm+\int_{0}^\infty b'db'\left[E_0(b-T_2(b,b',b_1))+T_0(b,b',b_1)\right]V_{1,1}^{(q)}(b')\ .
\end{split}
\end{equation}
We expand the sum over stable configurations as follows:
\begin{equation}
    \sum_{stable}V^{(q)}_{h_1,|B_1|+1}(b',B_1)V^{(q)}_{h_2,|B_2|+1}(b'',B_2)=\sum_{stable}V^{(q)}_{h_1,1}(b')V^{(q)}_{h_2,2}(b'',b_1)+\sum_{stable}V^{(q)}_{h_1,2}(b',b_1)V^{(q)}_{h_2,1}(b'')=0\ .
\end{equation}
The reason it is zero is because $h_1+h_2=1$ so we will always have terms that are genus zero with fewer than three boundaries, and so all the terms are thrown away. It follows then that:
\begin{equation}
\begin{split}
    &bV_{1,2}^{(q)}(b,b_1)=\frac{1}{2}\int_0^\infty b'db'b''db''\left[E_0D_2(b,b',b'')-D_0(b,b',b'')\right] \times\left[V^{(q)}_{0,3}(b',b'',b_1)\right]\\
    &\hskip 3cm+\int_{0}^\infty b'db'\left[E_0(b-T_2(b,b',b_1))+T_0(b,b',b_1)\right]V_{1,1}^{(q)}(b')\ .
\end{split}
\end{equation}
Substituting in the expression for the initial condition volumes we wrote in equation (\ref{initalConditionN=2Vol}) and evaluating the integrals using the results in Appendix \ref{KernelIntegralsN=2}, we arrive at the result:
\begin{equation}
\label{V12N=2}
    V_{1,2}^{(q)}(b',b'')=\frac{1}{8}-\frac{1}{24}\left[3(b'^2+b''^2)+14\pi^2\right]E_0+\frac{1}{192}\left(4\pi^2+b'^2+b''^2\right)\left(12\pi^2+b'^2+b''^2\right)E_0^2\ .
\end{equation}
It is interesting to note that highest order term in $E_0$ is actually the volume one gets for the bosonic JT case. This was mentioned in \cite{Turiaci:2023jfa} that in all computations the highest order in $E_0$ term will contain the bosonic volume. This result will be used to compute the volume $V_{2,1}^{(q)}$ in the next section.

\subsubsection*{$\bullet$ Computing $V^{(q)}_{2,1}(b)$ and $V^{(q)}_{3,1}(b)$, and others along the way.}
For the computation of $V_{2,1}^{(q)}$ we set $g=2$ and $n=0$ $\Rightarrow B=\varnothing$ and $|B|=0$. Then we can write:
\begin{equation}
\begin{split}
     &bV_{2,1}^{(q)}(b)=\frac{1}{2}\int_0^\infty b'db'b''db''\left[E_0D_2(b,b',b'')-D_0(b,b',b'')\right]\\
    &\hskip 3cm\times\left[V^{(q)}_{1,2}(b',b'')+\sum_{stable}V^{(q)}_{h_1,1}(b')V^{(q)}_{h_2,1}(b'')\right]\ .
\end{split}
\end{equation}
Consider the following sum:
\begin{equation}
    \sum_{h_1+h_2=2}V^{(q)}_{h_1,1}(b')V^{(q)}_{h_2,1}(b'')=V^{(q)}_{0,1}(b')V^{(q)}_{2,1}(b'')+V^{(q)}_{1,1}(b')V^{(q)}_{1,1}(b'')+V^{(q)}_{2,1}(b')V^{(q)}_{0,1}(b'')\ .
\end{equation}
The ``stable'' sum instructs us to discard any terms involving genus 0 volumes will less than 3 boundaries. So then we can write:
\begin{equation}
\begin{split}
     &bV_{2,1}^{(q)}(b)=\frac{1}{2}\int_0^\infty b'db'b''db''\left[E_0D_2(b,b',b'')-D_0(b,b',b'')\right]\\
    &\hskip 3cm \times\left[V^{(q)}_{1,2}(b',b'')+V^{(q)}_{1,1}(b')V^{(q)}_{1,1}(b'')\right]\ .
\end{split}
\end{equation}
As we can see it requires us to know the preceding volume $V_{1,2}^{(q)}$. We substitute in the expressions from equation (\ref{initalConditionN=2Vol}) and equation (\ref{V12N=2}) and using the results in Appendix~\ref{KernelIntegralsN=2} we will find {\it precisely} the result~(\ref{eq:new-V21}) that we computed earlier, thereby confirming  the ODE method!  

Moving to higher genus,  we will spare the reader the lengthy details of the computation and simply show the integrals that need to be evaluated and their results, but the processes involved are similar to those described above. Referring to  Table \ref{tab:volume-table} we can anticipate that we will require knowledge of $V_{0,4}^{(q)}$, $V^{(q)}_{1,3}$, and $V^{(q)}_{2,2}$ before we can compute $V_{3,1}^{(q)}$. The authors of \cite{Turiaci:2023jfa} compute $V_{0,4}^{(q)}$, and we confirmed by a quick trumpet computation here in~(\ref{eq:4-point-new}). 
All that is left for us are computations of the intermediate volumes $V_{1,3}^{(q)}$ and $V_{2,2}^{(q)}$. 
To obtain $V^{(q)}_{1,3}$ we must evaluate:
\begin{equation}
\begin{split}
    &bV_{1,3}^{(q)}(b,b_1,b_2)=\frac{1}{2}\int_0^\infty\int_0^\infty bdb b'db'\left[E_0D_2(b,b',b'')-D_0(b,b',b'')\right]\\
    &\hskip 3.5cm\times\left[V^{(q)}_{0,4}(b',b'',b_1,b_2)+V_{1,1}^{(q)}(b')V_{0,3}^{(q)}(b'',b_1,b_2)+V_{1,1}^{(q)}(b'')V_{0,3}^{(q)}(b',b_1,b_2)\right] \\
    &\hskip3cm +\int_0^\infty b'db'\left[E_0T_2(b,b',b_1)+T_0(b,b',b_1)\right]V_{1,2}^{(q)}(b',b_2)\\
    &\hskip3cm +\int_0^\infty b'db'\left[E_0T_2(b,b',b_2)+T_0(b,b',b_2)\right]V_{1,2}^{(q)}(b',b_1)\ .
\end{split}
\end{equation}
Using the results of Appendix \ref{KernelIntegralsN=2} we will arrive at the following expression for $V_{1,3}^{(q)}$:
\begin{eqnarray}
    &&V^{(q)}_{1,3}(b,b',b'') =-\frac{1}{4}+\left[\frac{41\pi^2}{12}+\frac{9}{16}\left(b^2+b'^2+b''^2\right)\right]E_0\nonumber\\
    &&\hskip2.6cm-\left[\frac{b^4+b'^4+b''^4+3\left(b^2b'^2+b^2b''^2+b'^2b''^2\right)+22\pi^2\left(b^2+b'^2+b''^2\right)+80\pi^4}{16}\right]E_0^2\nonumber\\
    &&\hskip2.6cm +V_{1,3}^{\text{Bos}}(b,b',b'')E_0^3\ ,
\end{eqnarray}
and to obtain $V^{(q)}_{2,2}$ we must evaluate:
\begin{equation}
    \begin{split}
        &bV_{2,2}^{(q)}(b,b_1)=\frac{1}{2}\int_0^\infty\int_0^\infty bdb b'db'\left[E_0D_2(b,b',b'')-D_0(b,b',b'')\right]\\
    &\hskip 5cm\times\left[V_{1,3}^{(q)}(b',b'',b_1)+V_{1,2}(b',b_1)V_{1,1}^{(q)}(b'')+V_{1,2}(b'',b_1)V_{1,1}^{(q)}(b')\right] \\
    &\hskip2.2cm +\int_0^\infty b'db'\left[E_0T_2(b,b',b_1)+T_0(b,b',b_1)\right]V_{2,1}^{(q)}(b')\ .
    \end{split}
\end{equation}
Upon evaluation of the integrals we  find:
\begin{equation}
\begin{split}
    &V^{(q)}_{2,2}(b,b')=\left[\frac{3\pi^2}{16}+\frac{9\left(b^2+b'^2\right)}{256}\right]-\left[\frac{14092\pi^4+3660\pi^2\left(b^2+b'^2\right)+5\left(29b^4+29b'^4+96b^2b'^2\right)}{3840}\right]E_0\\
    &\hskip1.8cm +\frac{\bar{A}(b,b')}{46080}E_0^2-\frac{\bar{B}(b,b')}{552960}E_0^3+V_{2,2}^{\text{Bos}}(b,b')E_0^4\ ,
\end{split}
\end{equation}
where:
\begin{eqnarray}
        &&\bar{A}(b,b')=139200 \pi ^4 \left(b^2+b'^2\right)+5 \left(31 b^4+173 b^2 b'^2+31 b'^4\right) \left(b^2+b'^2\right)\\
   && \hskip 6cm +40 \pi ^2 \left(233 b^4+738 b^2 b'^2+233
   b'^4\right)+442784 \pi ^6\ ,\nonumber\\
     &&   \bar{B}(b,b')=33 b^8+348 b^6 b'^2+756 b^4 b'^4+348 b^2 b'^6+1434240 \pi ^6 \left(b^2+b'^2\right)\nonumber\\
        &&\hskip2.6cm   +96 \pi ^4 \left(1297 b^4+4000 b^2 b'^2+1297 b'^4\right) +40
   \pi ^2 \left(b^2+b'^2\right) \left(91 b^4+491 b^2 b'^2+91 b'^4\right)\nonumber\\
        &&\hskip2.6cm  +33 b'^8+4052800 \pi ^8\ ,\nonumber\\
        &&V_{2,2}^{\text{Bos}}(b,b')=\frac{b^{10}+b'^{10}}{4423680}+\frac{b^8b'^2+b^2b'^8}{294912}+\frac{29\left(b^6b'^4+b^4b'^6\right)}{2211840}+\frac{11\pi^2\left(b^8+b'^8\right)}{276480}+\frac{29\pi^2\left(b^6b'^2+b^2b'^6\right)}{69120}\nonumber\\
            &&\hskip2.5cm   +\frac{7\pi^2b^4b'^4}{7680}+\frac{19\pi^4\left(b^6+b'^6\right)}{7680}+\frac{181\pi^4\left(b^4b'^2+b^2b'^4\right)}{11520}+\frac{551\pi^6\left(b^4+b'^4\right)}{8640}+\frac{7\pi^6b^2b'^2}{36}\nonumber\\
        &&   \hskip2.5cm
        +\frac{1085\pi^8\left(b^2+b'^2\right)}{1728}+\frac{787\pi^{10}}{480}\ ,\nonumber\\
        &&V_{1,3}^{\text{Bos}}(b,b',b'')=\frac{b^6+b'^6+b''^6}{1152}+\frac{b^4 b'^2+b^4 b''^2+b^2 b'^4+b^2 b''^4+b'^4 b''^2+b'^2 b''^4}{192}+\frac{b^2b'^2b''^2}{96}\nonumber\\
        &&  \hskip2.6cm +\frac{\pi^2\left(b^4+b'^4+b''^4\right)}{24}+\frac{\pi^2\left(b^2 b'^2+b^2 b''^2+b'^2 b''^2\right)}{8}+\frac{13\pi^4\left(b^2+b'^2+b''^2\right)}{24}+\frac{14\pi^6}{9}\ ,
        \nonumber
\end{eqnarray}
where we can recognize $V_{2,2}^{\text{Bos}}$ and $V_{1,3}^{\text{Bos}}$ exactly matches with bosonic volumes given in \cite{do2011modulispaceshyperbolicsurfaces}, as it should. 
Finally, after all these intermediate results, we computed $V^{(q)}_{3,1}$ by evaluating: 
\begin{equation}
    \begin{split}
        &bV_{3,1}^{(q)}(b)=\frac{1}{2}\int_0^\infty\int_0^\infty bdb b'db'\left[E_0D_2(b,b',b'')-D_0(b,b',b'')\right]\\
    &\hskip 5cm\times\left[V^{(q)}_{2,2}(b',b'')+V_{1,1}^{(q)}(b')V_{2,1}^{(q)}(b'')+V_{1,1}^{(q)}(b'')V_{2,1}^{(q)}(b')\right]\ ,
    \end{split}
\end{equation}
and doing so yielded an expression in precise agreement with our  result in~(\ref{eq:new-V31})!

\section{Volumes for  $\mathcal{N}{=}4$ JT supergravity}
\label{sec:ODE-approach-Neq4}
Emboldened by our success, it is time to venture into completely new territory. Since multicritical matrix models have been constructed~\cite{Johnson:2024tgg,Johnson:2025oty} for small and large ${\cal N}{=}4$ JT supergravity, it would be highly neglectful to not compute Weil-Petersson volumes for these cases. These constitute {\it predictions} for other approaches.

\subsection{Small $\mathcal{N}=4$}
The string equation will be the same in structure as the $\mathcal{N}=2$ case and the perturbative solution will have the same structure. We just have a different set of ${t}_k$ which are worked out in \cite{Johnson:2024tgg} and given by\footnote{In order to get the spectral density in ref.~\cite{Turiaci:2023jfa} we must rescale both $\widetilde{\Gamma}$ and the $t_k$ in \cite{Johnson:2024tgg} by a factor of 2.}:
\begin{equation}
{t}_k=\frac{8\pi^{k+1}}{J^kk!(4k^2+8k+3)}J_{k+1}(2\pi J)\ ,
\end{equation}
where $J_{k}(x)$ is the $k$-th Bessel function of the first kind not to be confused with the parameter $J\in \frac{1}{2}\mathbb{Z}_{\neq 0}$\footnote{It is absolutely essential to enforce the condition that $J$ is a half integer or integer not equal to zero otherwise we cannot have a match of the spectral density for the given ${t}_k$'s.}. Explicitly the tree level string equation is given by:
\begin{equation}
\begin{split}
&u_0\left(G_0(u_0)+x\right)^2=\widetilde{\Gamma}^2=4\ .
\end{split}
\end{equation}
All we need to do computations of the volumes is information of the solution to the string equation at the Fermi surface $x={\mu}$, where $u_0({\mu})=J^2$. Rearranging the string equation we can write:
\begin{equation}
    {\mu}=\frac{2(-1)^{2J+1}}{J}-G_0(J^2)=\frac{8\pi}{3} J_1(2\pi J)={t}_k|_{k=0}\ .
\end{equation}
The expression for ${\mu}$ is exact and is obtained by re-summing the expression for $G_0(J^2)$ which is done in detail in Appendix \ref{N=4G0Calcs}. To get the full perturbative solution to the string equation we only need to know $u_0$ and all its derivatives. This information is is conveniently encoded in the derivatives of $G_0$. In Appendix \ref{N=4G0Calcs} we have an exact expression for the $p$-th derivative (w.r.t $u_0$) of $G_0$ at the Fermi surface. It is exactly given in terms of hypergeometric functions:
\begin{eqnarray}
        &&
        G_0^{(p)}(u_0)\vert_{u_0=J^2}=\frac{16J\pi^{2p+2}(2p)!}{p!(3+2p)!}\left[_{1}F_2\left(1;p+2,p+\frac{5}{2};-(\pi J)^2\right)\right.\\
        &&\hskip8.0cm\left.-\frac{2\pi^2J^2 }{(p+2)(5+2p)} {}_1F_2\left(2;p+3,p+\frac{7}{2};-(\pi J)^2\right)\right]\ . \nonumber
\end{eqnarray}
One can check that the $p$-th derivative of $G_0$ has the following structure:
\begin{equation}
    G_0^{(p)}(u_0)\vert_{u_0=J^2}=\frac{(-1)^p(2p-1)!!}{2^{p-1}J^{2p+1}}\left[-\cos(2\pi J)+\frac{(2p+1)\sin(2\pi J)}{2\pi J}\right]+\sum_{k=1}^{p}\frac{c_k^{(p)}}{J^{2k+1}}\ ,
\end{equation}
where $c_k^{(p)}$ are some constant coefficients. Due to this fact we can immediately conclude that because $J\in\frac{1}{2}\mathbb{Z}_{\neq0}$ the $p$-th derivative of $G_0$ will contain a finite number of terms in inverse odd powers of $J$. For example we have:
\begin{equation}
\begin{split}
    &\dot{G}_0(J^2)=\frac{2+(-1)^{2J}}{J^3}\\
    &\ddot{G}_0(J^2)=\frac{2\pi^2}{J^3}-\frac{3(4+(-1)^{2J})}{2J^5}\\
    &\dddot{G}_0(J^2)=\frac{\pi^4}{J^3}-\frac{10\pi^2}{J^5}+\frac{15(6+(-1)^{2J})}{4J^7}\ ,
\end{split}
\end{equation}
where we explicitly use the condition that $2J\in \mathbb{Z}_{\neq0}$.

Using the derivatives of $G_0$ we can construct higher order derivatives of $u_0$ by differentiating the following string equation (the signs on the last term were deduced in the computations in Appendix \ref{N=4G0Calcs}):
\begin{equation}
    G_0(u_0(x))+x+\frac{2\text{sgn}(J)(-1)^{2J}}{u_0(x)^{1/2}}=0\ .
\end{equation}
Below we give the results for the first few derivatives:
\begin{equation}
\begin{split}
    &u_0'({\mu})=-\frac{J^3}{2}\\
    &u_0''({\mu})=\frac{3J^4-\pi^2J^6}{4}\\
    &u_0'''({\mu})=-\frac{63}{32}J^5+\frac{13\pi^2}{8}J^7-\frac{5\pi^4}{16}J^9\\
    &u_0^{(4)}(\mu)=\frac{15}{2}J^6-\frac{675\pi^2}{64}J^8+\frac{147\pi^4}{32}J^{10}-\frac{61\pi^6}{96}J^{12}\\
    &u_0^{(5)}(\mu)=-\frac{19305}{512}J^7+\frac{9855\pi^2}{128}J^9-\frac{14051\pi^4}{256}J^{11}+\frac{1057\pi^6}{64}J^{13}-\frac{1379\pi^8}{768}J^{15}\ .
\end{split}
\end{equation}
Using these we can readily compute predictions for the genus zero volumes  as was done in the bosonic and ${\cal N}{=}2$ cases earlier (the trumpet evidently has a factor $\e^{-\beta J^2}$ now). Just as before, since it is a straightforward exercise to generate derivatives of $u_0^{(p)}(\mu)$,  equation~(\ref{eq:correlator}) together with the trumpet factors (\ref{eq:trumpet-voluntary}) (times $\e^{-\beta J^2}$) give a simple generating formula for the $V_{0,n}(\{b_i\})$ for small ${\cal N}{=}4$. For example ({\it c.f.} equations~(\ref{eq:3-point-new}) and~(\ref{eq:4-point-new})):
\begin{equation}
\label{eq:3-point-new-Neq4}
    V_{0,3}(b_1,b_2,b_3)= J^3\ ,
\end{equation}
and:
\begin{equation}
\label{eq:4-point-new-Neq4}
      V_{0,4}(b_1,b_2,b_3,b_4)=-6J^4+\frac{J^6}{2}\left(4\pi^2+b^2_1+b^2_2+b^2_3+b_4^2\right)\ .
\end{equation}
Our main goal is really higher genus results. The volume $V_{1,1}^{(J)}(b)$ can be  computed by using the formula in equation (\ref{eq:Volume-Expression}) (remembering to set $E_0=J^2$) with the following result:
\begin{equation}
    V_{1,1}^{(J)}(b)=-\frac{1}{4}J+\frac{b^2+4\pi^2}{48}J^3\ .
\end{equation}
Computing higher order derivatives we can also compute higher genus results as well. Below we give the genus 2 and 3 predictions:
\begin{equation}
\begin{split}
    &V_{2,1}^{(J)}(b)=\frac{19}{32}J-\frac{29(3b^2+16\pi^2)}{192}J^3+\frac{1719b^4+36120\pi^2b^2+122896\pi^4}{46080}J^5\\
    &\hskip1.5cm -\frac{(b^2+4\pi^2)(29b^4+1320\pi^2b^2+12592\pi^4)}{46080}J^7\\
    &\hskip1.5cm +\frac{(b^2+4\pi^2)(b^2+12\pi^2)(5b^4+384\pi^2b^2+6960\pi^4)}{2211840}J^9\ .
\end{split}
\end{equation}

\begin{equation}
\begin{split}
&V_{3,1}^{(J)}(b)=-\frac{631 J}{128}+\frac{\left(13331 b^2+90664 \pi ^2\right) J^3}{1536}-\frac{\left(325305 b^4+9180000 \pi ^2 b^2+38443904 \pi ^4\right) J^5}{184320}\\
    &\hskip1.5cm +\frac{\left(1380675 b^6+92976660 \pi ^2 b^4+1516583712 \pi ^4 b^2+5052918208 \pi ^6\right) J^7}{15482880}\\
    &\hskip1.5cm -\frac{\left(35125 b^8+4424460 \pi ^2 b^6+168075432 \pi ^4 b^4+2087461952 \pi ^6 b^2+6108024352 \pi ^8\right) J^9}{23224320}\\
    &\hskip1.5cm +\frac{J^{11}}{1857945600}\widetilde{A}(b)-\frac{(b^2+4\pi^2)J^{13}}{22295347200}\widetilde{B}(b)+V_{3,1}^{\text{Bos}}(b)J^{15}\ ,
\end{split}
\end{equation}
where:
\begin{eqnarray}
        &&\widetilde{A}(b)=18159 b^{10}+3730300 \pi ^2 b^8+261496640 \pi ^4 b^6\\\nonumber
        &&\hskip6cm+7431462720 \pi ^6 b^4+78378647680 \pi ^8 b^2+211957124096 \pi ^{10}\ ,\\\nonumber
        &&\widetilde{B}(b)=539 b^{10}+163236 \pi ^2 b^8+18091216 \pi ^4 b^6\\\nonumber
        &&\hskip 6cm+900291776 \pi ^6 b^4+19557966336 \pi ^8 b^2+141438717952 \pi ^{10}\ ,\nonumber
\end{eqnarray}
and $V_{3,1}^{\text{Bos}}(b)$ is given in equation~(\ref{eq:V31-bos}). Very  similarly to the $\mathcal{N}=2$ case is the highest order term in~$J$, up to a factor, is the bosonic volume. 

\subsection{Large $\mathcal{N}=4$}
As was shown in \cite{Heydeman:2025vcc} leading order disk density of states for the large $\mathcal{N}=4$ theory can be expressed as:
\begin{equation}
    \tilde{\rho}_{j_+,j_-}=\frac{1}{2\pi\hbar}A_{j_+,j_-}\frac{\sinh\left(2\pi\sqrt{E-E_0}\right)}{(E-E^-)(E-E^+)}\Theta(E-E_0)\ , 
\end{equation}
where:
\begin{eqnarray}
    A_{j_+,j_-}=\frac{\alpha^{3/2}j_+j_-\sqrt{2\pi}}{16(1+\alpha)^3}\ ,\qquad E^\pm=\frac{\alpha(j_+\pm j_-)^2}{(1+\alpha)^2}\ , \qquad E_0=\frac{\alpha j_+^2+j_-^2}{(1+\alpha)}\ .
\end{eqnarray}
It was noticed in ref.~\cite{Johnson:2025oty} that this density does not have an analytic structure that can be incorporated into the class of matrix models that have  a string equations of the form~(\ref{eq:scaled-big-string-equation}). It was suggested that instead the density of states should be thought of as being generated by two statistically independent matrix ensembles, each of which has a form that has the structure of an $\mathcal{N}=2$ theory. This first with the observation that $\tilde{\rho}_{j_+,j_-}$ can be decomposed as $\tilde{\rho}_{j_+,j_-}=\rho_{j_+,j_-}^+-\rho_{j_+,j_-}^-$, where:
\begin{equation}
\begin{split}
    &\rho_{j_+,j_-}^\pm=\frac{\omega_\alpha}{2\pi\hbar}\frac{\sinh\left(2\pi\sqrt{E-E_0}\right)}{E-E^\pm}\Theta(E-E_0)\ , \qquad \omega_\alpha=\frac{\sqrt{2\pi\alpha}}{64\pi(1+\alpha)}\ .\\
\end{split}
\end{equation}
The natural conjecture~\cite{Johnson:2025oty} is that there is a further statistical independence between the sectors represented by these two densities, and hance they care captured by different matrix ensembles.

Densities $\rho_{j_+,j_-}^\pm$ are each consistent with the analytic structure of the $\mathcal{N}{=}2$ solution of the string equation with a minor modification which involves replacing every instance of $u(x)$ with $u^\pm(x){=}u(x){-}E^\pm$ so that the resulting string equation can be written as:
\begin{equation}
    u^\pm\left(\mathcal{R}^\pm\right)^2-\frac{\hbar^2}{2}\left(\mathcal{R}^\pm\right)\left(\mathcal{R}^\pm\right)''+\frac{\hbar^2}{4}\left(\mathcal{R}^\pm\right)'^2=\left(\widetilde{\Gamma}^{\pm}\right)^2\ .
\end{equation}
where $\mathcal{R}^\pm(x)=G(u^\pm(x))+x$, and $G(u^\pm(x))=\sum_{k=1}^\infty t_k^\pm R_k[u^\pm(x)]$. The tree level string equation will read:
\begin{equation}
    \left(u_0(x)-E^\pm\right)\left(G_0^\pm(u_0(x))+x\right)^2=\left(\widetilde{\Gamma}^\pm\right)^2\ , 
\end{equation}
where $G_0^{\pm}(u_0(x))=\sum_{k=1}^\infty t_k^{\pm}(u_0(x)-E^\pm)^k$. The set of $t_k^\pm$ and $\widetilde{\Gamma}^\pm$ which gives $\rho_{j_+,j_-}^\pm$ are derived in \cite{Johnson:2025oty} and are written as:
\begin{equation}
\begin{split}
    &t_k^\pm=4\pi^2\omega_\alpha\left[\frac{\pi^{k-1}J_{k}\left(2\pi\sqrt{E_0-E^\pm}\right)}{2(2k+1)\left(E_0-E^\pm\right)^{k/2}k!}\right]\ ,\\
&|\widetilde{\Gamma}^\pm|=4\pi^2\omega_\alpha\left[\frac{\sin\left(2\pi\sqrt{E_0-E^\pm}\right)}{4\pi^2}\right]\ .\\
\end{split}
\end{equation}
We wrote the expressions above in a way that makes it clear that the results are the same as $\mathcal{N}=2$ up to a rescaling by $4\pi^2\omega_\alpha$ and a replacement of $E_0-E^\pm\to E_0$. This provides a very clean way to understand the results that will follow. Essentially the volumes we will obtain using these results will simply correspond to a appropriate rescaling by $4\pi^2\omega_\alpha$ of the old $\mathcal{N}=2$ volumes we already obtained along with a redefinition of $E_0$. In particular, let $V_{g,1}^{\mathcal{N}=2}(b)$ and let $V_{g,1}^{\pm\mathcal{N}=4}(b)$ be the 1 geodesic boundary volumes obtained for the $\mathcal{N}=2$ and decomposed large $\mathcal{N}=4$ theory respectively. Then we have the exact relation:
\begin{equation}
\label{LargeN=4CleanResult}
V_{g,1}^{\pm\mathcal{N}=4}(b)=\frac{V_{g,1}^{\mathcal{N}=2}(b)}{\left(4\pi^2\omega_\alpha\right)^{2g-1}}\bigg\vert_{E_0\to E_0-E^\pm}\ .
\end{equation}
Now that we have neatly summarized the result let us go over the details to obtain it. 

As usual all we need to do is compute derivatives of the tree level string equation at the Fermi surface $\mu^\pm$ where $u_0(\mu^\pm)=E_0$. To do this we must repeatedly take derivatives of the tree level string equation. This will involve taking derivatives with respect to $x$ of $G^\pm(u_0(x))$ as a result we will be needing to do sums of the form:
\begin{equation}
\begin{split}
    &G^\pm_0|_{u_0=E_0}=\sum_{k=1}^\infty t_k^\pm \left(E_0-E^\pm\right)^k\ ,\\
    &(G^\pm_0)^{(p)}|_{u_0=E_0}=\sum_{k=1}^\infty t_k^\pm \frac{k!}{(k-p)!}\left(E_0-E^\pm\right)^{k-p}\ .\\
\end{split}
\end{equation}
Due to our observation that the $t_k$s for the decomposed large $\mathcal{N}=4$ theory is just the $\mathcal{N}=2$ ones with a rescaling we can just use our old results for $\mathcal{N}=2$ and immediately write:
\begin{equation}
    \begin{split}
    &G^\pm_0|_{u_0=E_0}=\frac{4\pi^2\omega_\alpha}{2\pi}\left[\frac{\sin\left(2\pi\sqrt{E_0-E^\pm}\right)}{2\pi\sqrt{E_0-E^\pm}}-J_0\left(2\pi\sqrt{E_0-E^\pm}\right)\right]\ ,\\
    &(G^\pm_0)^{(p)}|_{u_0=E_0}=4\pi^2\omega_\alpha\left[\frac{\pi^{2p-1}(2p)!}{2p!(2p+1)!}{}_1F_2\left(1;p+1,p+3/2;-\pi^2(E_0-E^\pm)\right)\right]\ .\\
    \end{split}
\end{equation}
As we can clearly see $G^\pm_0$ and its derivatives have the same functional for as $\mathcal{N}=2$ up to the factor of $4\pi^2\omega_\alpha$ and a replacement of the argument of the functions. Using the following tree level string equation:
\begin{equation}
    G_0^\pm+x-\frac{|\widetilde{\Gamma}^\pm|}{\sqrt{u_0-E^\pm}}=0\ ,
\end{equation}
we can compute derivatives at the Fermi surface and find the following:
\begin{equation}
    u_0^{(p)}(\mu^\pm)=\frac{u_{0,\mathcal{N}=2}^{(p)}(\mu)}{(4\pi\omega_\alpha)^p}\bigg\vert_{E_0\to E_0-E^\pm}\ ,
\end{equation}
where $u_{0,\mathcal{N}=2}^{(p)}(\mu)$ are the $\mathcal{N}=2$ derivatives (see for example the expressions in equation (\ref{deru0atFermi}).). Due to this simple rescaling when we compute volumes using the $\widehat{Q}_g$s we derived we can clearly see that the volumes will exactly rescale as shown in equation (\ref{LargeN=4CleanResult}).\footnote{{A subtle point to note is that the $\widehat{Q}_g$ we derived should also involve $u^\pm=u_0(x)-E^\pm$. Although this does not change too much (because almost everything in the $\widehat{Q}_g$s involve derivatives of $u_0^\pm$ which just become derivatives of $u_0$ itself) there is a subtle change in the denominators of $\widehat{Q}_g$, $X$, which is now replaced by $X^\pm=u_0-E^\pm-E$. So now before taking an inverse Laplace transform to define volumes we need to redefine the uniformizing variable $z$, as $E_0-E^\pm-E=z^2$ to eliminate the branch structure which comes from denominators of $\widehat{Q}_g$.}}

\section{Closing remarks}
\label{sec:closing-remarks}

In this paper we have extensively explored the ODE approach of ref.~\cite{Johnson:2024bue} for computing Weil-Petersson volumes, which takes the solution $u(x)$ to the random matrix model string equation and uses it to seed the Gel’fand-Dikii resolvent equation, readily yielding (after an integral and an inverse Laplace transform), the volumes $V_{g,1}(b)$. A key feature that we confirmed to high orders (and developed useful expressions for) is the fact that when~$u(x)$ is {\it specifically} a solution to the string equation, the resolvent, order by order in genus, becomes a total derivative, a necessary condition for the volumes (with their known properties) to emerge.

There's a notable simplicity or universality to the method. It can be used for a wide variety of models, whose  details  are captured by the function $u(x)$ (a solution of a string equation). The method takes that $u(x)$ (perturbatively) as input, and the methodology for then yielding the volumes is essentially the same regardless of the model. Many models in the JT (super)gravity class are described by different $u(x)$s, as are many kinds of string theory.

 This approach to volumes is also remarkably efficient, in that as a recursive procedure for constructing the volume  at genus $g$, only the volumes with the same number of boundaries (one) but  genus less than~$g$ are involved. This is rather different from the typical recursion methods used for Weil-Petersson volumes, such as that due to  Mirzakhani~\cite{Mirzakhani:2006fta} (equivalently the matrix model loop equations\cite{Eynard:2007fi}). Another reason for the relative efficiency is simply that differentiating (at the heart of the method) is much more straightforward than integrating (which is a common feature  of more standard topological recursion methods~\cite{Eynard:2014zxa}).
 On the other hand, the method really only works for volumes  with one boundary, {\it i.e.,}~$V_{g,1}( b)$, so currently the other methods ultimately have greater reach.

It is of course natural to wonder if this ODE method can be extended to give the full $V_{g,n}(\{ b_i\})$, providing an alternative or complement to standard topological recursion techniques. Perhaps the start of the answer to the question is to find a more general equation that replaces the Gel’fand-Dikii ODE. It should still take the function $u(x)$ as the seed, of course, but now the equation should yield the object that can be used to build $n$-point energy (or loop) correlates in the matrix model. A recent such generalization was proposed and explored in ref.~\cite{Johnson:2025dyb} with this application (among others) in mind. Whether the nimble perturbative analysis seen here (that readily yields $V_{g,1}(b)$) will have a counterpart for that equation has yet to be fully explored. Work on that is underway. Note also that refs.~\cite{Lowenstein:2024gvz,Lowenstein:2024fji} has developed string equation methods for deriving some formulae for volumes,  for $n>1$. For $n{=}1$ they are perturbatively equivalent to the methods used here.

While not the subject of this paper, it is worthwhile also exploring the rich non-perturbative physics that this formulation gives ready access to, and its implications for Weil-Petersson volumes. In addition to the  non-perturbative physics for $u(x)$ contained in the string equation, there is non-perturbative physics from the Gel’fand-Dikii ODE itself that is used to extract the volumes. Both sources should furnish important data concerning, {\it e.g.} asymptotic behavior of the volumes in limits such as large $g$, and beyond. This is well worth pursuing.

Beyond exploring the method itself, the main application in this paper was to explore and define Weil-Petersson volumes for JT supergravity with extended supersymmetry~\cite{Forste:2017apw,Stanford:2017thb,Mertens:2017mtv,Heydeman:2020hhw,Heydeman:2025vcc}. Volumes in this class are natural gravity/string observables that are relatively new on the market, but only a few of them having been explored in the literature~\cite{Turiaci:2023jfa} prior to our work. We were able to quickly compute several examples for the ${\cal N}{=}2$ case, finding agreement of those presented in ref.~\cite{Turiaci:2023jfa} as well as producing new ones\footnote{Ref.~\cite{Lowenstein:2024gvz} included some  string equation based computations for some low order ${\cal N}=2$ volumes, finding only leading order (in $E_0$) agreement. This seems likely due to an unfortunate choice of Laplace transform variable.}. We used \cite{Turiaci:2023jfa}’s topological recursion method to compute more volumes in order to see that the methods agree when they overlap and indeed they do. The volumes we have generated for the (small and large) ${\cal N}{=}4$ JT supergravity models~\cite{Heydeman:2020hhw,Turiaci:2023jfa,Heydeman:2025vcc} are entirely new, and we are confident that our methods are robust enough to have correctly defined these whole new classes of object that are of both physical and mathematical interest. 

A novel property, that the ${\cal N}{=}2$ and ${\cal N}{=}4$ volumes have a sub-component that is precisely the {\it bosonic}  JT gravity volumes, appearing for each genus at highest order in $E_0$ (the threshold energy), is confirmed by this method. (It was first observed for the $N{=}2$ case in ref.~\cite{Turiaci:2023jfa}, and we see it naturally emerge here for ${\cal N}{=}4$.)
It is a curious and interesting property that deserves better understanding. In this approach, it boils down to the fact that certain key  structural features of the perturbative expansion for $u(x)$  are shared by both the bosonic and extended supersymmetric JT models (when they have a threshold energy). In this regard, the standard ${\cal N}{=}1$ supersymmetric case is a special in that it does not have a threshold energy and correspondingly perturbation theory has a somewhat different organization. (This difference could disappear for ${\cal N}{=}1$ solutions representing backgrounds with a large (O$(\e^{S_0})$) number of Ramond punctures, which are readily constructed in this framework by employing the kinds of solutions employed in refs.~\cite{Johnson:2023ofr,Johnson:2025oty}.)

A better mathematical understanding the appearance of the bosonic sector more broadly ({\it e.g.,} directly as a feature of the moduli space geometry) seems like an interesting avenue of investigation, especially in light of the highly active interest in this area among the mathematics and physics community.

\section*{Acknowledgements}

We thank Maciej Kolanowski for helpful remarks.  This work  was supported by US Department of Energy grant  \#DE-SC 0011687, and by the University of California.  CVJ also thanks Amelia for her support and patience.

\appendix
\bigskip 
\bigskip 
\bigskip

\noindent {\bf \Large Appendices}

\section{Structure of Gel'fand-Dikii polynomials in the genus expansion}
\label{GelfandPolyAppendix}
In this Appendix we will describe some interesting and useful aspects of the structure of Gel'fand Dikii polynomials in a genus expansion in $\hbar$. The central result of the Appendix is our prescription to write expressions for $R_k[u]$ for arbitrary $k$ accurate to an arbitrary genus $g$ given $R_1,..,R_{3g}$ as ``initial'' data. We also provide a Mathematica algorithm which makes use of the prescription we describe to obtain~$R_k[u]$ up to genus 7 ({\it  i.e.}, to $\mathcal{O}(\hbar^{14})$). 

\bigskip
\noindent{\bf $\bullet$  Gel'fand-Dikii polynomials at fixed genus }
\bigskip

\noindent The Gel'fand-Dikii polynomials satisfy the following recursion relation\cite{Gelfand:1975rn}:
\begin{equation}
    R_{k+1}^{\prime} = \frac{2k + 2}{2k + 1} \left( \frac{1}{2} u^{\prime} R_k + u R_k^{\prime} - \frac{\hbar^2}{4} R_k^{\prime\prime\prime} \right)\ ,
\end{equation}
with an initial condition of $R_0[u]{=}1$. (Our conventions are different from those of ref.~\cite{Gelfand:1975rn} in that $R_k[u]{=}u^k{+}\cdots$.) Note that the appearance of the $\hbar$ in the recursion relation implies that the polynomials themselves will admit an expansion in $\hbar$. For a given $k$ the expansion is in even power of $\hbar$ and terminates at order $\hbar^{2(k-1)}$, in particular we can write:
\begin{equation}
\label{hbarExpGelfand}
\begin{split}
    &R_k[u]=u^k+\sum_{g=1}^{k-1}\hbar^{2g}r_{g,k}[u]\ ,
\end{split}
\end{equation}
the coefficients, $r_{g,k}[u]$, in the $\hbar$ expansion can be easily understood we can write  as a sum of terms involving a string of $k-g$, $u$s acted on by $2g$ $\frac{d}{dx}$ operators (alternatively all such terms can be explicitly generated by expanding $\frac{d^{2g}}{dx^{2g}}u^{k-g}$ with $k\gg g$)  in all possible combinations weighted by some coefficients which are fixed by the recursion relation. 

More explicitly we can write:
\begin{equation}
\begin{split}
    &r_{g,k}[u]=\sum_{n_1,n_2,\ldots  ,n_{2g}} c_{g,k,\{n,N\}}\left(\prod_{i,n_i\neq 0}^{2g}u^{(n_i)}\right)u^{k-3g+N}\\
    &\sum_{i=1}^{2g}n_i=2g\\
    &u^{(p)}=\frac{d^p}{dx^p}u(x)\\
    &N=\text{Number of $n_i$'s that are zero }=0,1,2,..,2g-1\\
    &n=1,2,3,..,\text{Number of distinct terms}\ .
\end{split}
\end{equation}
Basically the expression tells us that the $k$-th Gel'fand Dikii polynomial at $\mathcal{O}(\hbar^{2g})$ (i.e. $r_{g,k}$) is given by some linear combination of terms weighted by the coefficient $c_{g,k,\{n,N\}}$. The individual terms can be written as $2g$ $x$-derivatives acting on a string of $u$'s of length $k-g$. For example, when $g=1$ we will consider terms that are built from acting with 2 $x$-derivatives on a sting of $k-1$ $u$'s. With this we can write the following expansion:
\begin{equation}
    r_{1,k}[u]=c_{1,k,\{1,1\}}u''u^{k-3+1}+c_{1,k,\{2,0\}}u'^2u^{k-3+0}\ ,
\end{equation}
where we adopt the convention that $n=1$ corresponds to $u''$ term and $n=2$ corresponds to $u'^2$ term. The only nontrivial thing to do is to determine the coefficients. This can be done from the recursion relation. What one will generally find is that the coefficient is exactly given by:
\begin{equation}
    c_{g,k,\{n,N\}}=\left(\frac{k!}{(3g-N)!(k-3g+N)!}\right)\frac{\partial}{\partial (u^{\{n\}_N})}r_{g,3g-N}[u]\ ,
\end{equation}
where the notation $\frac{\partial}{\partial (u^{\{n\}_N})}$ means to take a partial derivative with respect to $u^{\{n\}_N}$, where:
\begin{equation}
    u^{\{n\}_N}=\frac{u^{(n_1)}u^{(n_2)}\cdot\cdot\cdot u^{(n_{2g})}}{u^{N}}=\prod_{i,n_i\neq 0}^{2g}u^{(n_i)}\ .
\end{equation}

As an example, in an application of these formulas, let us consider fixing $g=1$. Using the formula above we have that:
\begin{equation}
c_{1,k,\{1,1\}}=\left(\frac{k!}{(3(1)-1)!(k-3(1)+1)!}\right)\frac{\partial}{\partial\left(\frac{u''u}{u^1}\right)}r_{1,3-1}[u]=\frac{k!}{2!(k-2)!}\frac{\partial}{\partial u''}r_{1,2}[u]\ . 
\end{equation}
We note that $r_{1,2}$ is exactly the set of terms of $R_2[u]$ at order $\hbar^2$. It is given by $r_{1,2}[u]=-u''/3$. Using this we can write:
\begin{equation}
    c_{1,k,\{1,1\}}=\frac{k!}{2!(k-2)!}\frac{\partial}{\partial u''}r_{1,2}[u]=\frac{k!}{2!(k-2)!}\left(-\frac{1}{3}\right)=-\frac{1}{6}k(k-1)\ .
\end{equation}
Similarly, we can find the other coefficient as well which is readily extracted from $R_3[u]$ to order $\hbar^2$:
\begin{equation}
    c_{1,k,\{2,0\}}=\frac{k!}{3!(k-3)!}\frac{\partial}{\partial u'^2}r_{1,3}[u]=-\frac{1}{2}\frac{k!}{3!(k-3)!}=-\frac{1}{12}k(k-1)(k-2)\ .
\end{equation}
It follows that:
\begin{equation}
    r_{1,k}[u]=-\frac{1}{6}k(k-1)u''u^{k-2}-\frac{1}{12}k(k-1)(k-2)u'^2u^{k-3}\ ,
\end{equation}
which exactly agrees with what is known (see {\it e.g.,} ref~\cite{Johnson:2021owr}). From this we can see to compute $R_k[u]$ to order $\hbar^{2g}$ we actually need to explicitly know $R_1,R_2,\ldots  ,R_{3g}$ ({\it i.e.,} the first $3g$) Gel'fand-Dikii polynomials to order $\hbar^{2g}$ to fix our coefficients.  

To further demonstrate the procedure, let us also construct the second genus contribution to $R_k[u]$. It will contain terms that can be generated by acting with 4 $x$-derivatives on a string of $k-2$ $u$'s. From this there will be precisely 5 possible combination of terms which can be expressed as:
\begin{equation}
    \begin{split}
        r_{2,k}[u]=c_{2,k,(1,3)}u^{(4)}u^{k-3}+c_{2,k,(2,2)}u^{(3)}u'u^{k-4}+c_{2,k,(3,2)}u''^2u^{k-4}+c_{2,k,(4,1)}u''u'^2u^{k-5}+c_{2,k,(5,0)}u'^4u^{k-6}\ .
    \end{split}
\end{equation}
Using our prescription to compute coefficients we write:
\begin{equation}
    \begin{split}
        &c_{2,k,(1,3)}=\frac{k!}{3!(k-3)!}\frac{\partial}{\partial u^{(4)}}r_{2,3}[u]=\frac{k!}{3!(k-3)!}\frac{1}{10}\\
        &c_{2,k,(2,2)}=\frac{k!}{4!(k-4)!}\frac{\partial}{\partial u'''u'}r_{2,4}[u]=\frac{k!}{4!(k-4)!}\frac{4}{5}\\
        &c_{2,k,(3,2)}=\frac{k!}{4!(k-4)!}\frac{\partial}{\partial u''^2}r_{2,4}[u]=\frac{k!}{4!(k-4)!}\frac{3}{5}\\
        &c_{2,k,(4,1)}=\frac{k!}{5!(k-5)!}\frac{\partial}{\partial u''u'^2}r_{2,5}[u]=\frac{k!}{5!(k-5)!}\frac{11}{3}\\
        &c_{2,k,(5,0)}=\frac{k!}{6!(k-6)!}\frac{\partial}{\partial u'^4}r_{2,6}[u]=\frac{k!}{6!(k-6)!}\frac{5}{2}\ . \\
    \end{split}
\end{equation}
We conclude that:
\begin{equation}
    r_{2,k}[u]=\binom{k}{3}\frac{1}{10}u^{(4)}u^{k-3}+\binom{k}{4}\frac{4}{5}u^{(3)}u'u^{k-4}+\binom{k}{4}\frac{3}{5}u''^2u^{k-4}+\binom{k}{5}\frac{11}{3}u''u'^2u^{k-5}+\binom{k}{6}\frac{5}{2}u'^4u^{k-6}\ .
\end{equation}
This gives us the Gel'fand-Dikii polynomial for any $k$ to order $\hbar^4$. This same procedure can in principle be used to arbitrary order in the genus expansion but will become tedious to do by hand at higher genus\footnote{For genus 3 we will have 11 terms which is still quite manageable by ``hand'' however, one will quickly find that the number of terms grows quite quickly with genus. For example, genus 4,6,8,10 will involve fixing coefficients of 22,77,231,627 terms respectively. Therefore, computations at higher genus is better suited using a computer algorithm. We provide one such algorithm which can effectively find $R_k[u]$ to genus 7 in a reasonable amount of run time.}. For this reason it is interesting to formulate a computer algorithm that may handle higher genus computations of Gel'fand-Dikii polynomials.

Since we are interested in working at a particular genus order it will be more effective to work directly with $r_{g,k}[u]$ rather than $R_k[u]$ itself. By inserting the Anzatz \ref{hbarExpGelfand} into the recursion relation for the Gel'fand-Dikii polynomials we obtain a recursion relation between the $r_{g,k}$ of the form:
\begin{equation}
\label{littlerRecursion}
    r'_{g,k+1}-\frac{2(k+1)}{2k+1}\left[\frac{1}{2}u'r_{g,k}+ur'_{g,k}-\frac{1}{4}r_{g-1,k}'''\right]=0\ .
\end{equation}
One can check that the $r_{2,k}$ and $r_{1,k}$ we explicitly wrote down satisfy the recursion relation with initial condition $r_{0,k}=u^k$. At first glance the recursion relation looks like a first order differential equation for $r_{g,k}$ but actually, it is more complicated due to the extra $k$ index which varies among the terms. We can relate $r_{g,k}$ to $r_{g,k+1}$ through the following expression:
\begin{equation}
    \frac{\partial}{\partial u}r_{g,k+1}=\partial_ur_{g,k+1}=(k+1)r_{g,k}\ ,
\end{equation}
where we see is that $\partial_u$ acts as a lowering operator on the index $k$ for $r_{g,k}$. We can act with $\partial_u$ repeatedly to obtain:
\begin{equation}
    \partial_u^p r_{g,k}=k(k-1)(k-2)\cdot\cdot\cdot(k-p+1)r_{g,k-p}=p!\binom{k}{p}r_{g,k-p}\ .
\end{equation}
Note that $r_{g,k\leq g}{=}0$ which implies that to get a result that is non-zero we must have $p{<}k{-}g$. Now lets try to better understand how to apply the recursion relation to obtain coefficients needed for calculations at any order in the genus expansion. Start by noting that for a genus $g$ calculation we are specifically interested in $r_{g,3g-N}$ for $N=0,1,2,..,2g-1$. The recursion relation for these are given as:
\begin{equation}
\label{RecursionForInitialCondr}
    r'_{g,3g-N}=\frac{2(3g-N)}{2(3g-N)-1}\left[\frac{1}{2}u'r_{g,3g-1-N}+ur'_{g,3g-1-N}-\frac{1}{4}r'''_{g-1,3g-1-N}\right]\ .
\end{equation}
We can solve this recursively starting at $N{=}2g{-}1$ going towards $N{=}0$. In particular, for $N{=}2g{-}1$ the recursion relation will read:
\begin{equation}
\begin{split}
     &0=r'_{g,g+1}-\frac{2(g+1)}{2g+1}\left[\frac{1}{2}u'r_{g,g}+ur'_{g,g}-\frac{1}{4}r_{g-1,g}'''\right]\\
     &\hskip 0.5cm =r'_{g,g+1}+\frac{g+1}{2(2g+1)}r_{g-1,g}'''\\
    &\hskip 1cm\Rightarrow r_{g,g+1}=-\frac{g+1}{2(2g+1)}r_{g-1,g}''\ , 
\end{split}
\end{equation}
where we used the fact that $r_{g,g}=0$. This nice recursion relation can actually be easily solved. To see how we can consider the first few values of $g$.
For $g=1$ the relation gives:
\begin{equation}
    r_{1,2}=-\frac{1}{3}\frac{d^2}{dx^2}\left[u\right]=-\frac{1}{3}u''\ .
\end{equation}
For $g=2$ we have:
\begin{equation}
    r_{2,3}=-\frac{3}{10}r''_{1,2}=-\frac{3}{10}(-\frac{1}{3}u^{(4)})=\frac{1}{10}u^{(4)}\ .
\end{equation}
Proceeding iteratively in this way we will eventually realize that these terms can be obtained by the general expression:
\begin{equation}
    r_{g,g+1}=\frac{(-1)^{g}(g+1)!}{2^{g}(2g+1)!!}u^{(2g)}\ .
\end{equation}
Motivated by our success in finding a closed form we proceed further. Taking $N{=}2g{-}2$,  now the recursion relation is:
\begin{equation}
    r'_{g,g+2}=\frac{2(g+2)}{2g+1}\left[\frac{1}{2}u'r_{g,g+1}+ur'_{g,g+1}-\frac{1}{4}r'''_{g-1,g+1}\right]\ .
\end{equation}
Unfortunately, we no longer have a simplification as before and each term will be non-zero but it can still be used since we know what $r_{g,g+1}$ is from the $N=2g-1$ term. In the previous genus we need to know $r_{g-1,g+1}$. Continuing iteratively for a fixed $g$ all the way to the $N=0$ recursion relation we obtain:
\begin{equation}
    r'_{g,3g}=\frac{2(3g)}{2(3g)-1}\left[\frac{1}{2}u'r_{g,3g-1}+ur'_{g,3g-1}-\frac{1}{4}r'''_{g-1,3g-1}\right] \ .
\end{equation}
What we see is that to solve the recursion at genus $g$ we need $r_{g-1,g},r_{g-1,g+1}\cdots r_{g-1,3g-1}$ i.e. $2g$ $r$'s from genus $g-1$. In the following we describe a computer algorithm to solve the tower of recursion relations.

\bigskip
\noindent{\bf $\bullet$ A Computer algorithm for analytic expression for $R_k$ to genus $g$}
\bigskip

To solve the tower of recursion relations we mentioned previously we can set up a $(g+1)\times(2g+1)$ matrix with elements $r_{i,i+j}$ where $i=0,1,2,3,\ldots ,g$ and $j=0,1,2,3,\ldots ,2g$ (note the index $i,j=0$ represent initial conditions which we input at the beginning). The way to use the recursion is to start at $i=1$ and populate the row using the recursion relation. Then you go to the second row and start populating it using the results from previous row as initial conditions. This will generate the pieces of $R_k$'s that we need to determine the $R_k$ for any $k$ up to genus $g$ (or order $\hbar^{2g}$). 

Explicitly we setup a $g+1\times 2g+1$ matrix given as:
\begin{equation}
    [r]_{a,b}=\begin{bmatrix}
r_{0,0}&r_{0,1} & r_{0,2} &\cdot\cdot\cdot &r_{0,2g}\\
r_{1,1}&r_{1,2} & r_{1,3} & \cdot\cdot\cdot&r_{1,2g+1}\\
r_{2,2}&r_{2,3} & r_{2,4} & \cdot\cdot\cdot&r_{2,2g+2}\\
\vdots&\vdots & \vdots & \vdots&\vdots\\
r_{g,g}&r_{g,g+1}&r_{g,g+2}&\cdot\cdot\cdot&r_{g,3g}\\
\end{bmatrix}
=\begin{bmatrix}
1&u & u^2 &\cdot\cdot\cdot &u^{2g}\\
0&r_{1,2} & r_{1,3} & \cdot\cdot\cdot&r_{1,2g+1}\\
0&r_{2,3} & r_{2,4} & \cdot\cdot\cdot&r_{2,2g+2}\\
\vdots&\vdots & \vdots & \vdots&\vdots\\
0&r_{g,g+1}&r_{g,g+2}&\cdot\cdot\cdot&r_{g,3g}\\
\end{bmatrix}\ .
\end{equation}
We will adopt conventions that matrix element index $a,b$ begin at $1$ and not zero (as is the case in {\tt Mathematica}). The recursion relation to populate the matrix is then given by:
\begin{equation}
    [r]_{a,b}=\frac{2(a+b-2)}{2(a+b-2)-1}\int\left[\frac{1}{2}\frac{du}{dx}[r]_{a,b-1}+u\frac{d}{dx}\left([r]_{a,b-1}\right)-\frac{1}{4}\frac{d^3}{dx^3}\left([r]_{a-1,b}\right)\right]\ dx \ .
\end{equation}
This can be implemented in {\tt Mathematica} using a double for loop with in inner for loop running from $b=2,3,..,2g+1$ and the outer for loop running from $a=2,3,..,g+1$. Doing this will populate $[r]_{2,2}$ then $[r]_{2,3}$, and so on. Next define the following coefficient matrix [c] with elements:
\begin{equation}
    [c]_{a,b}=\binom{k}{a+b-2}u^{k-(a+b-2)}\ .
\end{equation}
Define $[r]\vert_{u=0}=[r_0]$ then we take dot products of the rows of $[r_0]$ with the rows of $[c]$ this will give us a vector $[v(k)]$ that has $g+1$ rows. Let $[v(k)]_p$ be the p-th row of the vector where $p=1,2,3,\ldots  ,g+1$. Then we have that:
\begin{equation}
    r_{g,k}[u]=[v(k)]_{g+1}\ .
\end{equation}
We can then write:
\begin{equation}
    R_k[u]=\sum_{p=1}^{g+1}\hbar^{2(p-1)}[v(k)]_p \ .
\end{equation}
This will completely determine $R_k$ for any $k$ to genus $g$. Below we provide an explicit {\tt Mathematica} realization of the algorithm which can successfully compute $R_k[u]$ to genus 7 (i.e. accurate to $\hbar^{14}$):

\begin{center}
\begin{tcolorbox}[colback=gray!5!white, colframe=black, width=0.95\textwidth, boxrule=0.8pt, arc=2mm, breakable]

\begin{doublespace}
\noindent\(\pmb{r[0,\text{k$\_$},x]\text{:=}u[x]{}^{\wedge}(k)}\)
\end{doublespace}

\begin{doublespace}
\noindent\(\pmb{r[\text{k$\_$},\text{k$\_$},x]\text{:=}0}\)
\end{doublespace}

\begin{doublespace}
\noindent\(\pmb{\text{genus}\text{:=}7}\)
\end{doublespace}

Setup the recursion matrix

\begin{doublespace}
\noindent\(\pmb{\text{rmat}=\text{Table}[r[i,j+i,x],\{i,0,\text{genus}\},\{j,0,2*\text{genus}\}];}\)
\end{doublespace}

Use a double for loop to implement the recursion and populate the elements of the matrix

\begin{doublespace}
\noindent\(\pmb{\text{AbsoluteTiming}[}\\
\pmb{\text{For}[a=2,a<\text{genus}+2,a\text{++},}\\
\pmb{\text{For}[b=2,\,b<2*\text{genus}+2,\,b\text{++},}\\
\pmb{\quad \text{rmat}[[a,b]]=\text{FullSimplify}[}\\
\pmb{\quad\quad \frac{2(a+b-2)}{2(a+b-2)-1}\left(u[x]*\text{rmat}[[a,b-1]] - \frac{1}{4}D[\text{rmat}[[a-1,b]],\{x,2\}]\right) }\\
\pmb{\quad\quad - \frac{(a+b-2)}{2(a+b-2)-1} \text{Integrate}[u'[x]*\text{rmat}[[a,b-1]],x] ];]]]}\)
\end{doublespace}

\begin{doublespace}
\noindent\(\{79.0712,\text{Null}\}\)
\end{doublespace}

Define the ``coefficient matrix'' c

\begin{doublespace}
\noindent\(\pmb{\text{cmat}\text{:=}\text{Table}[\text{Binomial}[k,a+b-2]*u[x]{}^{\wedge}(k-a-b+2),\{a,1,\text{genus}+1\},}\\
\noindent\pmb{\{b,1,2*\text{genus}+1\}]}\)
\end{doublespace}

Define the t-th Gel'fand Dikii polynomial accurate to genus specified

\begin{doublespace}
\noindent\(\pmb{R[\text{t$\_$}]\text{:=}\text{Sum}[h{}^{\wedge}(2*(p-1))*\text{cmat}[[p]].(\text{rmat}[[p]]\text{/.}u[x]\text{-$>$}0),\{p,1,\text{genus}+1\}]\text{/.}k\text{-$>$}t}\)
\end{doublespace}

Check the recursion relation is satisfied to order of the genus specified (i.e. CheckRecursion[g] = 0 to specified genus g)

\begin{doublespace}
\noindent\(\pmb{\text{CheckRecursion}[\text{g$\_$}]\text{:=} \text{SeriesCoefficient}[}\\
\pmb{\quad D[R[k+1],x] - \frac{2k+2}{2k+1} \left( \frac{1}{2}u'[x]*R[k] + u[x]*D[R[k],x] \right. }\\
\pmb{\quad \left. - \frac{h^2}{4}D[R[k],\{x,3\}] \right), \{h,0,2*g\} ]}\)
\end{doublespace}

\begin{doublespace}
\noindent\(\pmb{\text{Table}[\text{FullSimplify}[\text{CheckRecursion}[g]],\{g,0,\text{genus}\}]}\)
\end{doublespace}

\begin{doublespace}
\noindent\(\{0,0,0,0,0,0,0,0\}\)
\end{doublespace}

\end{tcolorbox}
\end{center}


\section{Derivation of recursion relation for $\widehat{R}_g$}
\label{RecursionRelDerRhatsAppendix}
In this Appendix we will write the solution to the equation (\ref{GDResDiffEq}) perturbatively in $\hbar$. We demonstrate the solution at any particular order in the genus expansion can be constructed recursively using lower the order solutions. 

We begin by defining the expansions we will be using to define the perturbative solution:
\begin{equation}
    \widehat{R}=\sum_{g=0}^\infty\hbar^{2g}\widehat{R}_g\ ,
\end{equation}
and
\begin{equation}
    u=\sum_{g=0}^\infty\hbar^{2g}u_{2g}\ .
\end{equation}
Then equation (\ref{GDResDiffEq}) becomes:
\begin{equation}
    \begin{split}
        &0=4(u_0-E)\sum_{k=0}^\infty\sum_{p=0}^\infty \hbar^{2(k+p)}\widehat{R}_k\widehat{R}_p+4\sum_{g=1}^\infty\sum_{k=0}^\infty\sum_{p=0}^\infty \hbar^{2(k+p+g)}u_{2g}\widehat{R}_k\widehat{R}_p\\
        &\hskip 2cm -2\sum_{k=0}^\infty\sum_{p=0}^\infty \hbar^{2(k+p+1)}\widehat{R}_k\widehat{R}_p''+\sum_{k=0}^\infty\sum_{p=0}^\infty \hbar^{2(k+p+1)}\widehat{R}_k'\widehat{R}_p'-1\ .\\
    \end{split}
\end{equation}
The solution to this can be expressed as a recursion relation which we can derive by organizing everything in powers of $\hbar$. Start by noting that:
\begin{equation}
    \begin{split}
        &\sum_{k=0}^\infty\sum_{p=0}^\infty \hbar^{2(k+p)}\widehat{R}_k\widehat{R}_p=\sum_{J=0}^\infty\sum_{p=0}^J\hbar^{2J}\widehat{R}_p\widehat{R}_{J-p}\\
        &\sum_{k=0}^\infty\sum_{p=0}^\infty \hbar^{2(k+p+1)}\widehat{R}_k\widehat{R}_p''=\sum_{J=1}^\infty\sum_{p=0}^{J-1}\hbar^{2J}\widehat{R}_p''\widehat{R}_{J-p-1}\\
        &\sum_{k=0}^\infty\sum_{p=0}^\infty \hbar^{2(k+p+1)}\widehat{R}_k'\widehat{R}_p'=\sum_{J=1}^\infty\sum_{p=0}^{J-1}\hbar^{2J}\widehat{R}_p'\widehat{R}_{J-p-1}'\\
        &\sum_{g=1}^\infty\sum_{k=0}^\infty\sum_{p=0}^\infty\hbar^{2(k+p+g)}u_{2g}\widehat{R}_p\widehat{R}_k=\sum_{J=1}^\infty\sum_{L=0}^{J-1}\sum_{p=0}^L\hbar^{2J}u_{2(J-L)}\widehat{R}_{L-p}\widehat{R}_p \ .\\
    \end{split}
\end{equation}
Then we can write:
\begin{equation}
\begin{split}
     &0=\left(4X\widehat{R}_0^2-1\right)+\sum_{J=1}^\infty\sum_{p=0}^J\hbar^{2J}4X\widehat{R}_p\widehat{R}_{J-p}+\sum_{J=1}^\infty\sum_{p=0}^{J-1}\hbar^{2J}\left[-2\widehat{R}_p''\widehat{R}_{J-p-1}+\widehat{R}_p'\widehat{R}'_{J-p-1}\right]\\
     &\hskip 1.5cm+\sum_{J=1}^\infty\sum_{L=0}^{J-1}\sum_{p=0}^L\hbar^{2J}4u_{2(J-L)}\widehat{R}_{L-p}\widehat{R}_p\ , 
\end{split}
\end{equation}
where $X\equiv u_0-E$.
At zeroth order in $\hbar$, $4X\widehat{R}_0^2-1=0$, leading to:
\begin{equation}
    \widehat{R}_0=-\frac{1}{2X^{1/2}}\ ,
\end{equation}
where the sign choice matches our earlier conventions.

At order $\hbar^{2J}$ with $J\geq1$ we have the following recursion relation:
\begin{equation}
    \sum_{p=0}^J4X\widehat{R}_p\widehat{R}_{J-p}+\sum_{p=0}^{J-1}\left[-2\widehat{R}_p''\widehat{R}_{J-p-1}+\widehat{R}_p'\widehat{R}'_{J-p-1}\right]+\sum_{L=0}^{J-1}\sum_{p=0}^L4u_{2(J-L)}\widehat{R}_{L-p}\widehat{R}_p=0\ .
\end{equation}
In the first sum we can isolate the terms for $p=0,J$ and write:
\begin{equation}
    8X\widehat{R}_0\widehat{R}_J+\sum_{p=1}^{J-1}4X\widehat{R}_p\widehat{R}_{J-p}+\sum_{p=0}^{J-1}\left[-2\widehat{R}_p''\widehat{R}_{J-p-1}+\widehat{R}_p'\widehat{R}'_{J-p-1}\right]+\sum_{L=0}^{J-1}\sum_{p=0}^L4u_{2(J-L)}\widehat{R}_{L-p}\widehat{R}_p=0\ .
\end{equation}
Now we can clearly see the sums only involve $\widehat{R}_p$ with $p<J$ so we have successfully written $\widehat{R}_J$ in terms of lower $\widehat{R}_p$'s. We can manipulate a bit further by also isolating for terms that will involve $\widehat{R}_0$ and write:
\begin{eqnarray} 
&&\hskip-0.75cm 8X\widehat{R}_0\widehat{R}_J+2\left[\widehat{R}_0'\widehat{R}_{J-1}'-\widehat{R}_0''\widehat{R}_{J-1}-\widehat{R}_{J-1}''\widehat{R}_0+2u_{2J}\widehat{R}_0^2+\delta_{J,1}\left(\widehat{R}_0''\widehat{R}_0-\frac{1}{2}(\widehat{R}_0')^{2}\right)\right]+8\sum_{L=1}^{J-1}u_{2(J-L)}\widehat{R}_0\widehat{R}_L\nonumber\\
        &&+4X\sum_{p=1}^{J-1}\widehat{R}_p\widehat{R}_{J-p}+\sum_{p=1}^{J-2}\left[\widehat{R}_p'\widehat{R}'_{J-p-1}-2\widehat{R}_p''\widehat{R}_{J-p-1}\right]+4\sum_{L=1}^{J-1}\sum_{\substack{p=1 \\ p\neq L}}^Lu_{2(J-L)}\widehat{R}_p\widehat{R}_{L-p}=0\ ,
\end{eqnarray}
which, after dividing through by $8X\widehat{R}_0$, will recover equation~(\ref{RecursionforRAnyg}).

\section{Procedure to explicitly construct $\widehat{Q}_g$}
\label{AppendixConstructQ}
In this Appendix we will explicitly demonstrate how to systematically determine $\widehat{Q}_g$ given $\widehat{R}_g$. Amusingly enough this procedure also provides a rather efficient way to obtaining the solution to the string equation in the perturbative expansion (i.e. $u_2,u_4,\ldots ,\text{\it etc.}$).  

Generally we will have the following expression for the resolvent in the genus expansion:
\begin{equation}
    \widehat{R}(x,E)=\sum_{g=0}^\infty \hbar^{2g}\widehat{R}_g(x,E)\ ,
\end{equation}
where we write:
\begin{equation}
\label{RgExpansion}
    \widehat{R}_g=\sum_{k=1}^{3g}\frac{\widehat{R}_{g,k}(x)}{(u_0-E)^{k+\frac{1}{2}}}\ .
\end{equation}
The anzatz for $\widehat{Q}_g$ will take the form:
\begin{equation}
\label{QgExpansion}
    \widehat{Q}_g=\sum_{k=1}^{3g-1}\frac{\widehat{Q}_{g,k}(x)}{(u_0-E)^{k+\frac{1}{2}}}\ .
\end{equation}
We can find explicit expressions for $\widehat{R}_{g,k}$ using these we can find we can express $\widehat{Q}_g$ in terms of $\widehat{R}_g$ by requiring that $\widehat{R}_g=\frac{d}{dx}\widehat{Q}_g$. The recursion relation we need to use is given by a rearrangement of equation~(\ref{RecursionForsmallq}):
\begin{equation}
    \widehat{Q}_{g,k}=\frac{\widehat{Q}'_{g,k+1}-\widehat{R}_{g,k+1}}{\left(k+\frac{1}{2}\right)u_0'}\ ,
\end{equation}
where we begin at $k=3g-1$ and proceed to $k=1$. Lets demonstrate our construction for $g=1,2$ given that: 
\begin{equation}
    \begin{split}
        &\widehat{R}_1=\frac{u_2}{4X^{3/2}}+\frac{u_0''}{16 X^{5/2}}-\frac{5u_0'^2}{64X^{7/2}}\\
        &\widehat{R}_2=\frac{u_4}{4X^{3/2}}+\frac{u_2''-3u_2^2}{16X^{5/2}}+\frac{u_0^{(4)}-10(u_0'u_2'+u_2u_0'')}{64X^{7/2}}+\frac{7(10u_2u_0'^2-3u_0''^2-4u_0'u_0''')}{256X^{9/2}}\\
        &+\frac{231u_0'^2u_0''}{512 X^{11/2}}-\frac{1155u_0'^4}{4096 X^{13/2}}\ ,
    \end{split}
\end{equation}
where $X\equiv u_0-E$. This can be obtained in a straightforward manner from the recursion relation given in equation (\ref{RecursionforRAnyg}).

For $g=1$ we begin with:
\begin{equation}
    \widehat{Q}_{1,2}=\frac{-\widehat{R}_{1,3}}{(2+\frac{1}{2})u_0'}=\frac{1}{32}u_0'\ ,
\end{equation}
where we used the fact that $\widehat{Q}_{1,3}=0$ and $\widehat{R}_{1,3}=-\frac{5}{64}u_0'^2$. Using this result we can proceed to compute:
\begin{equation}
    \widehat{Q}_{1,1}=\frac{\widehat{Q}'_{1,2}-\widehat{R}_{1,2}}{(1+\frac{1}{2})u_0'}=-\frac{u_0''}{48u_0'}\ ,
\end{equation}
where we used $\widehat{R}_{1,2}=\frac{1}{16}u_0''$. This completes the construction of $\widehat{Q}_1$ which is given in equation (\ref{ExpressionForQs}).

The final recursion relation will reproduce the genus 1 solution to the string equation as follows:
\begin{equation}
    0=\widehat{Q}_{1,0}=\frac{\widehat{Q}'_{1,1}-\widehat{R}_{1,1}}{\frac{1}{2}u_0'}\Rightarrow\widehat{R}_{1,1}=\widehat{Q}'_{1,1}\ .
\end{equation}
Using the fact that $\widehat{R}_{g,1}=\frac{1}{4}u_{2g}$ we arrive at the following expression for $u_2$:
\begin{equation}
    u_2=4\widehat{Q}'_{1,1}=4\frac{d}{dx}\left[-\frac{u_0''}{48u_0'}\right]=\frac{u_0''^2-u_0'u_0'''}{12u_0'^2}\ ,
\end{equation}
which exactly reproduces the solution to the genus 1 solution to the string equation. 

Now lets apply the same procedure for $g=2$. This time we start with:
\begin{equation}
    \widehat{Q}_{2,5}=\frac{-\widehat{R}_{2,6}}{\left(5+\frac{1}{2}\right)u_0'}=\frac{105 u_0'^3}{2048}\ .
\end{equation}
Using the result above we can compute:
\begin{equation}
    \widehat{Q}_{2,4}=\frac{\widehat{Q}'_{2,5}-\widehat{R}_{2,5}}{\left(4+\frac{1}{2}\right)u_0'}=-\frac{203u_0'u_0''}{3072}\ .
\end{equation}
Proceeding inductively we will have:
\begin{equation}
    \begin{split}
        &\widehat{Q}_{2,3}=\frac{\widehat{Q}'_{2,4}-\widehat{R}_{2,4}}{\left(3+\frac{1}{2}\right)u_0'}=\frac{-120u_2u_0'+19u_0'''+\frac{7u_0''^2}{u_0'}}{1536}\\
        \end{split}
        \end{equation}
        \begin{equation}
            \begin{split}
                &\widehat{Q}_{2,2}=\frac{\widehat{Q}'_{2,3}-\widehat{R}_{2,3}}{\left(2+\frac{1}{2}\right)u_0'}=\frac{120u_0'^3u_2'-7u_0''^3+14u_0'u_0''u_0'''+5u_0'^2\left(24u_2u_0''-u_0''''\right)}{3840u_0'^3}\\
            \end{split}
        \end{equation}
        \begin{equation}
            \begin{split}
        &\widehat{Q}_{2,1}=\frac{\widehat{Q}'_{2,2}-\widehat{R}_{2,2}}{\left(1+\frac{1}{2}\right)u_0'}\\
        &=\frac{720u_2^2u_0'^4+21u_0''^4-120u_0'^4u_2''-49u_0'u_0''^2u_0'''+120u_2u_0'^2\left(-u_0''^2+u_0'u_0'''\right)+u_0'^2\left(14u_0'''^2+19u_0''u_0^{(4)}\right)}{5760u_0'^5}\\
        &+\frac{5u_0'^3\left(24 u_2'u_0''-u_0^{(5)}
    \right)}{5760u_0'^5}\ .\\
    \end{split}
\end{equation}
This will complete the construction of $\widehat{Q}_2$ which after substituting into the expression for $u_2$ will give:
\begin{equation}
    \begin{split}
        &\widehat{Q}_2=\frac{105u_0'^3}{2048X^{11/2}}-\frac{203u_0'u_0''}{3072X^{9/2}}+\frac{29u_0'u_0'''-3u_0''^2}{1536u_0'X^{7/2}}-\frac{17u_0''^3-34u_0'u_0''u_0'''+15u_0'^2u_0^{(4)}}{3840u_0'^3X^{5/2}}\\
        &-\frac{64u_0''^4-111u_0'u_0''^2u_0'''+21u_0'^2u_0'''^2+31u_0'^2u_0''u_0^{(4)}-5u_0'^3u_0^{(5)}}{5760u_0'^5X^{3/2}}\ ,\\
    \end{split}
\end{equation}
which gives the expression in equation (\ref{ExpressionForQs}). Similar to before one can check that the genus 2 solution to the string equation is exactly given by:
\begin{equation}
\begin{split}
    &u_4=4\widehat{Q}_{2,1}'=4\frac{d}{dx}\left[-\frac{64u_0''^4-111u_0'u_0''^2u_0'''+21u_0'^2u_0'''^2+31u_0'^2u_0''u_0^{(4)}-5u_0'^3u_0^{(5)}}{5760u_0'^5}\right]\\
    &=\frac{d^2}{dx^2}\left[\frac{u_0''^3}{90u_0'^4}-\frac{7u_0''u_0'''}{480u_0'^3}+\frac{u_0^{(4)}}{288u_0'^2}\right]\ .
\end{split}
\end{equation}
Although we do not go over the details of obtaining the expression for $\widehat{Q}_3$ given in equation~(\ref{eq:Q3}) it follows from an analogous procedure as the others we reviewed here.

\section{Evaluations  at the Fermi surface}
A central aspect of calculating volumes using the ODE method is the ability to compute arbitrary order derivatives of the tree level solution to the string equation at the Fermi surface $x=\mu$. We accomplish this by repeatedly taking derivatives of the tree level string equation. This will involve arbitrary order derivatives of $G_0(u_0(x)){\equiv}\sum_{k=1}^\infty t_k u_0(x)^k$ with respect to $x$. When we take such derivatives we will inevitably end up having to deal with series expressions corresponding to $\frac{d^p}{du_0^p}G_0\equiv G_0^{(p)}(u_0)=\sum_{k=1}^\infty \frac{k!}{(k-p)!}t_ku_0^{k-p}$. As we will demonstrate in this Appendix, for the theories considered in this work, such series can be re-summed exactly at the Fermi surface ($x=\mu$).

\subsection{$\mathcal{N}=2$}

\label{GExactatE0}
Here we will provide closed form expression for $G_0^{(p)}(u_0({\mu}))=G_0^{(p)}(E_0)$ where $G_0^{(p)}(E_0)$ is the $p$-th derivative of $G_0$ at $u_0$ evaluated at $x={\mu}$ for $\mathcal{N}=2$. 

Lets start with the $p=0$ case:
We can write:
\begin{equation}
    G_0(E_0)=\sum_{k=1}^\infty{t}_kE_0^k=\sum_{k=1}^\infty\frac{\pi^{k-1}J_k(2\pi\sqrt{E_0})}{2(2k+1)k!E_0^{k/2}}E_0^k\ .
\end{equation}
The next step is to use the series representation of the Bessel function given by:
\begin{equation}
    J_k(x)=\sum_{n=0}^\infty\frac{(-1)^n}{n!(k+n)!}\left(\frac{x}{2}\right)^{k+2n}\ .
\end{equation}
Using this we arrive at the following double sum for $G_0(E_0)$:
\begin{equation}
    G_0(E_0)=\sum_{n=0}^\infty\sum_{k=1}^\infty\frac{(-1)^n\pi^{2n+2k-1}}{2(2k+1)n!k!(n+k)!}E_0^{n+k}\ .
\end{equation}
We can rewrite the double sum in a form that makes it more natural in the $E_0$ expansion by defining $J=n+k$ where $J=1,2,3,\ldots$ Then we can rewrite the double sum as:
\begin{equation}
    G_0(E_0)=\sum_{J=1}^\infty\sum_{n=0}^{J-1}\frac{(-1)^n\pi^{2J-1}}{2(2J-2n+1)n!J!(J-n)!}E_0^J \ .
\end{equation}
The nice thing about this representation is we can now explicitly evaluate the inner sum from $n=0$ to $J-1$ the result is:
\begin{equation}
    \sum_{J=1}^\infty\frac{(-1)^J}{2\pi}\left[\frac{(2\pi)^{2J}}{(2J+1)!}-\frac{\pi^{2J}}{J!^2}\right]E_0^J \ .
\end{equation}
The sum above can be evaluated exactly as well to give the following final result:
\begin{equation}
    G_0(E_0)=\frac{1}{2\pi}\left[-J_0(2\pi\sqrt{E_0})+\frac{\sin(2\pi\sqrt{E_0})}{2\pi\sqrt{E_0}}\right]\ .
\end{equation}
A similar procedure can be done for higher derivatives of $G_0$ in particular we will have
\begin{equation}
    G_0^{(p)}(E_0)=\sum_{k=1}^\infty\frac{k!}{(k-p)!}{t}_kE_0^{k-p}\ .
\end{equation}
This time the double sum will take the form:
\begin{equation}
    G_0^{(p)}(E_0)=\sum_{k=p}^\infty\sum_{n=0}^\infty\frac{(-1)^n\pi^{2k+2n-1}}{2(2k+1)n!(k-p)!(k+n)!}E_0^{n+k-p}\ .
\end{equation}
This time we take $J=n+k-p$ and $J=0,1,2,3,\ldots  $ Then the double sum can be rewritten as:
\begin{equation}
    \sum_{J=0}^\infty\sum_{n=0}^J\frac{(-1)^n\pi^{2J+2p-1}}{2[2(J-n+p)+1]n!(J+p)!(J-n)!}E_0^J\ .
\end{equation}
Just as before the inner sum over $n$ can be done exactly this gives:
\begin{equation}
    G^{(p)}(E_0)=\sum_{J=0}^\infty\frac{(-1)^J(2\pi)^{2J+2p-1}(2p)!}{4^pp!(2J+2p+1)!}E_0^J\ .
\end{equation}
The sum over $J$ can be done exactly and expressed in terms of hypergeometric functions:
\begin{equation}
    G^{(p)}(E_0)=\frac{\pi^{2p-1}(2p)!}{2p!(1+2p)!}{}_1F_2(1;p+1,p+3/2;-\pi^2E_0) \ .
\end{equation}
For example for $p=1$ we have:
\begin{equation}
    \dot{G}_0(E_0)=\frac{1}{4\pi E_0}-\frac{\sin(2\pi\sqrt{E_0})}{8\pi^2E_0^{3/2}} \ .
\end{equation}
For $p=2$ we get:
\begin{equation}
    \ddot{G}_0(E_0)=-\frac{3}{8\pi E_0^2}+\frac{\pi}{4E_0}+\frac{3\sin(2\pi\sqrt{E_0})}{16\pi^2E_0^{5/2}} \ .
\end{equation}
For $p=3$ we have:
\begin{equation}
    \dddot{G}_0(E_0)=\frac{15}{16\pi E_0^3}-\frac{5\pi}{8E_0^2}+\frac{\pi^3}{8E_0}-\frac{15\sin(2\pi\sqrt{E_0})}{32\pi^2 E_0^{7/2}} \ .
\end{equation}
etc.

Note that these results allow us to prove the expressions we had for ${\mu}(E_0)$ as well as the results we had for $u_0'({\mu})$ and $u_0''({\mu})$. It useful to write the term involving $\sin$ in terms of $\widetilde{\Gamma}$ so that we can write:
\begin{equation}
    \begin{split}
        &\dot{G}_0(E_0)=\frac{1}{4\pi E_0}-\frac{\widetilde{\Gamma}}{2E_0^{3/2}}\\
        &\ddot{G}_0(E_0)=-\frac{3}{8\pi E_0^2}+\frac{\pi}{4E_0}+\frac{3\widetilde{\Gamma}}{4E_0^{5/2}}\\
        &\dddot{G}_0(E_0)=\frac{15}{16\pi E_0^3}-\frac{5\pi}{8E_0^2}+\frac{\pi^3}{8E_0}-\frac{15}{8}\frac{\widetilde{\Gamma}}{E_0^{7/2}}\ .\\
    \end{split}
\end{equation}
This suggests that we can conjecture $G_0^{(p)}(E_0)$ to have the following form:
\begin{equation}
    G_0^{(p)}(E_0)=\widetilde{\Gamma}\frac{d^p}{dE_0^p}\left(\frac{1}{E_0^{1/2}}\right)+\sum_{k=1}^{p}\frac{c_k^{(p)}}{E_0^k}=\frac{\sqrt{\pi}}{\Gamma(1/2-p)}\frac{\widetilde{\Gamma}}{E_0^{p+1/2}}+\sum_{k=1}^{p}\frac{c_k^{(p)}}{E_0^k}\ .
\end{equation}
Note that $\Gamma$ in the denominator of the expression above  is the $\Gamma$-function not the constant in the string equation. Assuming that the conjecture is true we  arrive at:
\begin{equation}
    \sum_{k=1}^{p}\frac{c_k^{(p)}}{E_0^k}=G_0^{(p)}(E_0)-\frac{\sqrt{\pi}}{\Gamma(1/2-p)}\frac{\widetilde{\Gamma}}{E_0^{p+1/2}}\ .
\end{equation}

\subsection{$\mathcal{N}=4$}
\label{N=4G0Calcs}

The (small) $\mathcal{N}=4$ analysis is similar to the $\mathcal{N}=2$ computations. In this case we have:
\begin{equation}
{t}_k=\frac{8\pi^{k+1}}{J^kk!(4k^2+8k+3)}J_{k+1}(2\pi J)\ .
\end{equation}
We can use the series representation of Bessel function and rewrite the double sum for $G_0$ and its derivatives as follows:
\begin{equation}
\begin{split}
    &G_0(u_0)\vert_{u_0=J^2}=\sum_{L=1}^\infty\sum_{n=0}^{L-1} \frac{(-1)^n(2\pi)^22\pi^{2L}}{\left[4(L-n)^2+8(L-n)+3\right](L-n)!(L+1)!n!}J^{2L+1}\\
    &G_0^{(p)}(u_0)|_{u_0=J^2}=\sum_{L=0}^\infty\sum_{n=0}^L\frac{(-1)^n8\pi^{2L+2p+2}}{\left[4(L-n+p)^2+8(L-n+p)+3\right](L+p+1)!(L-n)!n!}J^{2L+1}\ .\\
\end{split}
\end{equation}
Doing the sum over $n$ first then doing the sum over $L$ gives us:
\begin{equation}
    \begin{split}
        &G_0(u_0)\vert_{u_0=J^2}=-\frac{8}{3}\pi J_1(2\pi J)+\frac{2}{J}\left[\frac{\sin(2\pi J)}{2\pi J}-\cos(2\pi J)\right]\\
        &G_0^{(p)}(u_0)\vert_{u_0=J^2}\\
        &=\frac{16J\pi^{2p+2}(2p)!}{p!(3+2p)!}\left[_{1}F_2\left(1;p+2,p+\frac{5}{2};-(\pi J)^2\right)-\frac{2\pi^2J^2 }{(p+2)(5+2p)} {}_1F_2\left(2;p+3,p+\frac{7}{2};-(\pi J)^2\right)\right]\ .\\
    \end{split}
\end{equation}
One can check that $G^{(p)}$ have the following structure for any given $p\in\mathbb{N}$:
\begin{equation}
    G_0^{(p)}(u_0)\vert_{u_0=J^2}=\frac{(-1)^p(2p-1)!!}{2^{p-1}J^{2p+1}}\left[-\cos(2\pi J)+\frac{(2p+1)\sin(2\pi J)}{2\pi J}\right]+\sum_{k=1}^{p}\frac{c_k^{(p)}}{J^{2k+1}}\ ,
\end{equation}
for some choice of constant coefficient $c_k^{(p)}$. Finally, using the fact that $J\in\frac{1}{2}\mathbb{Z}_{\neq0}$ we can simplify further and write it as:
\begin{equation}
    G_0^{(p)}(u_0)|_{u_0=J^2}=\frac{(-1)^{2J+p+1}(2p-1)!!}{2^{p-1}J^{2p+1}}+\sum_{k=1}^p\frac{c_k^{(p)}}{J^{2k+1}}\ .
\end{equation}
Recall the tree level string equation is given by:
\begin{equation}
    u_0(G_0+x)^2=\widetilde{\Gamma}^2\ ,
\end{equation}
where $|\widetilde{\Gamma}|=2$. We can set $x={\mu}$ and also use the fact that $u_0({\mu})=J^2$ and isolate for ${\mu}$ to get:
\begin{equation}
    {\mu}=\frac{2w}{|J|}-G_0(J^2)\ ,
\end{equation}
where $w=\pm 1$ which we will leave unfixed for now and the $|J|$ comes from the fact that $|J|=\sqrt{J^2}$. Explicitly we can write the following exact expression for ${\mu}$:
\begin{equation}
\begin{split}
    {\mu}&=\frac{2w}{|J|}+\frac{8\pi}{3}J_1(2\pi J)-\frac{2}{J}\left[\frac{\sin(2\pi J)}{2\pi J}-\cos(2\pi J)\right]\\
    &=\frac{8\pi}{3}J_1(2\pi J)+\frac{2w}{|J|}+\frac{2(-1)^{2J}}{J}\ ,\\
\end{split}
\end{equation}
where in the last line we used the fact that $2J$ is a non-zero integer. In order to get ${\mu}={t}_k|_{k=0}$ we must require that $w=\text{sgn}(J)(-1)^{2J+1}$. This suggests the string equation for different values of $J$ slightly changes, in particular, we will write the string equation in our calculations as:
\begin{equation}
    G_0(u_0(x))+x+\frac{2\text{sgn}(J)(-1)^{2J}}{u_0(x)^{1/2}}=0\ .
\end{equation}

\section{Integrals for evaluating $\mathcal{N}=2$ recursion relations}
\label{KernelIntegralsN=2}
In this Appendix we will list the expressions for:
\begin{equation}
\begin{split}
    &\mathcal{A}_{0}(b;p,q)=\int_0^{\infty}db'db''(b')^p(b'')^qD_0(b,b',b'')\ ,\quad\mathcal{A}_2(b;p,q)=\int_0^{\infty}db'db''(b')^p(b'')^qD_2(b,b',b'')\ ,\\
    &\mathcal{T}_0(b,b_1;p)=\int_0^\infty db'(b')^pT_{0}(b,b',b_1)\ , \quad
    \mathcal{T}_2(b,b_1;p)=\int_0^\infty db'(b')^p[b-T_{2}(b,b',b_1)]\ ,
\end{split}
\end{equation}
which are used to evaluate various integrals in the topological recursion formulas defining volumes in Section \ref{sec:TW-approach}.
For various values of $p,q=1,3,5,7,9, \text{etc.}$. Defining $B=b(b^2+4\pi^2)$ we have:
\begin{equation}
\small
\begin{split}
    &\mathcal{A}_0(b;1,1)=\frac{1}{6}B\\
&\mathcal{A}_0(b;1,3)=\frac{1}{60} \left(3 b^2+28 \pi ^2\right)B\\
    &\mathcal{A}_0(b;1,5)=\frac{1}{126} \left(3 b^4+72 \pi ^2 b^2+496 \pi ^4\right)B\\
    &\mathcal{A}_0(b;1,7)=\frac{1}{360} \left(5 b^6+220 \pi ^2 b^4+3824 \pi ^4 b^2+24384 \pi ^6\right) B\\
    &\mathcal{A}_0(b;1,9)=\frac{1}{330} \left(b^2+20 \pi ^2\right) \left(3 b^6+148 \pi ^2 b^4+3600 \pi ^4 b^2+32704 \pi ^6\right) B\\
    &\mathcal{A}_0(b;1,11)=\frac{\left(105 b^{10}+10500 \pi ^2 b^8+518560 \pi ^4 b^6+14948480 \pi ^6 b^4+233106688 \pi ^8
   b^2+1448424448 \pi ^{10}\right) B}{16380}\\
   \end{split}
\end{equation}
\begin{equation}
\small
\begin{split}
    &\mathcal{A}_0(b;3,3)=\frac{1}{420} \left(3 b^4+72 \pi ^2 b^2+496 \pi ^4\right) B\\
    &\mathcal{A}_0(b;3,5)=\frac{\left(5 b^6+220 \pi ^2 b^4+3824 \pi ^4 b^2+24384 \pi ^6\right) B}{2520}\\
    &\mathcal{A}_0(b;3,7)=\frac{\left(b^2+20 \pi ^2\right) \left(3 b^6+148 \pi ^2 b^4+3600 \pi ^4 b^2+32704 \pi ^6\right) B}{3960}\\
    &\mathcal{A}_0(b;3,9)=\frac{\left(105 b^{10}+10500 \pi ^2 b^8+518560 \pi ^4 b^6+14948480 \pi ^6 b^4+233106688 \pi ^8 b^2+1448424448 \pi ^{10}\right) B}{300300}\\
     \end{split}
\end{equation}
\begin{equation}
\small
\begin{split}
    &\mathcal{A}_0(b;5,5)=\frac{\left(b^2+20 \pi ^2\right) \left(3 b^6+148 \pi ^2 b^4+3600 \pi ^4 b^2+32704 \pi ^6\right) B}{8316}\\
    &\mathcal{A}_0(b;5,7)=\frac{\left(105 b^{10}+10500 \pi ^2 b^8+518560 \pi ^4 b^6+14948480 \pi ^6 b^4+233106688 \pi ^8 b^2+1448424448 \pi ^{10}\right) B}{1081080}\\
    &\mathcal{A}_0(b;5,9)=\frac{\left(3 b^{12}+408 \pi ^2 b^{10}+28944 \pi ^4 b^8+1302784 \pi ^6 b^6+36631808 \pi ^8 b^4+567728128 \pi ^{10} b^2+3522785280 \pi ^{12}\right) B}{90090}\\
     \end{split}
\end{equation}
\begin{equation}
\small
\begin{split}
    &\mathcal{A}_0(b;7,7)=\frac{\left(3 b^{12}+408 \pi ^2 b^{10}+28944 \pi ^4 b^8+1302784 \pi ^6 b^6+36631808 \pi ^8 b^4+567728128 \pi ^{10} b^2+3522785280 \pi ^{12}\right) B}{154440}\\
    &\mathcal{A}_0(b;7,9)=\frac{A_{0;7,9} B}{2917200}\\
    &\mathcal{A}_0(b;9,9)=\frac{A_{0;9,9} B}{96996900}\ ,
\end{split}
\end{equation}
where
\begin{equation}
    \begin{split}
        &A_{0;7,9}=15 b^{14}+2660 \pi ^2 b^{12}+255920 \pi ^4 b^{10}+16514880 \pi ^6 b^8+724307200 \pi ^8 b^6+20231040000 \pi ^{10} b^4\\
        &\hskip2cm+313047289856 \pi ^{12} b^2+1941802827776 \pi ^{14}\\
        &A_{0;9,9}=105 b^{16}+23520 \pi ^2 b^{14}+2944704 \pi ^4 b^{12}+257370624 \pi ^6 b^{10}+16171771392 \pi ^8 b^8+704327852032 \pi ^{10} b^6\\
        &\hskip2cm+19639061233664 \pi ^{12} b^4+303764595933184 \pi ^{14}
   b^2+1884058929397760 \pi ^{16}\ .
    \end{split}
\end{equation}
\begin{equation}
\small
\begin{split}
    &\mathcal{A}_2(b;1,1)=\frac{1}{360} \left(3 b^2+28 \pi ^2\right) B\\
&\mathcal{A}_2(b;1,3)=\frac{\left(3 b^4+72 \pi ^2 b^2+496 \pi ^4\right) B}{2520}\\
&\mathcal{A}_2(b;1,5)=\frac{\left(5 b^6+220 \pi ^2 b^4+3824 \pi ^4 b^2+24384 \pi ^6\right) B}{15120}\\
&\mathcal{A}_2(b;1,7)=\frac{\left(b^2+20 \pi ^2\right) \left(3 b^6+148 \pi ^2 b^4+3600 \pi ^4 b^2+32704 \pi ^6\right) B}{23760}\\
&\mathcal{A}_2(b;1,9)=\frac{\left(105 b^{10}+10500 \pi ^2 b^8+518560 \pi ^4 b^6+14948480 \pi ^6 b^4+233106688 \pi ^8 b^2+1448424448 \pi ^{10}\right) B}{1801800}\\
&\mathcal{A}_2(b;1,11)=\frac{\left(3 b^{12}+408 \pi ^2 b^{10}+28944 \pi ^4 b^8+1302784 \pi ^6 b^6+36631808 \pi ^8 b^4+567728128
   \pi ^{10} b^2+3522785280 \pi ^{12}\right) B}{98280}\\
      \end{split}
\end{equation}
\begin{equation}
\small
\begin{split}
    &\mathcal{A}_2(b;3,3)=\frac{\left(5 b^6+220 \pi ^2 b^4+3824 \pi ^4 b^2+24384 \pi ^6\right) B}{50400}\\
     &\mathcal{A}_2(b;3,5)=\frac{\left(b^2+20 \pi ^2\right) \left(3 b^6+148 \pi ^2 b^4+3600 \pi ^4 b^2+32704 \pi ^6\right) B}{166320}\\
      &\mathcal{A}_2(b;3,7)=\frac{\left(105 b^{10}+10500 \pi ^2 b^8+518560 \pi ^4 b^6+14948480 \pi ^6 b^4+233106688 \pi ^8 b^2+1448424448 \pi ^{10}\right) B}{21621600}\\
      &\mathcal{A}_2(b;3,9)=\frac{\left(3 b^{12}+408 \pi ^2 b^{10}+28944 \pi ^4 b^8+1302784 \pi ^6 b^6+36631808 \pi ^8 b^4+567728128 \pi ^{10} b^2+3522785280 \pi ^{12}\right) B}{1801800}\\
      \end{split}
\end{equation}
\begin{equation}
\small
\begin{split}
      &\mathcal{A}_2(b;5,5)=\frac{\left(105 b^{10}+10500 \pi ^2 b^8+518560 \pi ^4 b^6+14948480 \pi ^6 b^4+233106688 \pi ^8 b^2+1448424448 \pi ^{10}\right) B}{45405360}\\
      &\mathcal{A}_2(b;5,7)=\frac{\left(3 b^{12}+408 \pi ^2 b^{10}+28944 \pi ^4 b^8+1302784 \pi ^6 b^6+36631808 \pi ^8 b^4+567728128 \pi ^{10} b^2+3522785280 \pi ^{12}\right) B}{6486480}\\
      &\mathcal{A}_2(b;5,9)=\frac{A_{2;5,9}B}{122522400}\\
      &\mathcal{A}_2(b;7,7)=\frac{A_{2;7,7}B}{210038400}\\
      &\mathcal{A}_2(b;7,9)=\frac{A_{2;7,9}B}{6983776800}\\
      &\mathcal{A}_2(b;9,9)=\frac{A_{2;9,9}B}{64017954000}\ ,
\end{split}
\end{equation}
where:
\begin{equation}
    \begin{split}
        &A_{2;5,9}=15 b^{14}+2660 \pi ^2 b^{12}+255920 \pi ^4 b^{10}+16514880 \pi ^6 b^8+724307200 \pi ^8 b^6+20231040000 \pi ^{10} b^4\\
        &\hskip 2cm+313047289856 \pi ^{12} b^2+1941802827776 \pi ^{14}\\
        &A_{2;7,7}=15 b^{14}+2660 \pi ^2 b^{12}+255920 \pi ^4 b^{10}+16514880 \pi ^6 b^8+724307200 \pi ^8 b^6+20231040000 \pi ^{10} b^4\\&
        \hskip2cm+313047289856 \pi ^{12} b^2+1941802827776 \pi ^{14}\\
        &A_{2;7,9}=105 b^{16}+23520 \pi ^2 b^{14}+2944704 \pi ^4 b^{12}+257370624 \pi ^6 b^{10}+16171771392 \pi ^8 b^8+704327852032 \pi ^{10} b^6\\
        &\hskip2cm+19639061233664 \pi ^{12} b^4+303764595933184 \pi ^{14}
   b^2+1884058929397760 \pi ^{16}\\
        &A_{2;9,9}=165 b^{18}+45540 \pi ^2 b^{16}+7191360 \pi ^4 b^{14}+817132800 \pi ^6 b^{12}+69506004480 \pi ^8 b^{10}+4336065607680 \pi ^{10} b^8\\
        &\hskip2cm+188505820119040 \pi ^{12} b^6+5253875647447040 \pi ^{14}
   b^4+81255368743518208 \pi ^{16} b^2\\
   &\hskip3cm+503964453960941568 \pi ^{18}\ .
    \end{split}
\end{equation}
Note that $\mathcal{A}_0(b,p,q)=\mathcal{A}_0(b,q,p)$ and $\mathcal{A}_2(b,p,q)=\mathcal{A}_2(b,q,p)$.
\begin{equation}
\begin{split}
    &\mathcal{T}_0(b,b_1;1)=-b\\
    &\mathcal{T}_0(b,b_1;3)=-b(b^2+3b_1^2+4\pi^2)\\
    &\mathcal{T}_0(b,b_1;5)=-\frac{b}{3}\left[3(b^4+10b^2b_1^2+5b_1^4)+40\pi^2(b^2+3b_1^2)+112\pi^4\right]\\
    &\mathcal{T}_0(b,b_1;7)=-\frac{b}{3}  \left(784 \pi ^4 \left(b^2+3 b_1^2\right)+84 \pi ^2 \left(b^4+10 b^2 b_1^2+5 b_1^4\right)+3 \left(b^6+21 b^4 b_1^2+35 b^2 b_1^4+7 b_1^6\right)+1984 \pi
   ^6\right)\\
    &\mathcal{T}_0(b,b_1;9)=-\frac{b}{5}  \left[39680 \pi ^6 \left(b^2+3 b_1^2\right)+4704 \pi ^4 \left(b^4+10 b^2 b_1^2+5 b_1^4\right)+5 \left(b^6+33 b^4 b_1^2+27 b^2 b_1^4+3 b_1^6\right)
   \left(b^2+3 b_1^2\right)\right. \\
   &\hskip3cm\left. +240 \pi ^2 \left(b^6+21 b^4 b_1^2+35 b^2 b_1^4+7 b_1^6\right)+97536 \pi ^8\right]\ .
\end{split}
\end{equation}
and:
\begin{equation}
\begin{split}
    &\mathcal{T}_2(b,b_1;1)=\frac{b}{6}(b^2+3b_1^2+4\pi^2)\\
    &\mathcal{T}_2(b,b_1;3)=\frac{b}{60}\left[3(b^4+10b^2b_1^2+5b_1^4)+40\pi^2(b^2+3b_1^2)+112\pi^4\right]\\
    &\mathcal{T}_2(b,b_1;5)=\frac{b}{126}  \left(784 \pi ^4 \left(b^2+3 b_1^2\right)+84 \pi ^2 \left(b^4+10 b^2 b_1^2+5 b_1^4\right)+3 \left(b^6+21 b^4 b_1^2+35 b^2 b_1^4+7 b_1^6\right)+1984
   \pi ^6\right)\\
    &\mathcal{T}_2(b,b_1;7)=\frac{b}{360}  \left[39680 \pi ^6 \left(b^2+3 b_1^2\right)+4704 \pi ^4 \left(b^4+10 b^2 b_1^2+5 b_1^4\right)+5 \left(b^6+33 b^4 b_1^2+27 b^2 b_1^4+3 b_1^6\right)
   \left(b^2+3 b_1^2\right)\right. \\
   &\hskip3cm \left. +240 \pi ^2 \left(b^6+21 b^4 b_1^2+35 b^2 b_1^4+7 b_1^6\right)+97536 \pi ^8\right]\\
    &\mathcal{T}_2(b,b_1;9)=\frac{b}{330}  \left[1072896 \pi ^8 \left(b^2+3 b_1^2\right)+130944 \pi ^6 \left(b^4+10 b^2 b_1^2+5 b_1^4\right) \right.\\
    &\hskip3cm\left.+220 \pi ^2 \left(b^6+33 b^4 b_1^2+27 b^2 b_1^4+3
   b_1^6\right) \left(b^2+3 b_1^2\right)+7392 \pi ^4 \left(b^6+21 b^4 b_1^2+35 b^2 b_1^4+7 b_1^6\right)\right.\\
   &\hskip 4cm\left. +3 \left(b^{10}+55 b^8 b_1^2+330 b^6 b_1^4+462 b^4
   b_1^6+165 b^2 b_1^8+11 b_1^{10}\right)+2616320 \pi ^{10}\right]\ .
\end{split}
\end{equation}

\bibliography{Ref.bib,N=2SJTVolumesOfModuliSpaces/extra_references,N=2SJTVolumesOfModuliSpaces/Fredholm_super_JT_gravity1,N=2SJTVolumesOfModuliSpaces/Fredholm_super_JT_gravity2}
\bibliographystyle{JHEP}

\end{document}